\newcommand*{\ATLASLATEXPATH}{./}
\documentclass[cernpreprint,txfonts,USenglish,texlive=2016]{atlasdoc}
\pdfoutput=1
\DeclareOldFontCommand{\it}{\normalfont\itshape}{\mathit}

\usepackage[subfigure]{atlaspackage}
\usepackage{atlasbiblatex}
\usepackage{atlascontribute}

\usepackage[BSM=false,math=false,misc=true,other=false,xref=false,process=true,journal=false,jetetmiss=true]{atlasphysics}

\usepackage{JES2015-defs}

\AtlasTitle{Jet energy scale measurements and their systematic uncertainties in proton--proton collisions at $\sqrt{s} = 13$~\ensuremath{\text{Te\kern -0.1em V}} with the ATLAS detector}
\author{The ATLAS Collaboration}
\AtlasRefCode{PERF-2016-04}
\AtlasNote{ATLAS-COM-PHYS-2016-213}
\PreprintIdNumber{CERN-EP-2017-038}
\AtlasJournal{Phys.\ Rev.\ D.}
\AtlasAbstract{%
Jet energy scale measurements and their systematic uncertainties are reported for jets measured with the ATLAS detector using proton--proton
collision data with a center-of-mass energy of $\sqrt{s} = 13$~\ensuremath{\text{Te\kern -0.1em V}}, corresponding to an integrated luminosity of 3.2~fb$^{-1}$ collected during 2015 at the LHC.
Jets are reconstructed from energy deposits forming topological clusters of calorimeter cells, using the anti-$k_{t}$ algorithm with radius parameter $R = 0.4$.
Jets are calibrated with a series of simulation-based corrections and \textit{in situ} techniques.
\textit{In situ} techniques exploit the transverse momentum balance between a jet and a
reference object such as a photon, $Z$~boson, or multijet system for jets with $20 < p_\text{T} < 2000$~\ensuremath{\text{Ge\kern -0.1em V}} and pseudorapidities of $|\eta|<4.5$, using both data and simulation.
An uncertainty in the jet energy scale of less than 1\% is found in the central calorimeter region ($|\eta| < 1.2$) for jets with $100 < p_\text{T} < 500$~\ensuremath{\text{Ge\kern -0.1em V}}.
An uncertainty of about 4.5\% is found for low-$p_\text{T}$ jets with $p_\text{T} = 20$~\ensuremath{\text{Ge\kern -0.1em V}} in the central region,
dominated by uncertainties in the corrections for multiple proton--proton interactions.
The calibration of forward jets ($|\eta| > 0.8$) is derived from dijet $p_\text{T}$ balance measurements.
For jets of $p_\text{T} = 80$~\ensuremath{\text{Ge\kern -0.1em V}}, the additional uncertainty for the forward jet calibration reaches its largest value of about 2\% in the range $|\eta| > 3.5$ and in a narrow slice of $2.2 < |\eta| < 2.4$.
}

\hypersetup{pdftitle={ATLAS document},pdfauthor={The ATLAS Collaboration}}

\begin{document}

\maketitle

\tableofcontents

\clearpage

\section{Introduction}
\label{sec:introduction}

Jets are a prevalent feature of the final state in high-energy proton--proton (\pp) interactions at CERN's Large Hadron Collider (LHC).
Jets, made of collimated showers of hadrons, are important elements in many Standard Model (SM) measurements and in searches for new phenomena.
They are reconstructed using a clustering algorithm run on a set of input four-vectors, typically obtained from
topologically associated energy deposits, charged-particle tracks, or simulated particles.

This paper details the methods used to calibrate the four-momenta of jets in Monte Carlo (MC) simulation
and in data collected by the ATLAS detector~\cite{PERF-2007-01,PERF-2010-01} at a center-of-mass energy of $\sqrt{s}=13~\TeV$ during the 2015 data-taking period of Run~2 at the LHC.
The jet energy scale (JES) calibration consists of several consecutive stages derived from a combination of MC-based methods and \insitu techniques.
MC-based calibrations correct the reconstructed jet four-momentum to that found from the
simulated stable particles within the jet.
The calibrations account for features of the detector, the jet reconstruction algorithm, jet fragmentation,
and the busy data-taking environment resulting from multiple \pp interactions, referred to as pile-up.
\Insitu techniques are used to measure the difference in jet response between data and MC simulation,
with residual corrections applied to jets in data only.
The 2015 jet calibration builds on procedures developed for the 2011 data~\cite{PERF-2012-01} collected at $\sqrt{s}=7~\TeV$ during Run~1.
Aspects of the jet calibration, particularly those related to pile-up~\cite{PERF-2014-03}, were also developed on 2012 data collected at $\sqrt{s}=8~\TeV$ during Run~1.

This paper is organized as follows. \Secref{detector} describes the ATLAS detector, with an emphasis on the subdetectors relevant for jet reconstruction.
\Secref{reconstruction} describes the jet reconstruction inputs and algorithms, highlighting changes in 2015.
\Secref{dataMC} describes the 2015 data set and the MC generators used in the calibration studies.
\Secref{JES} details the stages of the jet calibration, with particular emphasis on the 2015 \insitu calibrations and their combination.
\Secref{SysUncert} lists the various systematic uncertainties in the JES and describes their combination into a reduced set of nuisance parameters.

\section{The ATLAS detector}
\label{sec:detector}

The ATLAS detector consists of an inner detector tracking system spanning the
pseudorapidity\footnote{The ATLAS reference system is a Cartesian right-handed coordinate system,
with the nominal collision point at the origin. The anticlockwise beam direction defines the positive $z$-axis, while the
positive $x$-axis is defined as pointing from the collision point to the center of the LHC ring and the positive $y$-axis points upwards.
The azimuthal angle $\phi$ is measured around the beam axis, and the polar angle $\theta$ is measured with respect to the $z$-axis.
Pseudorapidity is defined as $\eta = - \ln [\tan(\theta/2)]$, rapidity is defined as $y = 0.5\ \ln[(E + p_\text{z})/(E - p_\text{z})]$,
where $E$ is the energy and $p_\text{z}$ is the $z$-component of the momentum, and transverse energy is defined as $\ET = E \sin \theta$.}
range $|\eta|<2.5$,
sampling electromagnetic and hadronic calorimeters covering the range $|\eta|<4.9$, and a muon spectrometer spanning $|\eta|<2.7$.
A detailed description of the ATLAS experiment can be found in Ref.~\cite{PERF-2007-01}.

Charged-particle tracks are reconstructed in the inner detector (ID), which consists of three subdetectors:
a silicon pixel tracker closest to the beamline, a microstrip silicon tracker, and a straw-tube transition radiation tracker farthest from the beamline.
The ID is surrounded by a thin solenoid providing an axial magnetic field of 2~T, allowing the measurement of charged-particle momenta.
In preparation for Run~2, a new innermost layer of the silicon pixel tracker, the insertable B-layer (IBL)~\cite{ATLAS-TDR-19},
was introduced at a radial distance of 3.3~cm from the beamline to improve track reconstruction, pile-up mitigation, and the identification of jets initiated by $b$-quarks.

The ATLAS calorimeter system consists of inner electromagnetic calorimeters surrounded by hadronic calorimeters.
The calorimeters are segmented in $\eta$ and $\phi$, and each region of the detector has at least three calorimeter
readout layers to allow the measurement of longitudinal shower profiles.
The high-granularity electromagnetic calorimeters use liquid argon (LAr) as the active material with
lead absorbers in both the barrel  ($|\eta|$ < 1.475) and endcap (1.375 < $|\eta|$ < 3.2) regions.
An additional LAr presampler layer in front of the electromagnetic calorimeter within $|\eta|$ < 1.8
measures the energy deposited by particles in the material upstream of the electromagnetic calorimeter.
The hadronic Tile calorimeter incorporates plastic scintillator tiles and steel absorbers in the barrel ($|\eta|$ < 0.8)
and extended barrel (0.8 < $|\eta|$ < 1.7) regions, with photomultiplier tubes (PMT) aggregating signals from a group of neighboring tiles.
Scintillating tiles in the region between the barrel and the extended barrel of the Tile calorimeter serve a similar
purpose to that of the presampler and were extended to increase their area of coverage during the shutdown leading up to Run~2.
A LAr hadronic calorimeter with copper absorbers covers the hadronic endcap region (1.5 < $|\eta|$ < 3.2).
A forward LAr calorimeter with copper and tungsten absorbers covers the forward calorimeter region (3.1 < $|\eta|$ < 4.9).

The analog signals from the LAr detectors are sampled digitally once per bunch crossing over four bunch crossings.
Signals are converted to an energy measurement using an optimal digital filter, calculated from dedicated calibration runs~\cite{Abreu:2010zzc,Cleland:2002rya}.
The signal was previously reconstructed from five bunch crossings in Run~1, but the use of four bunch crossings was found to
provide similar signal reconstruction performance with a reduced bandwidth demand.
The LAr readout is sensitive to signals from the preceding 24 bunch crossings during 25~ns bunch-spacing operation in Run~2.
This is in contrast to the 12 bunch-crossing sensitivity during 50~ns operation in Run~1,
increasing the sensitivity to out-of-time pile-up from collisions in the preceding bunch crossings.
The LAr signals are shaped~\cite{Abreu:2010zzc} to reduce the measurement sensitivity to pile-up,
with the shaping optimized for the busier pile-up conditions at 25~ns.
In contrast, the fast readout of the Tile calorimeter~\cite{TCAL-2010-01} reduces the signal sensitivity
to out-of-time pile-up from collisions in neighboring bunch crossings.

The muon spectrometer (MS)~\cite{PERF-2007-01} surrounds the ATLAS calorimeters and measures muon tracks within $|\eta|$ < 2.7
using three layers of precision tracking chambers and dedicated trigger chambers.
A system of three superconducting air-core toroidal magnets provides a magnetic field for measuring muon momenta.

The ATLAS trigger system begins with a hardware-based Level 1 (L1) trigger followed by a
software-based high-level trigger (HLT)~\cite{TRIG-2016-01}.
The L1 trigger is designed to accept events at an average 100~kHz rate, and accepted a peak rate of 70~kHz in 2015.
The HLT is designed to accept events that are written out to disk at an average rate of 1~kHz and reached a peak rate of 1.4~kHz in 2015.
For the trigger, jet candidates are constructed from coarse calorimeter towers using a sliding-window algorithm at L1, and are fully reconstructed in the HLT.
Electrons and photons are triggered in the pseudorapidity range $|\eta|$ < 2.5, where the electromagnetic calorimeter is finely segmented
and track reconstruction is available.
Compact electromagnetic energy deposits triggered at L1 are used as the seeds for the HLT algorithms,
which are designed to identify electrons based on calorimeter and fast track reconstruction.
The muon trigger at L1 is based on a coincidence of trigger chamber layers.
The parameters of muon candidate tracks are then derived in the HLT by fast reconstruction algorithms in both the ID and MS.
Events used in the jet calibration are selected from regions of kinematic phase space where the relevant triggers are fully efficient.

\section{Jet reconstruction}
\label{sec:reconstruction}
The calorimeter jets used in the following studies are reconstructed at the electromagnetic energy scale (EM scale) with the \antikt algorithm~\cite{antikt}
and radius parameter $R = 0.4$ using the $\textsc{FastJet}$ 2.4.3 software package~\cite{fastjet}.
A collection of three-dimensional, massless, positive-energy topological clusters (topo-clusters)~\cite{PERF-2014-07,PERF-2011-03}
made of calorimeter cell energies are used as input to the \antikt algorithm.
Topo-clusters are built from neighboring calorimeter cells containing a significant energy above a noise threshold that is
estimated from measurements of calorimeter electronic noise and simulated pile-up noise.
The calorimeter cell energies are measured at the EM scale, corresponding to the energy deposited by electromagnetically interacting particles.
Jets are reconstructed with the \antikt algorithm if they pass a \pt threshold of $7~\GeV$.

In 2015 the simulated noise levels used in the calibration of the topo-cluster reconstruction algorithm were updated
using observations from Run~1 data and accounting for different data-taking conditions in 2015.
This results in an increase in the simulated noise at the level of 10\% with respect to the Run~1 simulation
in the barrel region of the detector, and a slightly larger increase in the forward region~\cite{PERF-2014-03}.
The noise thresholds of the topo-cluster reconstruction were increased accordingly.
The topo-cluster reconstruction algorithm was also improved in 2015,
with topo-clusters now forbidden from being seeded by the presampler layers.
This restricts jet formation from low-energy pile-up depositions that do not penetrate the calorimeters.

Jets referred to as \emph{truth} jets are reconstructed using the \antikt algorithm with $R = 0.4$ using stable, final-state particles from MC generators as input.
Candidate particles are required to have a lifetime of c$\tau > 10$~mm and muons, neutrinos, and particles from pile-up activity are excluded.
Truth jets are therefore defined as being measured at the particle-level energy scale.
Truth jets with $\pt > 7~\GeV$ and $\abseta < 4.5$ are used in studies of jet calibration using MC simulation.
Reconstructed calorimeter jets are geometrically matched to truth jets using the distance measurement\footnote{
The distance between two four-vectors is defined as $\DeltaR = \sqrt{(\Delta\eta)^{2}+(\Delta\phi)^{2}}$,
where $\Delta\eta$ is their distance in pseudorapidity and $\Delta\phi$ is their azimuthal distance.
The distance with respect to a jet is calculated from its principal axis.} $\DeltaR$.

Tracks from charged particles used in the jet calibration are reconstructed within the full acceptance of the ID ($|\eta|<2.5$).
The track reconstruction was updated in 2015 to include the IBL and uses a neural network clustering algorithm~\cite{PERF-2012-05},
improving the separation of nearby tracks and the reconstruction performance in the high-luminosity conditions of Run~2.
Reconstructed tracks are required to have a $\pt > 500~\MeV$ and to be associated with the hard-scatter vertex,
defined as the primary vertex with at least two associated tracks and the largest $\pt^2$ sum of associated tracks.
Tracks must satisfy quality criteria based on the number of hits in the ID subdetectors.
Tracks are assigned to jets using ghost association~\cite{Cacciari:2007fd},
a procedure that treats them as four-vectors of infinitesimal magnitude during the jet reconstruction
and assigns them to the jet with which they are clustered.

Muon track segments are used in the jet calibration as a proxy for the uncaptured jet energy
carried by energetic particles passing through the calorimeters without being fully absorbed.
The segments are partial tracks constructed from hits in the MS~\cite{PERF-2015-10} which serve as inputs to fully reconstructed tracks.
Segments are assigned to jets using the method of ghost association described above for tracks,
with each segment treated as an input four-vector of infinitesimal magnitude to the jet reconstruction.

\section{Data and Monte Carlo simulation}
\label{sec:dataMC}

Several MC generators are used to simulate \pp collisions for the various jet calibration stages and for estimating systematic uncertainties in the JES.
A sample of dijet events is simulated at next-to-leading-order (NLO) accuracy in perturbative QCD using \POWHEGBOX~2.0~\cite{Nason:2004rx,Frixione:2007vw,Alioli:2010xd}.
The hard scatter is simulated with a $2\rightarrow3$ matrix element that is interfaced with the CT10 parton distribution function (PDF) set~\cite{CT10}.
The dijet events are showered in \PYTHIAV{8.186}~\cite{pythia8}, with additional radiation simulated to the
leading-logarithmic approximation through \pt-ordered parton showers~\cite{Corke2010}.
The simulation parameters of the underlying event, parton showering, and hadronization are set according to the A14 event tune~\cite{ATL-PHYS-PUB-2014-021}.
For \insitu analyses, samples of $Z$ bosons with jets (\Zjet) are similarly produced
with \POWPYTHIA using the CT10 PDF set and the AZPHINLO event tune~\cite{ATL-PHYS-PUB-2013-017}.
Samples of multijets and of photons with jets (\gjet) are generated in \PYTHIA,
with the $2\rightarrow2$ matrix element convolved with the NNPDF2.3LO PDF set~\cite{NNPDF23},
and using the A14 event tune.

For studies of the systematic uncertainties, the \SHERPA~2.1~\cite{sherpa} generator is used to simulate all relevant processes in dijet, \Zjet, and \gjet events.
\SHERPA uses multileg $2\rightarrow N$ matrix elements that are matched to parton showers following the CKKW~\cite{Catani:2001cc} prescription.
The CT10 PDF set and default \SHERPA event tune are used.
The multijet systematic uncertainties are studied using the \HERWIGpp~2.7~\cite{Bahr:2008pv,herwig} generator,
with the $2\rightarrow2$ matrix element convolved with the CTEQ6L1 PDF set~\cite{Pumplin:2002vw}.
\HERWIGpp simulates additional radiation through angle-ordered parton showers, and is configured with the UE-EE-5 event tune~\cite{Gieseke:2012ft}.

Pile-up interactions can occur within the bunch crossing of interest (in-time) or in neighboring bunch crossings (out-of-time), altering
the measured energy of a hard-scatter jet or leading to the reconstruction of additional, spurious jets.
Pile-up effects are modeled using \PYTHIA, simulated with underlying-event characteristics using the NNPDF2.3LO PDF set and A14 event tune.
A number of these interactions are overlaid onto each hard-scatter event following a Poisson distribution about the mean number of
additional \pp collisions per bunch crossing ($\mu$) of the event.
The value of $\mu$ is proportional to the predicted instantaneous luminosity assigned to the MC event.
It is simulated according to the expected distribution in the 2015 data-taking period and subsequently reweighted to the measured distribution.
Events are overlaid both in-time with the simulated hard scatter and out-of-time for nearby bunches.
The number of in-time and out-of-time pile-up interactions associated with an event is correlated with the number of reconstructed primary vertices ($\Npv$) and with $\mu$,
respectively, providing a method for estimating the per-event pile-up contribution.

Generated events are propagated through a full simulation~\cite{SOFT-2010-01} of the ATLAS detector
based on Geant4~\cite{GEANT4} which describes the interactions of the particles with the detector.
Hadronic showers are simulated with the FTFP BERT model, whereas the QGSP BERT model was used in Run~1~\cite{PERF-2015-05}.
A parameterized simulation of the ATLAS calorimeter called Atlfast-II (AFII)~\cite{SOFT-2010-01}
is used for faster MC production, and a dedicated MC-based calibration is derived for AFII samples.

The data set used in this study consists of \intLumi of \pp collisions
collected by ATLAS between August and December of 2015 with all subdetectors operational.
The LHC was operated at $\sqrt{s}=13~\TeV$, with bunch crossing intervals of 25~ns.
The mean number of interactions per bunch crossing was estimated through luminosity measurements~\cite{DAPR-2013-01} to be on average $\avg{\mu}=13.7$.
The specific trigger requirements and object selections vary among the \insitu analyses and are described in the relevant sections.

\section{Jet energy scale calibration}
\label{sec:JES}

\begin{figure}[!ht]
\begin{center}
  \includegraphics[width=0.9\textwidth]{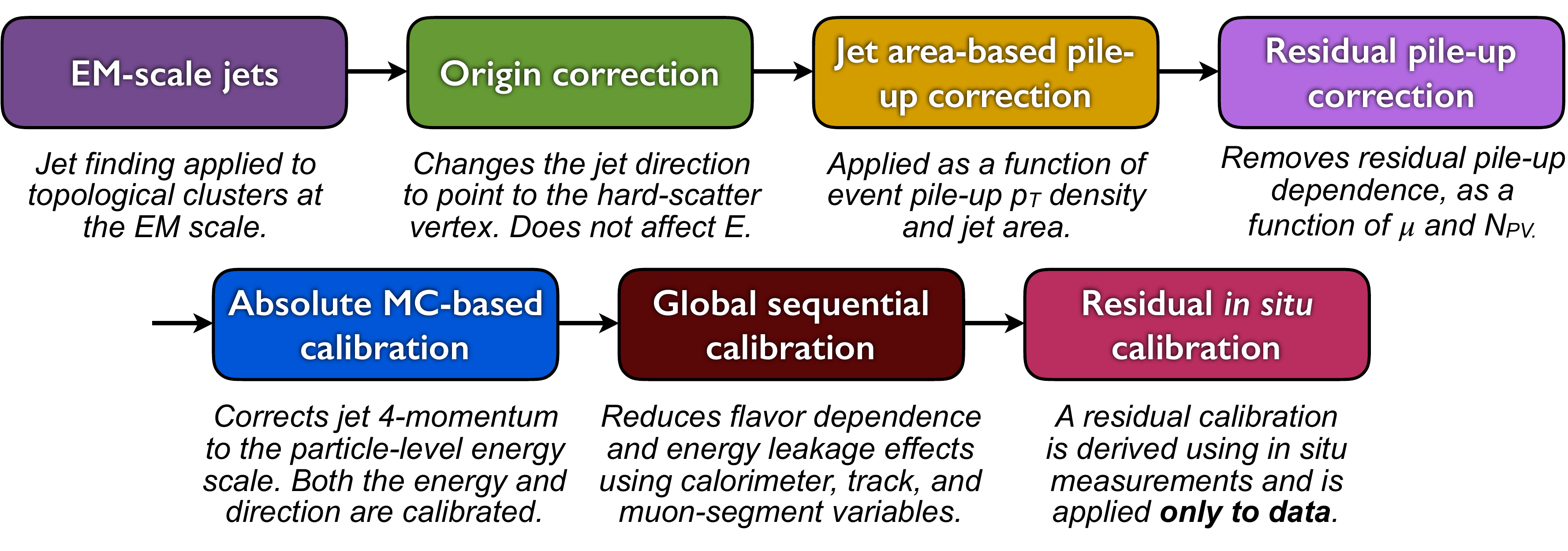}
  \caption{Calibration stages for EM-scale jets. Other than the origin correction, each stage of the calibration is applied to the four-momentum of the jet.}
  \label{fig:IntroCartoon}
\end{center}
\end{figure}

Figure~\ref{fig:IntroCartoon} presents an overview of the 2015 ATLAS calibration scheme for EM-scale calorimeter jets.
This calibration restores the jet energy scale to that of
truth jets reconstructed at the particle-level energy scale.
Each stage of the calibration corrects the full four-momentum unless otherwise stated, scaling the jet \pt, energy, and mass.

First, the origin correction recalculates the four-momentum of jets to point to the
hard-scatter primary vertex rather than the center of the detector, while keeping the jet energy constant.
This correction improves the $\eta$ resolution of jets, as measured from the difference between
reconstructed jets and truth jets in MC simulation.
The $\eta$ resolution improves from roughly 0.06 to 0.045 at a jet \pt of 20~\GeV\
and from 0.03 to below 0.006 above 200~\GeV.
The origin correction procedure in 2015 is identical to that used in the 2011 calibration~\cite{PERF-2012-01}.

Next, the pile-up correction removes the excess energy due to in-time and out-of-time pile-up.
It consists of two components; an area-based \pt density subtraction~\cite{Cacciari:2007fd}, applied at the per-event level,
and a residual correction derived from the MC simulation, both detailed in \secref{pileup}.
The \emph{absolute} JES calibration corrects the jet four-momentum to the particle-level energy scale, as
derived using truth jets in dijet MC events, and is discussed in \secref{etaJES}.
Further improvements to the reconstructed energy and related uncertainties
are achieved through the use of calorimeter, MS, and track-based variables in the global sequential calibration,
as discussed in \secref{GSC}.
Finally, a residual \insitu calibration is applied to correct jets in data using well-measured reference objects,
including photons, $Z$ bosons, and calibrated jets, as discussed in \secref{insitu}.
The full treatment and reduction of the systematic uncertainties is discussed in \secref{SysUncert}.

\subsection{Pile-up corrections}
\label{sec:pileup}

The pile-up contribution to the JES in the 2015 data-taking environment differs in several ways from Run~1.
The larger center-of-mass energy affects the jet \pt dependence on pile-up-sensitive variables,
while the switch from 50 to 25~ns bunch spacing increases the amount of out-of-time pile-up.
In addition, the higher topo-clustering noise thresholds alter the impact of pile-up on the JES.
The pile-up correction is therefore evaluated using updated MC simulations of the 2015 detector and beam conditions.
The pile-up correction in 2015 is derived using the same methods developed in 2012~\cite{PERF-2014-03}, summarized in the following paragraphs.

First, an area-based method subtracts the per-event pile-up contribution to the \pt of each jet according to its area.
The pile-up contribution is calculated from the median \pt density $\rho$ of jets in the $\eta$--$\phi$ plane.
The calculation of $\rho$ uses only positive-energy topo-clusters with $\abs\eta < 2$ that are clustered using the \kt algorithm~\cite{antikt,Ellis:1993tq}
with radius parameter $R=0.4$.
The \kt algorithm is chosen for its sensitivity to soft radiation, and is only used in the area-based method.
The central $\abs\eta$ selection is necessitated by the higher calorimeter occupancy in the forward region.
The \pt density of each jet is taken to be $\pt / A$, where the area $A$ of a jet
is calculated using ghost association. In this procedure, simulated ghost particles of infinitesimal momentum are
added uniformly in solid angle to the event before jet reconstruction.
The area of a jet is then measured from the relative number of ghost particles associated with a jet after clustering.
The median of the \pt density is used for $\rho$ to reduce the bias from hard-scatter jets which populate the high-\pt tails of the distribution.

\begin{figure}[!ht]
\begin{center}
  \includegraphics[width=0.47\textwidth]{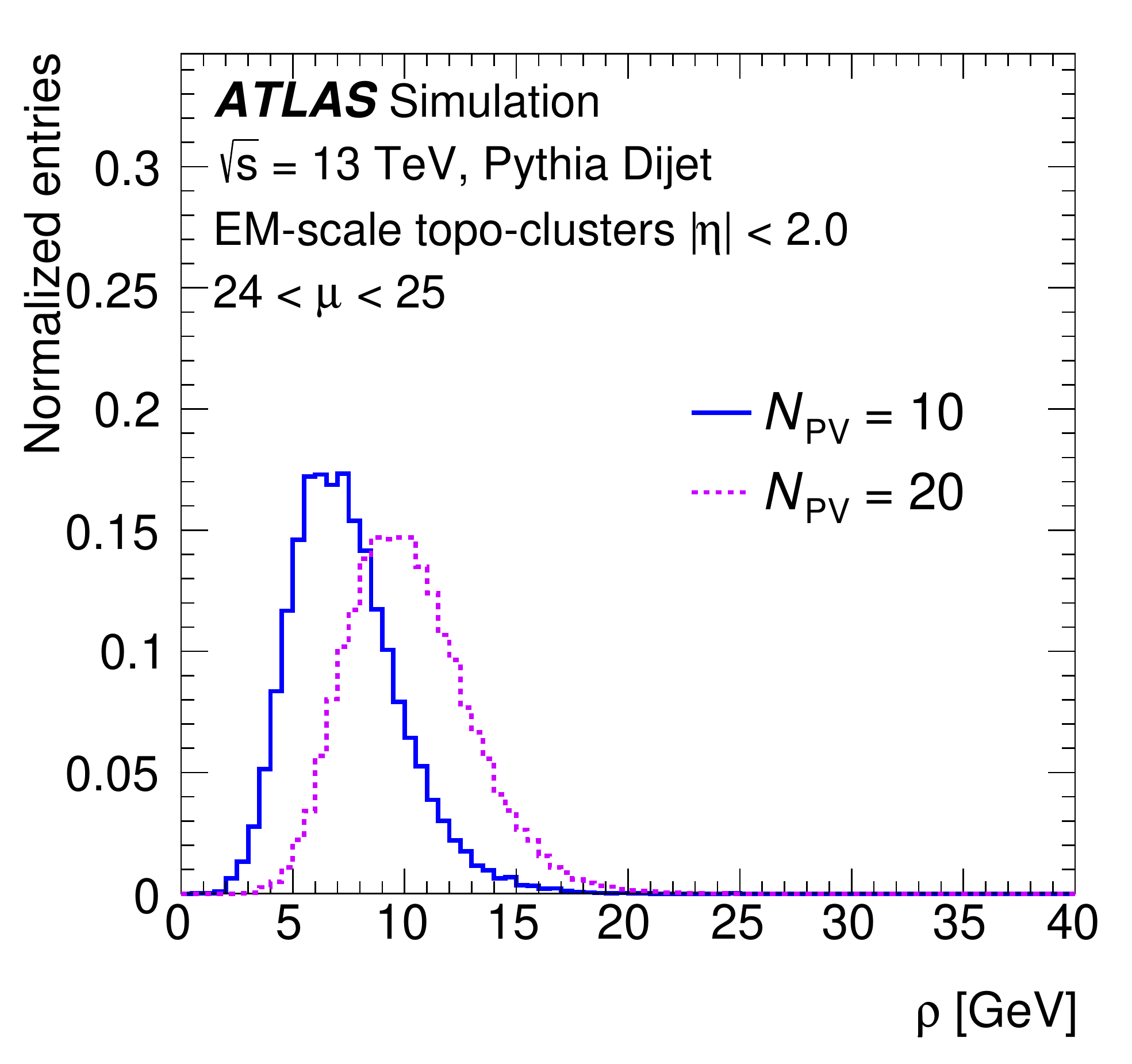}
  \caption{Per-event median \pt density, $\rho$, at $\Npv=10$ (solid) and $\Npv=20$ (dotted) for $24 < \mu < 25$ as found in MC simulation.
  }
  \label{fig:Density}
\end{center}
\end{figure}

The $\rho$ distribution of events with a given $\Npv$
is shown for MC simulation in \figref{Density}, and has roughly the same magnitude at 13~\TeV as seen at 8~\TeV.
At 13~\TeV the increase in the center-of-mass energy is offset by the higher noise
thresholds and the larger out-of-time pile-up, the latter reducing the average energy readout of any given cell
due to the inherent pile-up suppression of the bipolar shaping of LAr signals~\cite{Abreu:2010zzc}.
The ratio of the $\rho$-subtracted jet \pt to the uncorrected jet \pt is taken as a correction factor applied to the jet four-momentum,
and does not affect the jet $\eta$ and $\phi$ coordinates.

The $\rho$ calculation is derived from the central, lower-occupancy regions of the calorimeter,
and does not fully describe the pile-up sensitivity in the forward calorimeter region or in the higher-occupancy core of high-\pt jets.
It is therefore observed that after this correction some dependence of the \antikt jet \pt on the amount of pile-up remains,
and an additional residual correction is derived.
A dependence is seen on \Npv, sensitive to in-time pile-up, and $\mu$, sensitive to out-of-time pile-up.
The residual \pt dependence is measured as the difference between the reconstructed jet \pt and truth jet \pt, with the latter
being insensitive to pile-up.
Reconstructed jets with $\pt > 10~\GeV$ are geometrically matched to truth jets within $\DeltaR = 0.3$.

The residual \pt dependence on \Npv ($\alpha$) and on $\mu$ ($\beta$) are
observed to be fairly linear and independent of one another, as was found in 2012 MC simulation.
Linear fits are used to derive the initial $\alpha$ and $\beta$ coefficients separately in bins of $\pttrue$ and $\abs\eta$.
Both the $\alpha$ and $\beta$ coefficients are then seen to have a logarithmic dependence on $\pttrue$, and logarithmic fits are performed in the range $20 < \pttrue < 200~\GeV$ for each bin of $\abs\eta$.
In each $\abs\eta$ bin, the fitted value at $\pttrue = 25~\GeV$ is taken as the nominal $\alpha$ and $\beta$ coefficients,
reflecting the dependence in the \pt region where pile-up is most relevant.
The logarithmic fits over the full \pttrue range are used for a \pt-dependent systematic uncertainty in the residual pile-up dependence.
Finally, linear fits are performed to the binned coefficients as a function of \abs\eta in 4 regions,
$\abs\eta < 1.2$, $1.2<\abs\eta<2.2$, $2.2<\abs\eta<2.8$, and $2.8<\abs\eta<4.5$.
This reduces the effects of statistical fluctuations and allows the $\alpha$ and $\beta$ coefficients to be
smoothly sampled in \abs\eta, particularly in regions of varying dependence.
The pile-up-corrected \pt, after the area-based and residual corrections, is given by

\begin{equation*}
  \pt^\text{corr} = \pt^\text{reco} - \rho \times A - \alpha \times \left( \Npv - 1 \right) - \beta \times \mu,
\label{eq:residualPileup}
\end{equation*}
where $\pt^\text{reco}$ refers to the EM-scale \pt of the reconstructed jet before any pile-up corrections are applied.

\begin{figure}[!ht]
\begin{center}
  \subfigure[In-time pile-up dependence]{
    \label{fig:Area:EM_in}
    \includegraphics[width=0.47\textwidth]{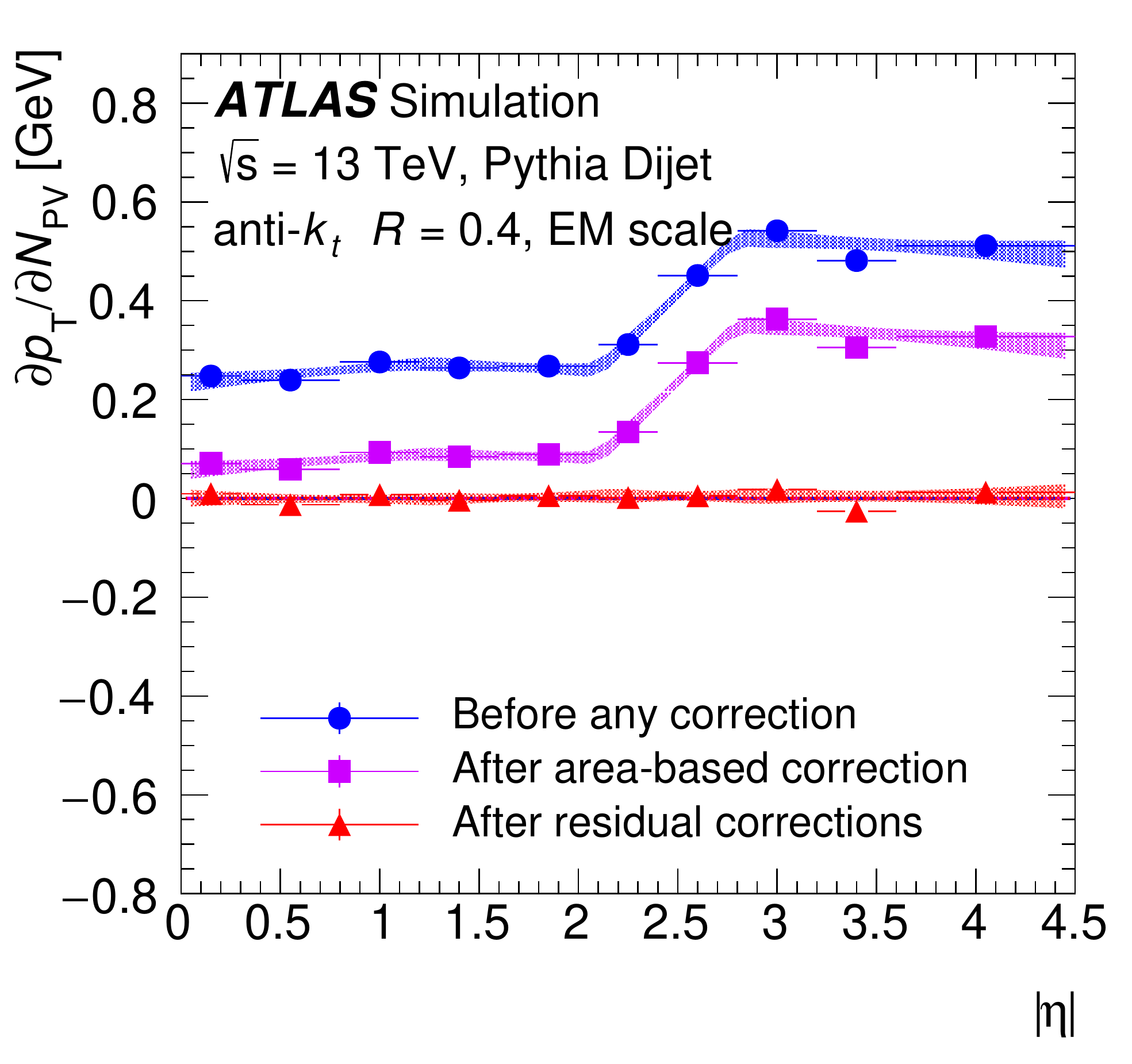}
  }
  \subfigure[Out-of-time pile-up dependence]{
    \label{fig:Area:EM_out}
    \includegraphics[width=0.47\textwidth]{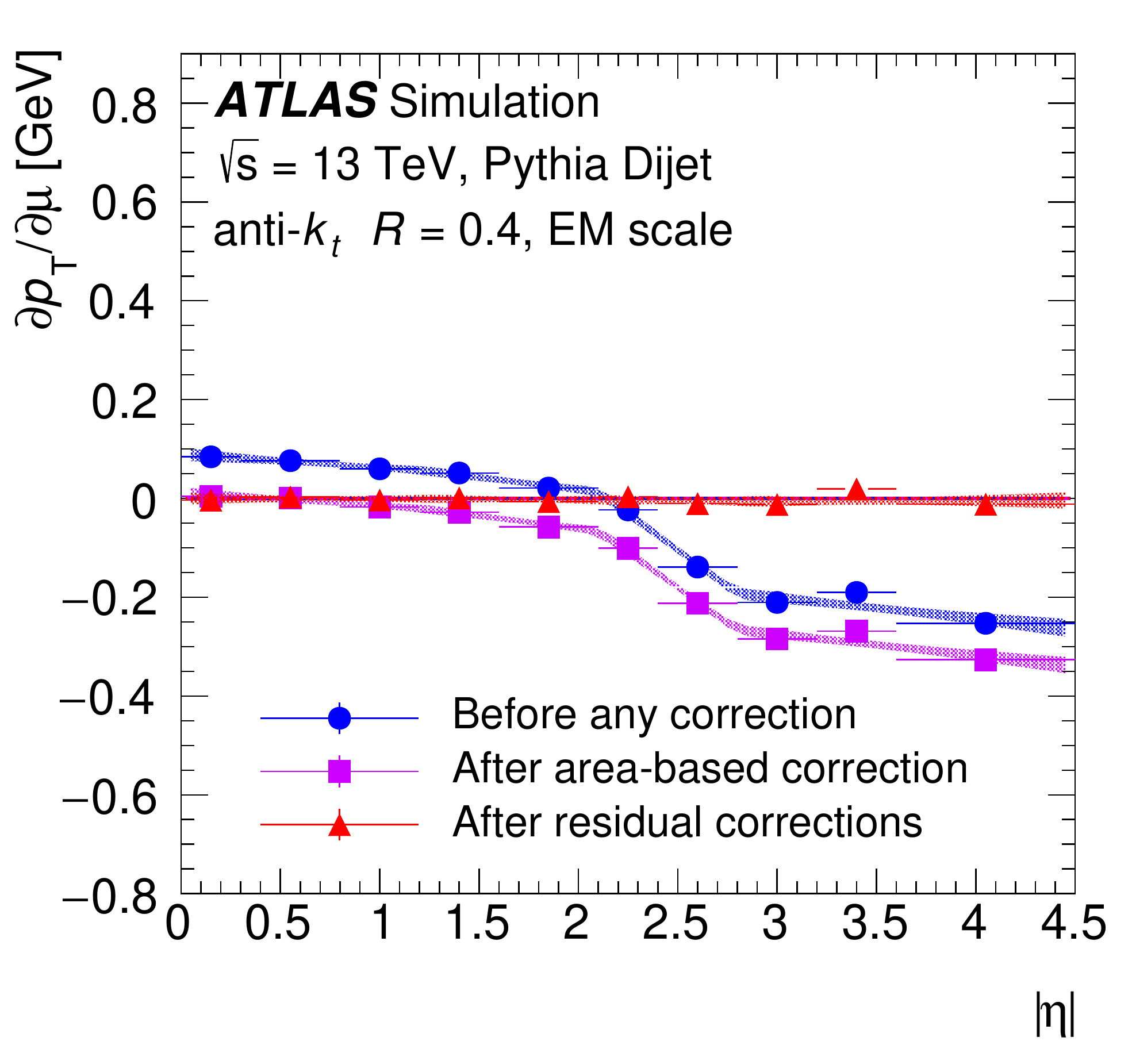}
  }
  \caption{ Dependence of EM-scale \antikt jet \pt on \subref{fig:Area:EM_in} in-time pile-up (\Npv averaged over $\mu$) and
    \subref{fig:Area:EM_out} out-of-time pile-up ($\mu$ averaged over \Npv) as a function of $\abs\eta$ for $\pttrue = 25~\GeV$.
    The dependence is shown in bins of \abs\eta before pile-up corrections (blue circle), after the area-based correction (violet square),
    and after the residual correction (red triangle).
    The shaded bands represent the 68\% confidence intervals of the linear fits in 4 regions of \abs\eta.
    The values of the fitted dependence on in-time and out-of-time pile-up after the area-based correction (purple shaded band) are taken as the residual correction factors $\alpha$ and $\beta$, respectively.
  \label{fig:Area}}
\end{center}
\end{figure}

The dependence of the area-based and residual corrections on \Npv and $\mu$
are shown as a function of \abs\eta in \figref{Area}.
The shape of the residual correction is comparable to that found in 2012 MC simulation, except in the forward
region ($\abs\eta > 2.5$) of \figref{Area:EM_in}, where it is found to be larger by 0.2~\GeV.
This difference in the in-time pile-up term is primarily caused
by higher topo-cluster noise thresholds, which are more consequential in the forward region.

Two \insitu validation studies are performed and no statistically significant difference is observed in the jet \pt dependence on \Npv or $\mu$ between 2015 data and MC simulation.
Four systematic uncertainties are introduced to account for MC mismodeling of \Npv, $\mu$, and the $\rho$ topology,
as well as the \pt dependence of the \Npv and $\mu$ terms used in the residual pile-up correction.
The $\rho$ topology uncertainty encapsulates the uncertainty in the underlying event contribution to $\rho$
through the use of several distinct MC event generators and final-state topologies.
The uncertainties in the modeling of \Npv and $\mu$ are taken as the difference between MC simulation and data in the \insitu validation studies.
The \pt-dependent uncertainty in the residual pile-up dependence is derived from the full logarithmic fits to $\alpha$ and $\beta$.
Both the \insitu validation studies and the systematic uncertainties are described in detail in Ref.~\cite{PERF-2014-03}.

\subsection{Jet energy scale and $\eta$ calibration}
\label{sec:etaJES}

The \emph{absolute} jet energy scale and $\eta$ calibration corrects the reconstructed jet four-momentum to the particle-level
energy scale and accounts for biases in the jet $\eta$ reconstruction.
Such biases are primarily caused by the transition between different calorimeter technologies and
sudden changes in calorimeter granularity.
The calibration is derived from the \PYTHIA MC sample using reconstructed jets after the application of the
origin and pile-up corrections.
The JES calibration is derived first as a correction of the reconstructed jet energy
to the truth jet energy~\cite{PERF-2012-01}.
Reconstructed jets are geometrically matched to truth jets within $\DeltaR = 0.3$.
Only isolated jets are used, to avoid any ambiguities in the matching of calorimeter jets to truth jets.
An isolated calorimeter jet is required to have no other calorimeter jet of $\pt > 7~\GeV$ within $\DeltaR = 0.6$,
and only one truth jet of $\pttruth > 7~\GeV$ within $\DeltaR = 1.0$.

\begin{figure}[!ht]
\begin{center}
  \subfigure[]{
    \label{fig:EtaJES:FullSim}
    \includegraphics[width=0.48\textwidth]{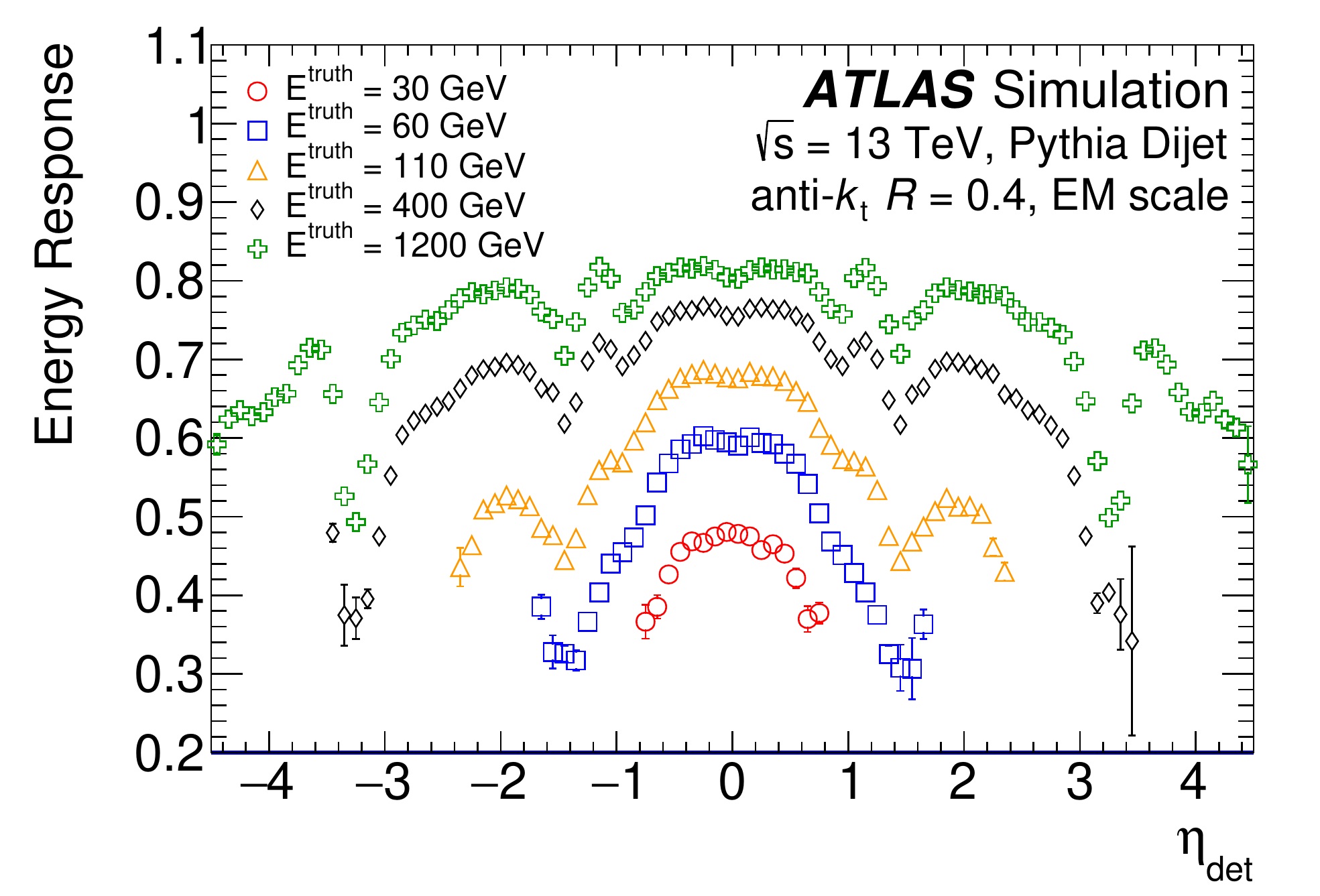}
  }
  \subfigure[]{
    \label{fig:EtaJES:Residual}
    \includegraphics[width=0.48\textwidth]{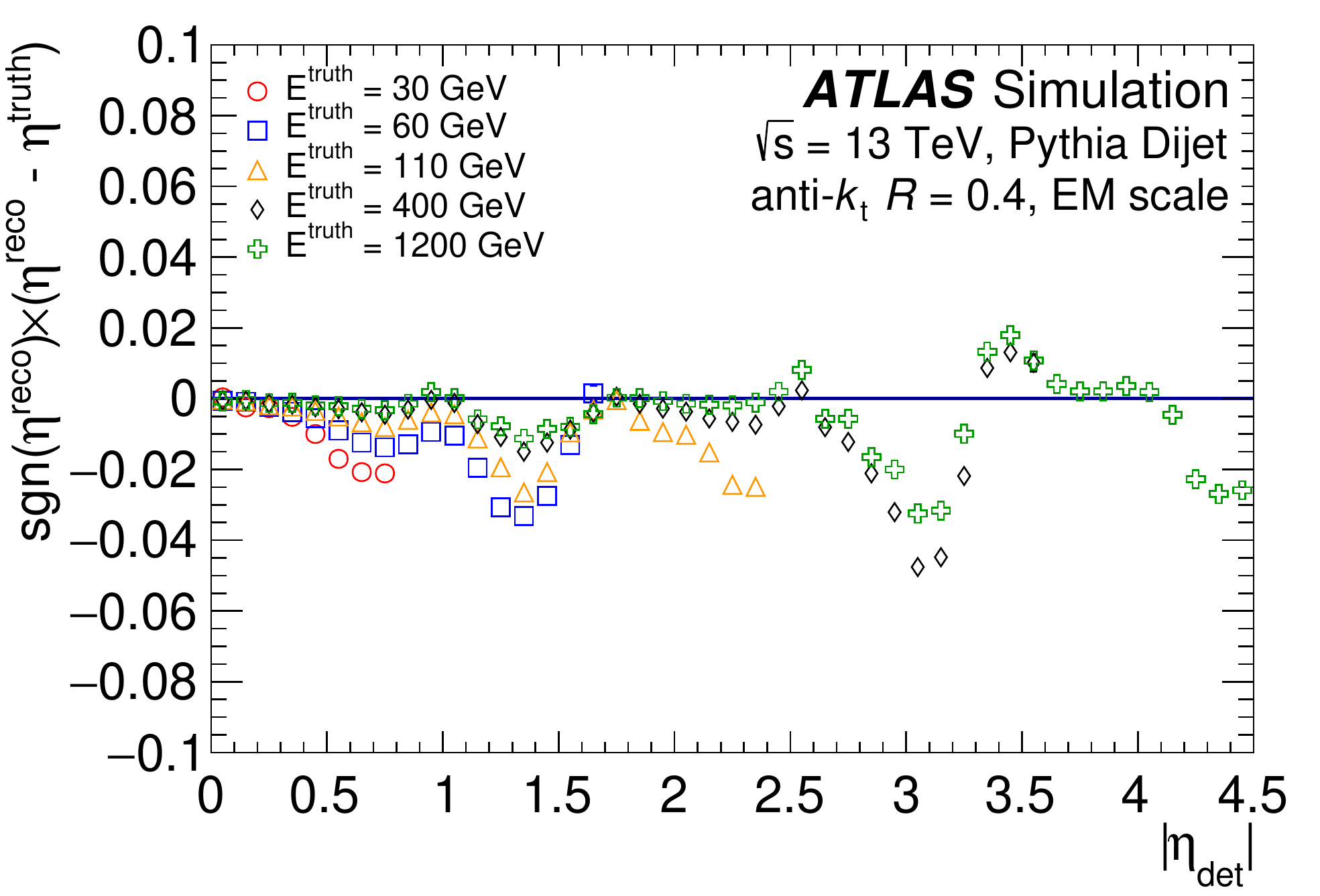}
  }
  \caption{\subref{fig:EtaJES:FullSim} The average energy response as a function of \etadet for jets of a truth energy of 30, 60, 110, 400, and 1200~\GeV.
    The energy response is shown after origin and pile-up corrections are applied.
    \subref{fig:EtaJES:Residual} The signed difference between the truth jet $\eta^\text{truth}$ and the reconstructed jet $\eta^\text{reco}$ due to biases in the jet reconstruction.
    This bias is addressed with an $\eta$ correction applied as a function of \etadet.
  \label{fig:EtaJES}}
\end{center}
\end{figure}

The average energy response is defined as the mean of a Gaussian fit to the core of the
$E^\text{reco}/\Etruth$ distribution for jets, binned in $\Etruth$ and $\etadet$.
The response is derived as a function of $\etadet$, the jet $\eta$ pointing from the geometric center of the detector,
to remove any ambiguity as to which region of the detector is measuring the jet.
The response in the full ATLAS simulation is shown in \figref{EtaJES:FullSim}.
Gaps and transitions between calorimeter subdetectors result in a lower energy response
due to absorbed or undetected particles, evident when parameterized by \etadet.
A numerical inversion procedure is used to derive corrections in $E^\text{reco}$ from \Etruth, as detailed in Ref.~\cite{PERF-2011-03}.
The average response is parameterized as a function of $E^\text{reco}$ and the jet calibration factor is taken as the inverse of the average energy response.
Good closure of the JES calibration is seen across the entire $\eta$ range, compatible with that seen in the 2011 calibration.
As in 2011, a small non-closure on the order of a few percent is seen for low-\pt jets due to
a slightly non-Gaussian energy response and jet reconstruction threshold effects, both of which impact the response fits.

A bias is seen in the reconstructed jet $\eta$, shown in \figref{EtaJES:Residual} as a function of \abs\etadet.
It is largest in jets that encompass two calorimeter regions with different energy responses caused by changes in calorimeter geometry or technology.
This artificially increases the energy of one side of the jet with respect to the other, altering the reconstructed four-momentum.
The barrel--endcap ($\abs\etadet \sim 1.4$) and endcap--forward ($\abs\etadet \sim 3.1$) transition regions can be clearly seen in \figref{EtaJES:Residual} as susceptible to this effect.
A second correction is therefore derived as the difference between the reconstructed $\eta^\text{reco}$ and truth $\eta^\text{truth}$, parameterized as a function of \Etruth and \etadet.
A numerical inversion procedure is again used to derive corrections in $E^\text{reco}$ from \Etruth.
Unlike the other calibration stages, the $\eta$ calibration alters only the jet $\pt$ and $\eta$, not the full four-momentum.
Jets calibrated with the full jet energy scale and $\eta$ calibration are considered to be at the EM+JES.

An \emph{absolute} JES and $\eta$ calibration is also derived for fast simulation samples using the same methods with a \PYTHIA MC sample simulated with AFII.
An additional JES uncertainty is introduced for AFII samples to account for a small non-closure in the calibration, particularly beyond $\abs\eta \sim 3.2$, due to the approximate treatment of hadronic showers in the forward calorimeters.
This uncertainty is about 1\% at a jet \pt of 20~\GeV and falls rapidly with increasing \pt.

\subsection{Global sequential calibration}
\label{sec:GSC}

Following the previous jet calibrations, residual dependencies of the JES on longitudinal and transverse features of the jet are observed.
The calorimeter response and the jet reconstruction are sensitive to fluctuations in the jet particle composition and the distribution of energy within the jet.
The average particle composition and shower shape of a jet varies between initiating particles, most notably between quark- and gluon-initiated jets.
A quark-initiated jet will often include hadrons with a higher fraction of the jet \pt that penetrate further into the calorimeter,
while a gluon-initiated jet will typically contain more particles of softer \pt, leading to a lower calorimeter response and a wider transverse profile.
Five observables are identified that improve the resolution of the JES through the
global sequential calibration (GSC), a procedure explored in the 2011 calibration~\cite{PERF-2011-03}.

For each observable, an independent jet four-momentum correction is derived as a function of \pttruth and \abs\etadet by inverting the reconstructed jet response in MC events.
Both the numerical inversion procedure and the method to geometrically match reconstructed jets to truth jets are outlined in \secref{etaJES}.
An overall constant is multiplied to each numerical inversion to ensure the average energy is unchanged at each stage.
The effect of each correction is therefore to remove the dependence of the jet response on each observable
while conserving the overall energy scale at the  EM+JES.
Corrections for each observable are applied independently and sequentially to the jet four-momentum, neglecting correlations between observables.
No improvement in resolution was found from including such correlations or altering the sequence of the corrections.

The five stages of the GSC account for the dependence of the jet response on (in order):

\begin{enumerate}
  \item $f_\text{Tile0}$, the fraction of jet energy measured in the first layer of the hadronic Tile calorimeter ($\abs\etadet < 1.7$);
  \item $f_\text{LAr3}$, the fraction of jet energy measured in the third layer of the electromagnetic LAr calorimeter ($\abs\etadet < 3.5$);
  \item $n_\text{trk}$, the number of tracks with $\pt > 1~\GeV$ ghost-associated with the jet ($\abs\etadet < 2.5$);
  \item $\mathcal{W}_\text{trk}$, the average \pt-weighted transverse distance in the $\eta$--$\phi$ plane between the jet axis
    and all tracks of $\pt > 1~\GeV$ ghost-associated to the jet ($\abs\etadet < 2.5$);
  \item \Nseg, the number of muon track segments ghost-associated with the jet ($\abs\etadet < 2.7$).
\end{enumerate}
The \Nseg correction reduces the tails of the response distribution caused by high-\pt jets that are not fully contained
in the calorimeter, referred to as punch-through jets.
The first four corrections are derived as a function of jet \pt, while the punch-through correction is derived as a function
of jet energy, being more correlated with the energy escaping the calorimeters.

The underlying distributions of these five observables are fairly well modeled by MC simulation.
Slight differences with data have a negligible impact on the GSC as long as the dependence of the average jet response on the observables is well modeled in MC simulation.
This average response dependence was tested using the dijet tag-and-probe method developed in 2011 and detailed in Section~12.1 of Ref.~\cite{PERF-2011-03}.
The average \pt asymmetry between back-to-back jets was again measured in 2015 data as a function of each observable and found to be compatible between data and MC simulation,
with differences small compared to the size of the proposed corrections.

\begin{figure}[!h]
\begin{center}
  \subfigure[]{
    \label{fig:GSC:Tile}
    \includegraphics[width=0.31\textwidth]{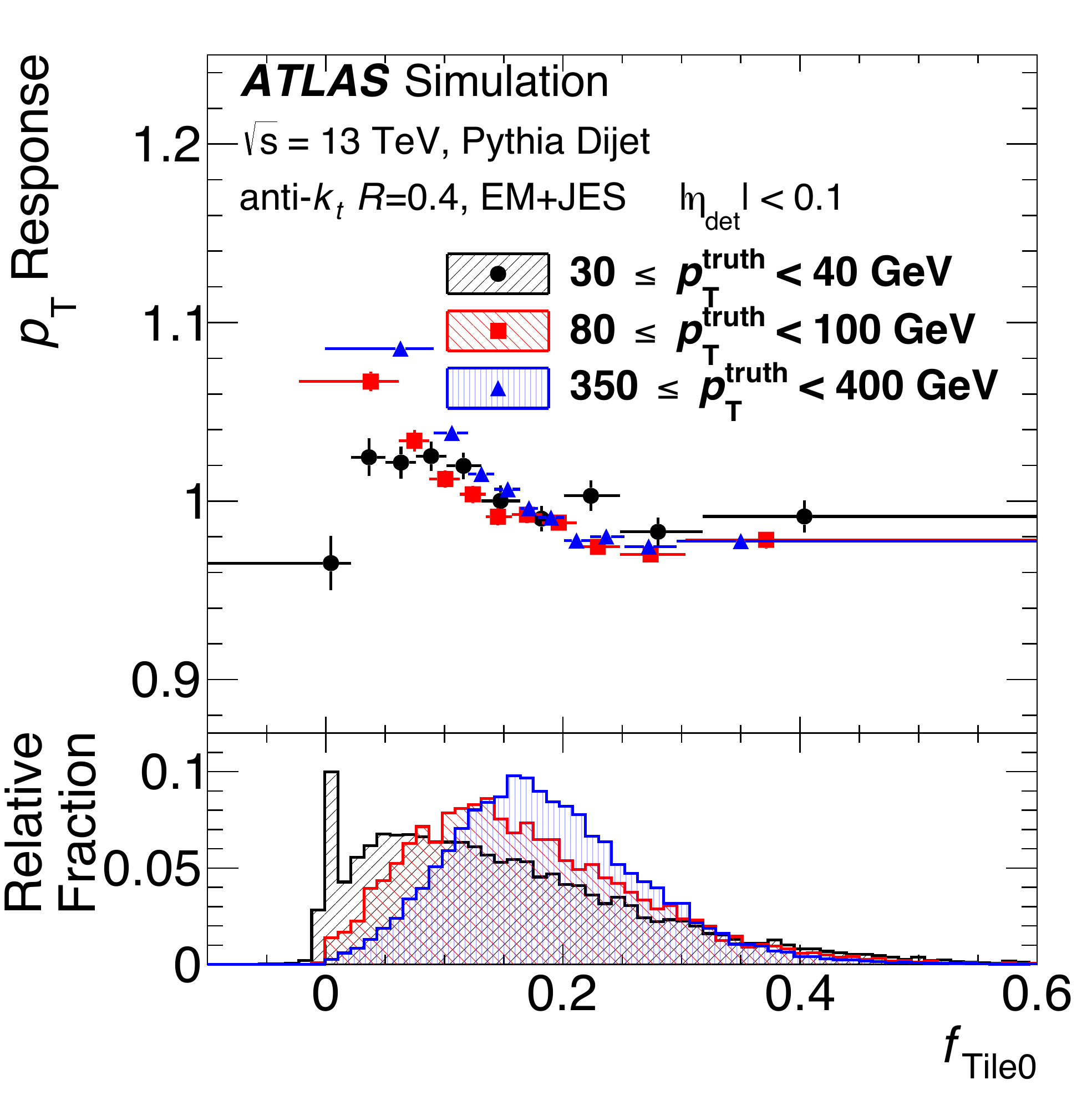}
  }
  \subfigure[]{
    \label{fig:GSC:LAr}
    \includegraphics[width=0.31\textwidth]{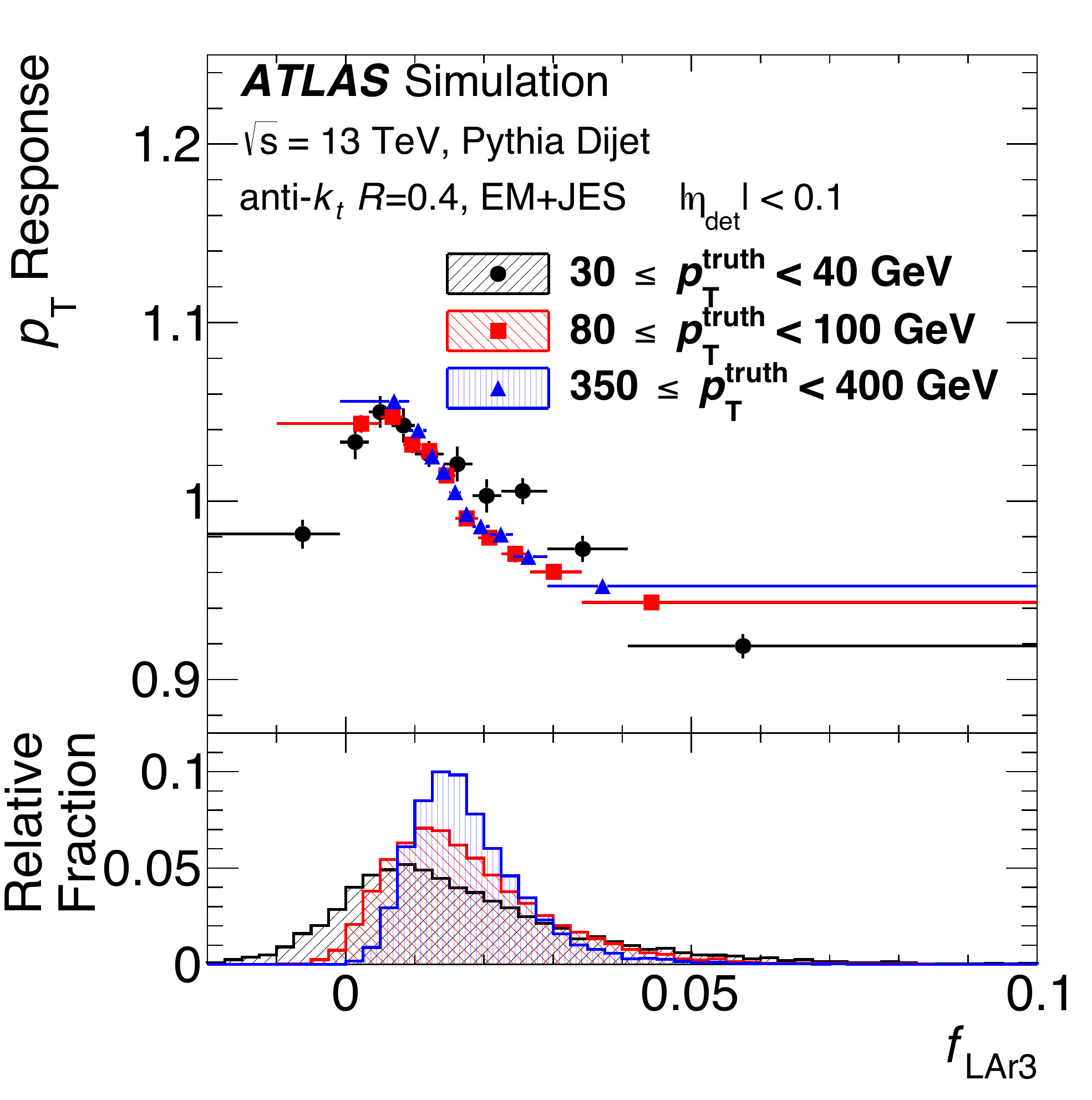}
  }
  \subfigure[]{
    \label{fig:GSC:nTrack}
    \includegraphics[width=0.31\textwidth]{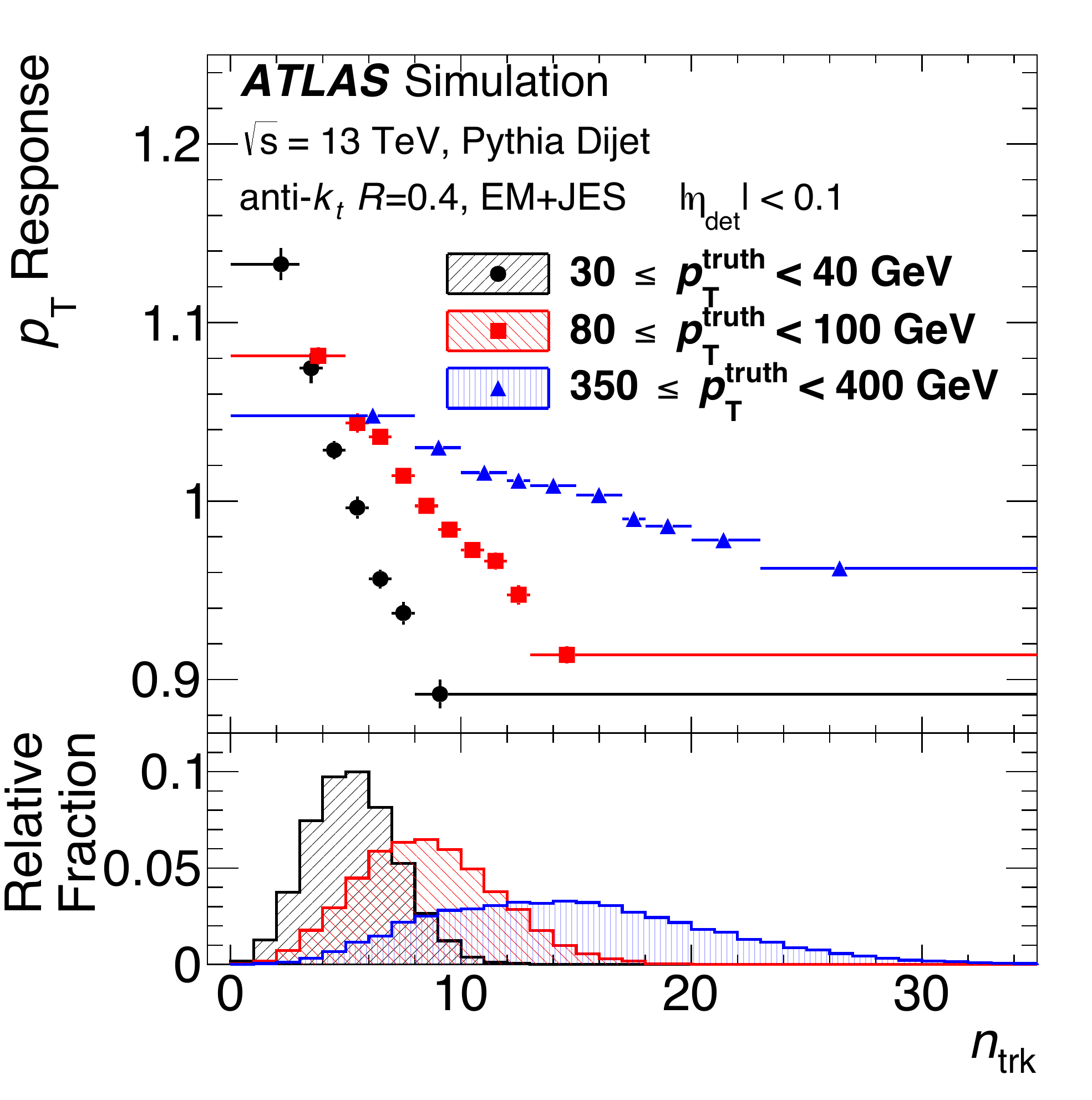}
  }
  \subfigure[]{
    \label{fig:GSC:trackWidth}
    \includegraphics[width=0.31\textwidth]{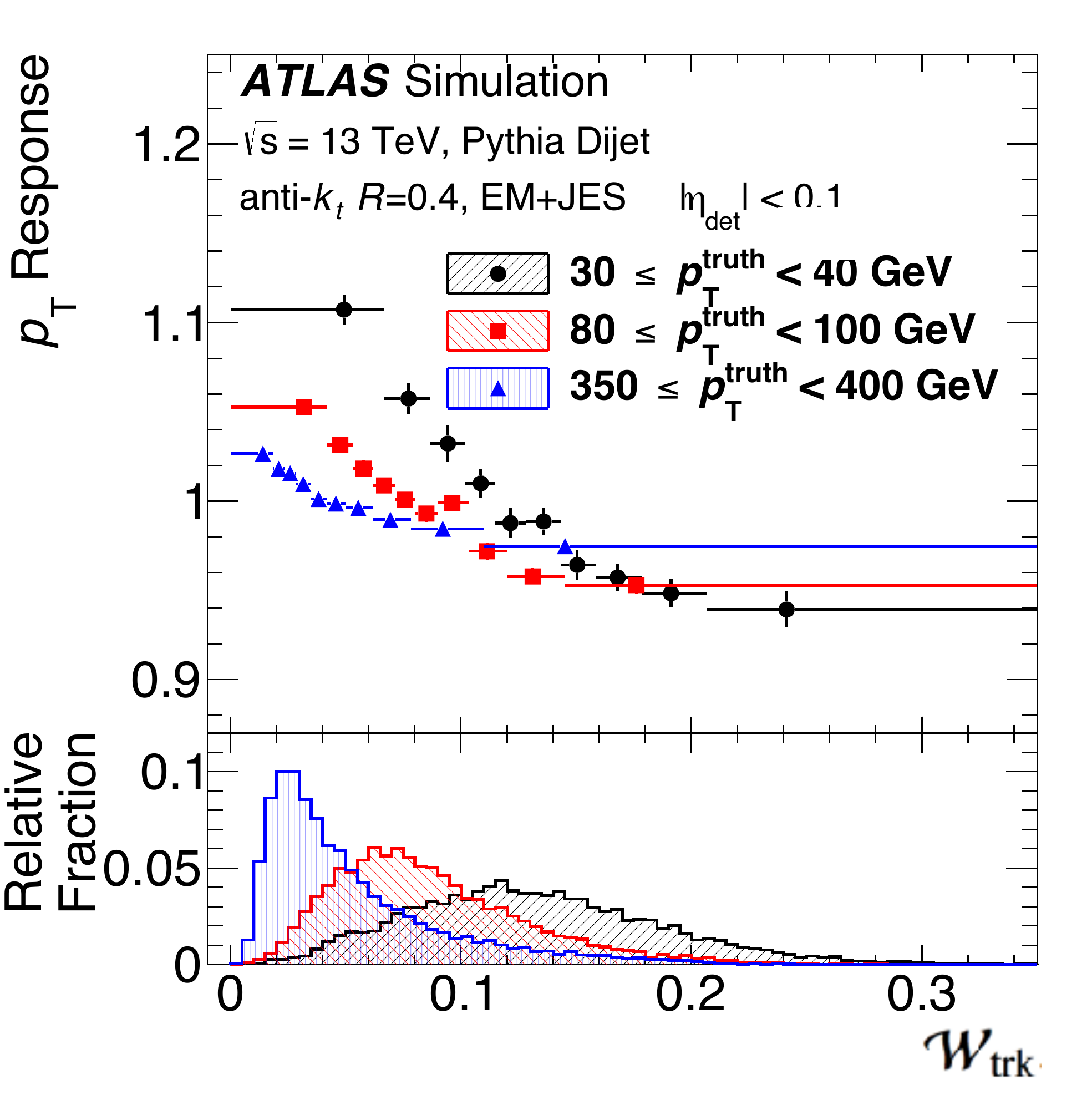}
  }
  \subfigure[]{
    \label{fig:GSC:Nseg}
    \includegraphics[width=0.31\textwidth]{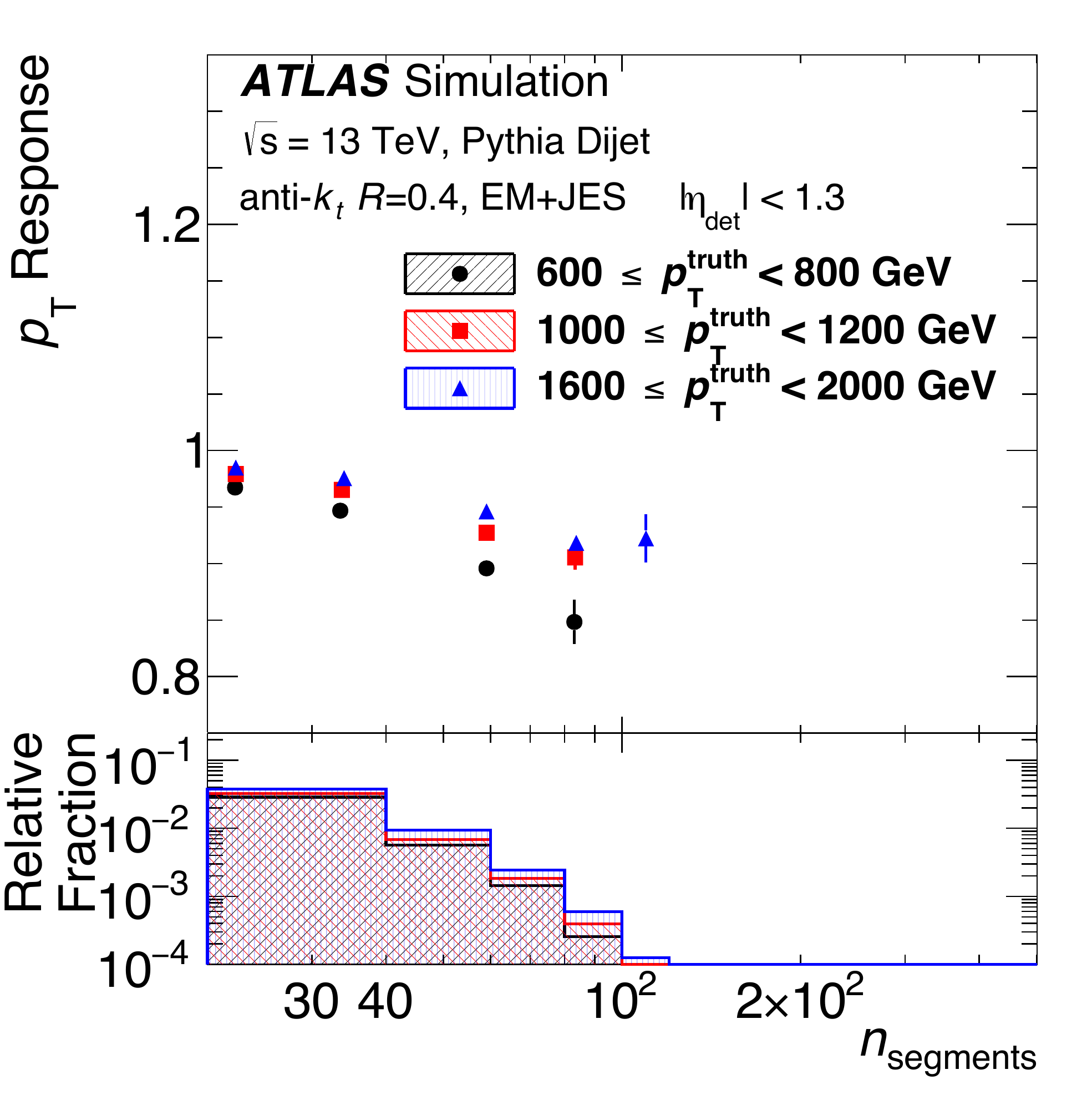}
  }
  \caption{
  The average jet response in MC simulation as a function of the GSC variables for three ranges of \pttruth.
  These include \subref{fig:GSC:Tile} the fractional energy in the first Tile calorimeter layer,
  \subref{fig:GSC:LAr} the fractional energy in the third LAr calorimeter layer,
  \subref{fig:GSC:nTrack} the number of tracks per jet, \subref{fig:GSC:trackWidth} the \pt-weighted track width,
  and \subref{fig:GSC:Nseg} the number of muon track segments per jet.
  Jets are calibrated with the EM+JES scheme and have GSC corrections applied for the preceding observables.
  The calorimeter distributions \subref{fig:GSC:Tile} and \subref{fig:GSC:LAr} are shown with no GSC corrections applied,
  the track-based distributions \subref{fig:GSC:trackWidth} and \subref{fig:GSC:nTrack} are shown with both preceding calorimeter corrections applied,
  and the punch-through distribution \subref{fig:GSC:Nseg} is shown with the four calorimeter and track-based corrections applied.
  Jets are constrained to $\abs\eta<0.1$ for the distributions of calorimeter and track-based observables and $\abs\eta<1.3$ for the muon \Nseg distribution.
  The distributions of the underlying observables in MC simulation are shown in the lower panels for each \pttruth region,
  normalized to unity.
  The shading in the legend reflects the shading of the distributions in the lower panel.}
  \label{fig:GSC}
\end{center}
\end{figure}

The jet \pt response in MC simulation as a function of each of these observables is shown in \figref{GSC} for several regions of \pttruth.
The distributions are shown at various stages of the GSC to reflect the relative disagreement at the stage when each correction is derived.
The dependence of the jet response on each observable is reduced to less than 2\% after the full GSC is applied, with
small deviations from unity reflecting the correlations between observables that are unaccounted for in the corrections.
The distribution of each observable in MC simulation is shown in the bottom panels in \figref{GSC}.
The spike at zero in the $f_\text{Tile0}$ distribution of \figref{GSC:Tile} at low \pttruth reflects jets that are
fully contained in the electromagnetic calorimeter and do not deposit energy in the Tile calorimeter.
The negative tail in the $f_\text{LAr3}$ distribution of \figref{GSC:LAr} (and, to a lesser extent, in the $f_\text{Tile0}$ distribution of \figref{GSC:Tile}) at low \pttruth reflects calorimeter noise fluctuations.

\subsection{\Insitu calibration methods}
\label{sec:insitu}

The last stages of the jet calibration account for differences in the jet response between data and MC simulation.
Such differences arise from the imperfect description of the detector response and detector material in MC simulation,
as well as in the simulation of the hard scatter, underlying event, pile-up, jet formation, and electromagnetic and hadronic interactions with the detector.
Differences between data and MC simulation are quantified by balancing the \pt of a jet against other well-measured reference objects.

The \etainter corrects the average response of forward jets to that of well-measured central jets using dijet events.
Three other \insitu calibrations correct for differences in the average response of central jets with respect to those of well-measured reference objects,
each focusing on a different \pt region using $Z$ boson, photon, and multijet systems.
For each \insitu calibration the response $\Response_\text{\insitu}$ is defined in data and MC simulation
as the average ratio of jet \pt to reference object \pt, binned in regions of the reference object \pt.
It is proportional to the response of the calorimeter to jets at the EM+JES, but
is also sensitive to secondary effects such as gluon radiation and the loss of energy outside of the jet cone.
Event selections are designed to reduce the impact of such secondary effects.
Assuming that these secondary effects are well modeled in the MC simulation, the ratio

\begin{equation}
  c = \frac{\Response^\text{data}_\text{\insitu}}{\Response^\text{MC}_\text{\insitu}}
\label{eq:doubleratio}
\end{equation}
is a useful estimate of the ratio of the JES in data and MC simulation.
Through numerical inversion a correction is derived to the jet four-momentum.
The correction is derived as a function of jet \pt, and also as a function of jet $\eta$ in the \etainter.

Events used in the \insitu calibration analyses are required to satisfy common selection criteria.
At least one reconstructed primary vertex is required with at least two associated tracks of \pt > 500 \MeV.
Jets are required to satisfy data-quality criteria that discriminate against
calorimeter noise bursts, cosmic rays, and other non-collision backgrounds.
Spurious jets from pile-up are identified and rejected through the exploitation of track-based variables by the jet vertex tagger
(JVT)~\cite{PERF-2014-03}. Jets with $\pt<50$~\GeV\ and $\abs\etadet < 2.4$ are required to be
associated with the primary vertex at the medium JVT working point, accepting 92\% of hard-scatter jets and rejecting 98\% of pile-up jets.

The \etainter corrects the jet energy scale of forward jets ($0.8 < \abs\etadet < 4.5$)
to that of central jets ($\abs\etadet < 0.8$) in a dijet system, and is discussed in \secref{etaInter}.
The \Zgjet balance analyses use a well-calibrated photon or $Z$ boson, the latter decaying into an electron or muon pair,
to measure the \pt response of the recoiling jet in the central region up to a \pt of about 950~\GeV, as
discussed in \secref{vjets}.
Finally, the multijet balance (MJB) analysis calibrates central ($\abs\eta < 1.2$), high-\pt jets ($300 < \pt < 2000~\GeV$) recoiling against
a collection of well-calibrated, lower-\pt jets, as discussed in \secref{MJB}.
While the \Zgjet and MJB calibrations are derived from central jets, their corrections are applicable to forward jets
whose energy scales have been equalized by the \etainter procedure.
The calibration constants derived in each of these analyses following \eref{doubleratio} are statistically combined into a
final \insitu calibration covering the full kinematic region, as discussed in \secref{insituCombo}.

The \etainter, \Zgjet, and MJB calibrations are derived and applied sequentially, with systematic uncertainties propagated through the chain.
Systematic uncertainties reflect three effects:
\begin{enumerate}
\item uncertainties arising from potential mismodeling of physics effects;
\item uncertainties in the measurement of the kinematics of the reference object;
\item uncertainties in the modeling of the \pt balance due to the selected event topology.
\end{enumerate}
Systematic uncertainties arising from mismodeling of certain physics effects are estimated
through the use of two distinct MC event generators.
The difference between the two predictions is taken as the modeling uncertainty.
Uncertainties in the kinematics of reference objects are propagated from the $1\sigma$ uncertainties in their own calibration.
Uncertainties related to the event topology are addressed by varying the event selections for each \insitu calibration and
comparing the effect on the \pt-response balance between data and MC simulation.

Systematic uncertainty estimates depend upon data and MC samples with event yields that fluctuate when applying the systematic uncertainty variations.
To obtain results that are statistically significant, the binning of
$\Response_\text{\insitu}$ in $\pt$ and $\eta$ is dynamically determined for each variation using a bootstrapping procedure~\cite{bohm2010introduction}.
In this procedure, pseudo-experiments are derived from the data or MC simulation by sampling each event with a weight taken from a Poisson distribution with a mean of one.
Each pseudo-experiment therefore emphasizes a unique subset of the data or MC simulation while maintaining statistical correlations between the nominal and varied samples.
The statistical uncertainty of the response variation between the nominal and varied configuration is then taken as the RMS across the pseudo-experiments,
and each varied configuration is rebinned until a target significance of a few standard deviations is achieved.
Bins are combined only in regions where the observed response in $\pt$ is nearly constant so that no significant features are removed.

\subsubsection{\etainter}
\label{sec:etaInter}

In the \etainter~\cite{PERF-2012-01},
well-measured jets in the central region of the detector ($\abs\etadet < 0.8$) are used to derive a residual calibration for jets in the forward region ($0.8 < \abs\etadet < 4.5$).
The two jets are expected to be balanced in \pt at leading order in QCD, and any imbalance can be attributed to
differing responses in the calorimeter regions, which are typically less understood in the forward region.
Dijet topologies are selected in which the two leading jets are back-to-back in $\phi$ and there is no substantial contamination from a third jet.
The calibration is derived from the ratio of the jet \pt responses in data and MC simulation in bins of \pt and \etadet.
Two distinct NLO MC event generators are used, \POWPYTHIA and \SHERPA, with the former taken as the nominal generator.
The events are generated with a $2\rightarrow3$ leading-order matrix element,
increasing the accuracy of the dijet balance for events sensitive to the rejection criteria for the third jet.

The jet \pt balance is quantified by the asymmetry

\begin{equation*}
  \asym = \frac{\ptprobe - \ptref }{ \ptavg },
\label{eq:etaAsym}
\end{equation*}
where $\ptprobe$ is the transverse momentum of the forward jet, \ptref is the transverse momentum of the jet in a
well-calibrated reference region, and \ptavg is the average \pt of the two jets.
The asymmetry is a useful quantity as the distribution is Gaussian in fixed bins of $\ptavg$, whereas $\ptprobe/\ptref$ is not.
Given that the asymmetry is Gaussian, the relative jet response with respect to the reference region may be written as

\begin{equation*}
  \left\langle\frac{\ptprobe }{ \ptref }\right\rangle \approx \frac{ 2 + \avg{\asym }}{ 2 - \avg{\asym } },
\label{eq:etaCorrection}
\end{equation*}
where $\avg{\asym}$ is the mean value of the asymmetry distribution for a bin of \ptavg and \etadet.

Events used in the \etainter follow from a combination of single-jet triggers with various \pt thresholds in regions of either $\abs\etadet < 3.1$ or $\abs\etadet > 2.8$.
Triggers are only used in regions of kinematic phase space in which they are 99\% efficient.
Triggers may also be prescaled, randomly rejecting a set fraction of events to satisfy bandwidth considerations, and the event weight is scaled proportionally.
Events are required to have at least two jets with $\pt > 25~\GeV$ and with $\abs\etadet < 4.5$.
Events that include a third jet with relatively substantial \pt, $\pt^\text{jet3} > 0.4 \ptavg$, are rejected.
The two leading jets are also required to be fairly back-to-back, such that $\Delta\phi > 2.5$ rad.

The residual calibration is derived from the ratio of the jet responses in data and the \POWPYTHIA sample.
The \SHERPA sample is used to provide a systematic uncertainty in the MC modeling.
The full range of $\abs\etadet < 4.5$ is used to derive calibrations for statistically significant regions of \ptavg,
offering an improvement on the 2011 calibration that extrapolated the measurement
from a constrained region $\abs\etadet < 2.7$ due to statistical considerations.
A two-dimensional sliding Gaussian kernel~\cite{PERF-2012-01} is used to reduce statistical fluctuations while preserving the shape of the MC-to-data ratio
and to extrapolate the average response into regions of low statistics.

Two \etainter methods are performed that provide complementary results.
In the central reference method, central regions ($ \abs\etadet < 0.8$) are used as references to measure the relative
jet response in the forward probe bins ($0.8 < \abs\etadet < 4.5$).
In the matrix method, numerous independent reference regions are chosen and the relative jet response in a given
forward probe bin is measured relative to all reference regions simultaneously.
The response relative to the central region is then obtained as a function of \ptavg and \etadet through a set of linear equations.
The matrix method takes advantage of a larger data set by allowing multiple reference regions, including forward ones,
increasing the statistical precision of the calibration.

The binning is chosen such that each reference region is statistically significant in data and \POWPYTHIA samples.
Some reference regions, particularly for forward probe bins, may not be statistically significant for the \SHERPA sample due to its smaller sample size.
Such regions are ignored in the combined fit of the response, leading to small fluctuations in the \SHERPA response,
which are smoothed in \pt and \etadet by the two-dimensional sliding Gaussian kernel.

The relative jet responses derived from the two methods show agreement at the level of 2\%, within the uncertainty of the methods.
A slightly larger response is seen in the most forward bins ($\abs\etadet > 2.5$) in the matrix method, as seen in 2011.
This difference exists in the response in both data and MC simulation, and the MC-to-data ratio is consistent between methods.
The matrix method is used to derive the nominal calibration in the following results, with the central reference method providing validation.
As in the 2011 calibration, \gjet events are also used to validate the response in the forward regions, and show good agreement between data and MC simulation in the forward region.

\begin{figure}[!ht]
\begin{center}
  \subfigure[]{
    \label{fig:EIC:etaLowPt}
    \includegraphics[width=0.47\textwidth]{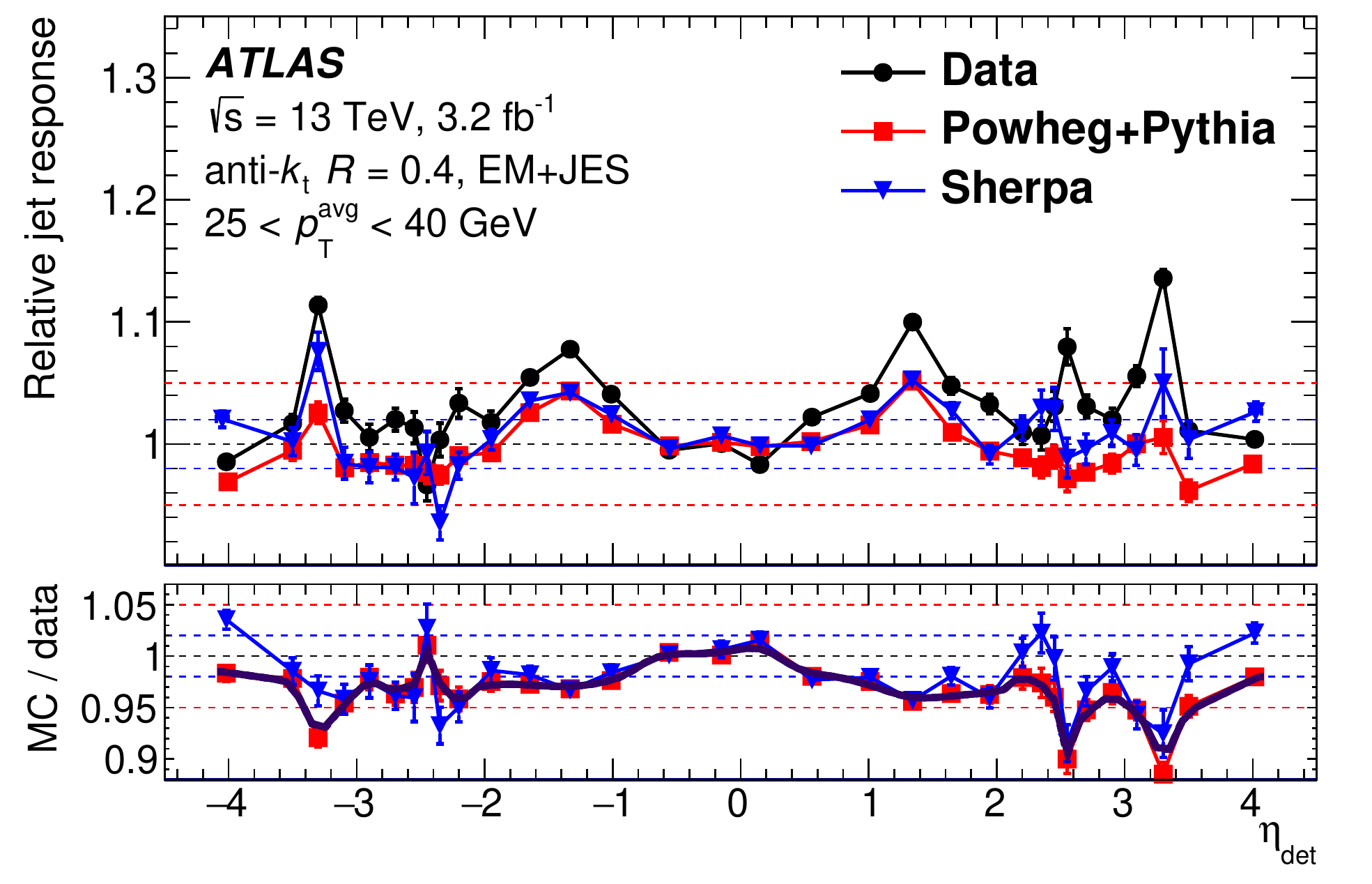}
  }
  \subfigure[]{
    \label{fig:EIC:etaHighPt}
    \includegraphics[width=0.47\textwidth]{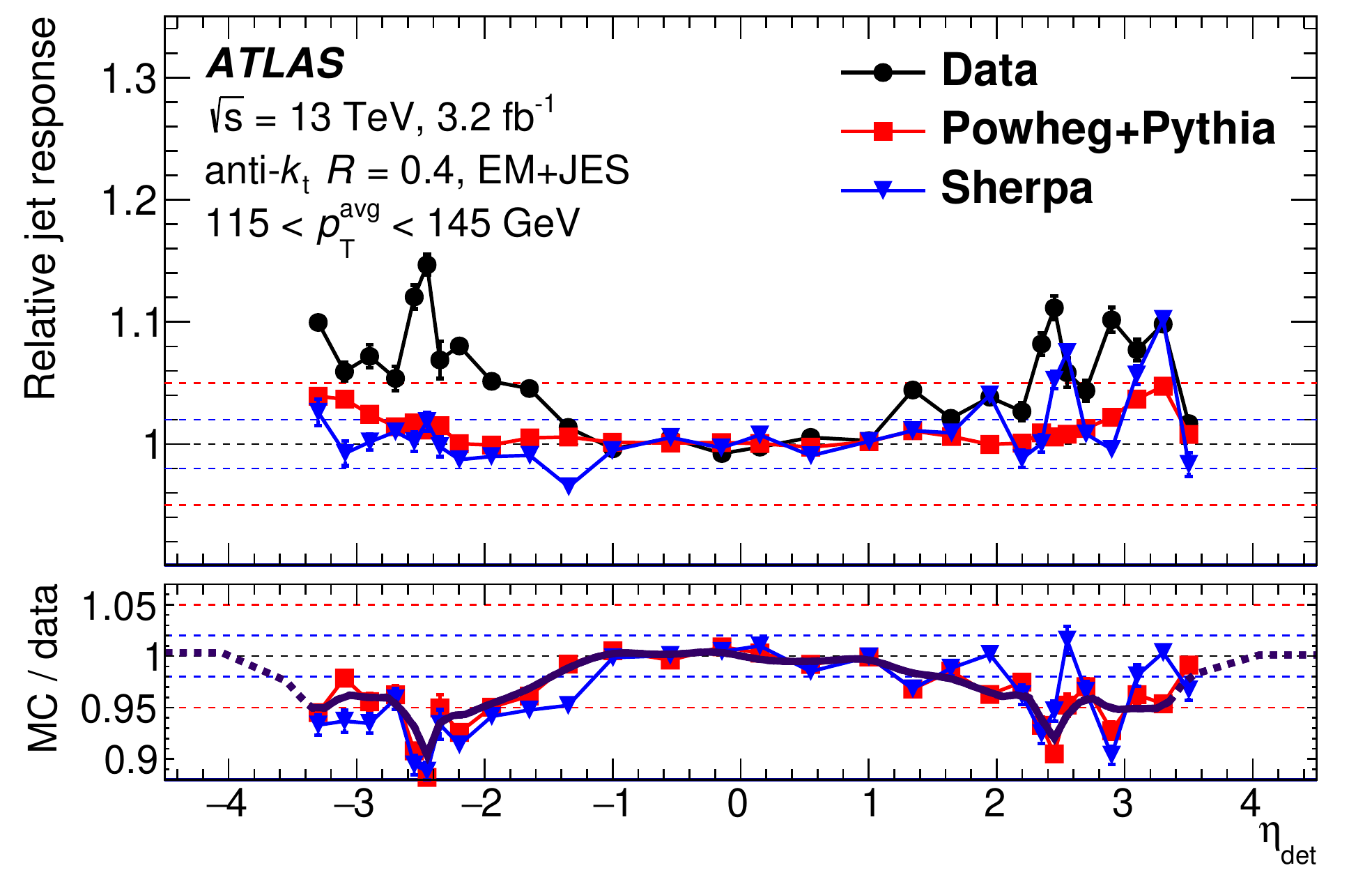}
  }
  \subfigure[]{
    \label{fig:EIC:ptLowEta}
    \includegraphics[width=0.47\textwidth]{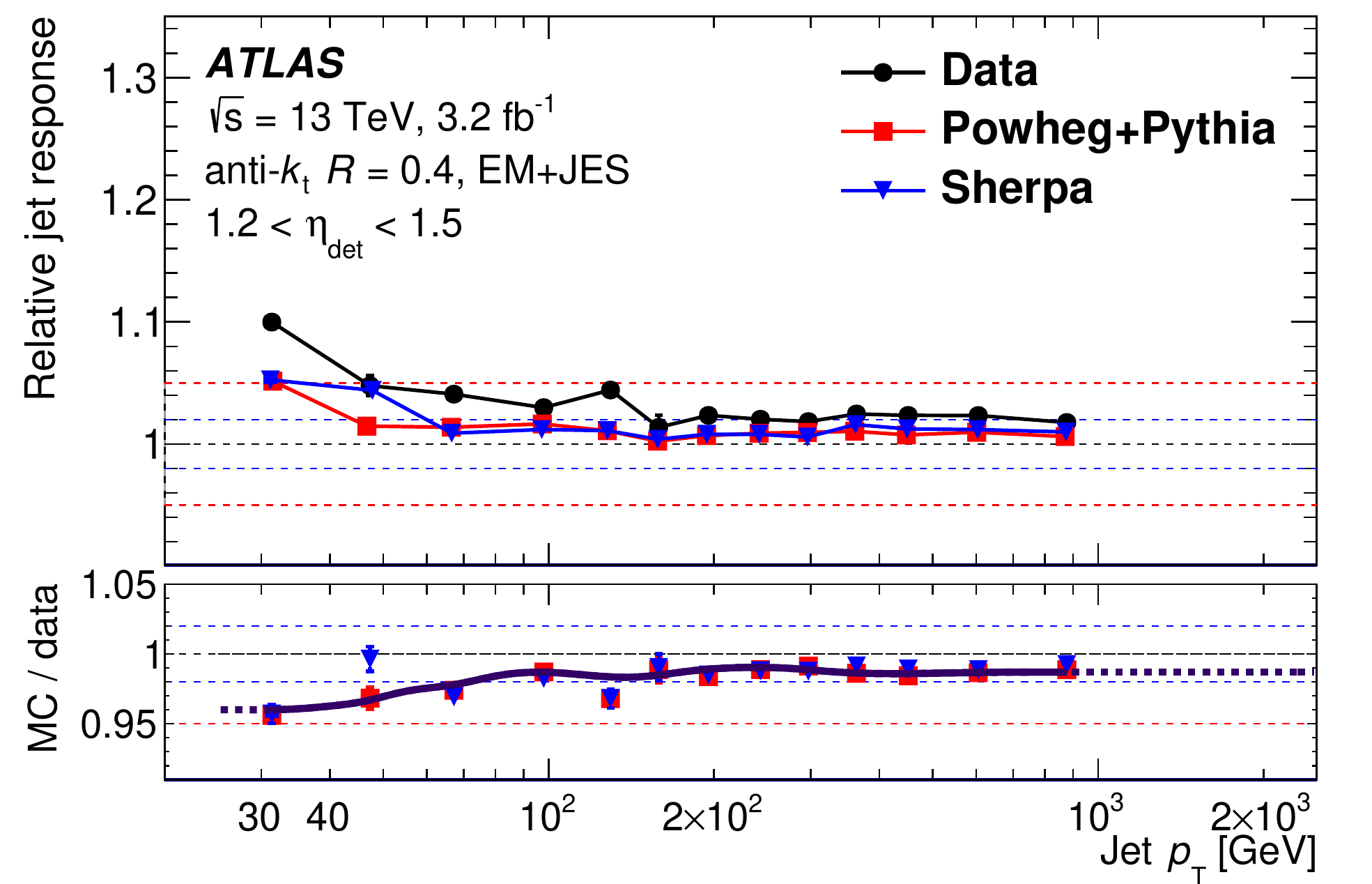}
  }
  \subfigure[]{
    \label{fig:EIC:ptHighEta}
    \includegraphics[width=0.47\textwidth]{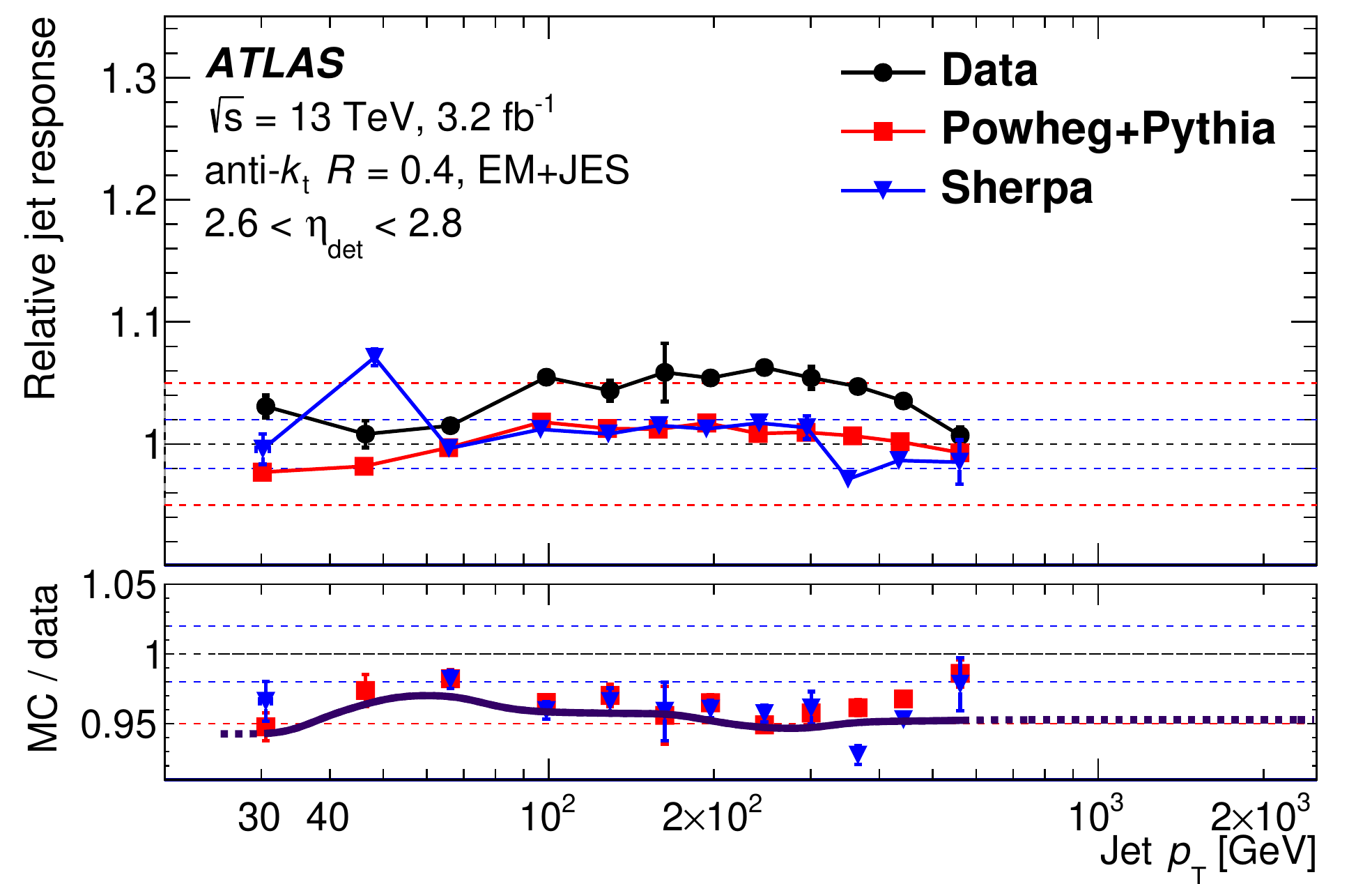}
  }
  \caption{Relative response of EM+JES jets as a function of $\eta$ at
           \subref{fig:EIC:etaLowPt} low \pt and \subref{fig:EIC:etaHighPt} high \pt,
           and as a function of jet \pt within the ranges of \subref{fig:EIC:ptLowEta} $1.2 < \etadet < 1.5$ and
           \subref{fig:EIC:ptHighEta} $2.6 < \etadet < 2.8$.
           The bottom panels show the MC-to-data ratios, and the overlayed curve reflects the smoothed \insitu correction,
           appearing solid in the regions in which it is derived and dotted in the regions to which it is extrapolated by the two-dimensional sliding Gaussian kernel.
           Results are obtained with the matrix method.
           The binning is optimized for data and \POWPYTHIA, and fluctuations in the response in \SHERPA are not statistically significant.
           Horizontal dotted lines are drawn in all at 1, $1\pm0.02$, and $1\pm0.05$ to guide the eye.
         }
  \label{fig:EIC}
\end{center}
\end{figure}

The relative jet response is shown in \figref{EIC} for both data and the two MC samples,
parameterized by \pt in two \etadet ranges and by \etadet in two \pt ranges.
The level of modeling agreement, taken between \POWPYTHIA and \SHERPA, is significantly better than in previous results and is generally within 1\%, with larger differences at low \pt and in forward \etadet regions.
This improved agreement is not due to any changes to the method but results from better overall particle-level agreement,
particularly the improved modeling of the third-jet radiation by the NLO \POWPYTHIA and \SHERPA generators over that of the LO \PYTHIA and \HERWIG generators used in the 2011 calibration.
The particle-level response was also measured with a \POWHEGBOX sample showered with \HERWIGpp, and shows a similar level of agreement as found between \POWPYTHIA and \SHERPA. 
Uncertainties are calculated in a given bin by shifting the observed asymmetry with all reference regions and recalculating the response.
While accurate for data and \POWPYTHIA, this can lead to statistical uncertainties that do not cover the observed fluctuations in \SHERPA,
but that do not affect the final systematic uncertainty derived from the smoothed difference between MC samples.

The response in data is consistently larger than that in both MC samples and in the 2011 data for the forward detector region for all \pt ranges.
This is due to the reduction in the number of samples used to reconstruct pulses in the LAr calorimeter from five to four, which is sensitive to differences in the pulse shape between data and MC simulation.
The reduction was predicted to increase the response in the forward region, as seen in a comparison of Run~1 data processed using both five and four samples.
The expected increase matches that seen in 2015 data, and is corrected for by the \etainter procedure.
The effect was predicted to be particularly large for $2.3 < \abs\etadet < 2.6$ due to details of the jet reconstruction
in calorimeter transition regions.
To fully account for this effect, a finer binning of $\Delta\etadet$ is used in this region.

\begin{figure}[!ht]
\begin{center}
  \subfigure[]{
    \label{fig:EIC_uncert:lowpt}
    \includegraphics[width=0.48\textwidth]{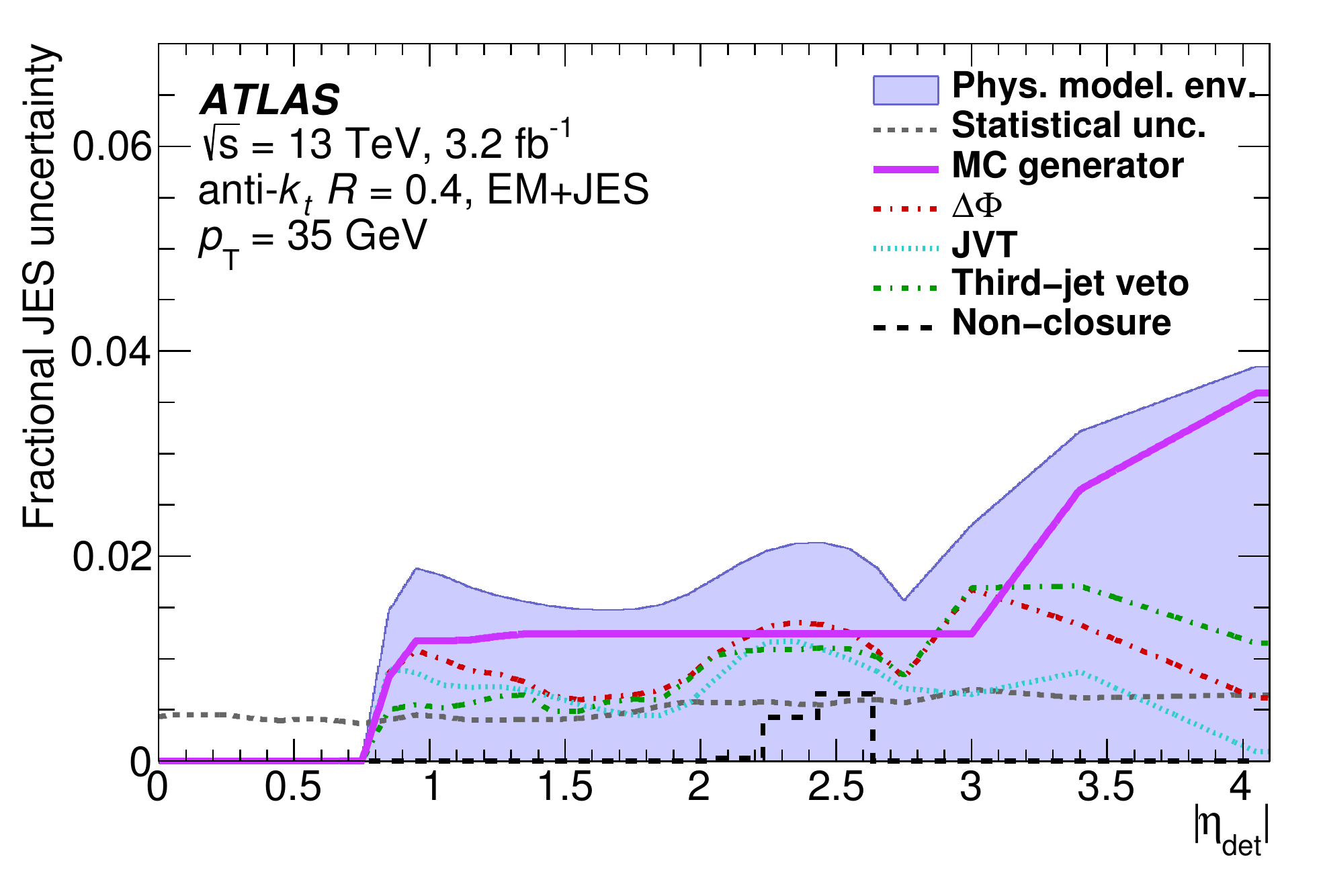}
  }
  \subfigure[]{
    \label{fig:EIC_uncert:highpt}
    \includegraphics[width=0.48\textwidth]{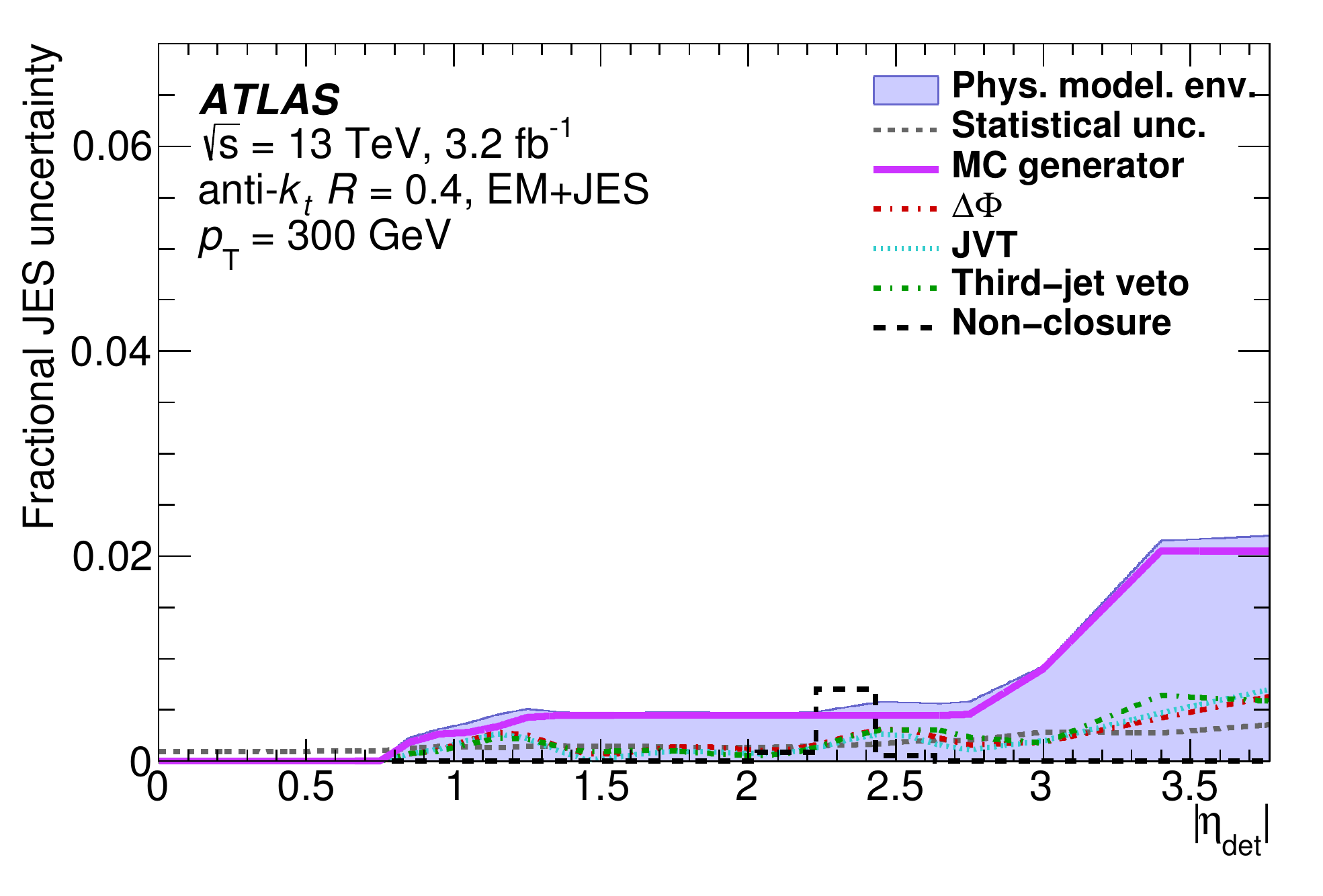}
  }
  \caption{
    Systematic uncertainties of EM+JES jets as a function of \abs\etadet at \subref{fig:EIC_uncert:lowpt} $\pt =35$~\GeV
    and at \subref{fig:EIC_uncert:highpt} $\pt =300$~\GeV in the \etainter. 
    The physics mismodeling envelope includes the uncertainty derived from the alternative MC event generator as well as
    the uncertainties of the JVT, $\Delta\phi$, and third-jet veto event selections. 
    Also shown are the statistical uncertainties of the MC-to-data response ratio and the localized non-closure uncertainty for $2.0 < \abs\etadet < 2.6$. 
    Small fluctuations in the uncertainties are statistically significant and are smoothed in the combination, described in \secref{insituCombo}.
  }
  \label{fig:EIC_uncert}
\end{center}
\end{figure}

The systematic uncertainties account for physics and detector mismodelings as well as the effect of the event topology on the modeling of the \pt balance.
They are derived as a function of \pt and $\abs\etadet$, with no statistically significant variations observed between positive and negative \etadet.
The dominant uncertainty due to MC mismodeling is taken as the difference in the smoothed jet response between \POWPYTHIA and \SHERPA.
The estimation of systematic uncertainties due to pile-up and the choice of event topology are similar to those of the 2011 calibration~\cite{PERF-2012-01},
but now use the bootstrapping procedure to ensure statistical significance.
These uncertainties, including those from varying the $\Delta\phi$ separation requirement between the two leading jets and the third-jet veto requirement,
are usually small compared to the MC uncertainty and are therefore summed in quadrature with it into a single physics mismodeling uncertainty, with a negligible loss of correlation information.
Two additional and separate uncertainties are derived to account for statistical fluctuations and the observed non-closure
of the calibration for $2.0 < \abs\etadet < 2.6$, primarily due to the LAr pulse reconstruction effects described above.
The latter is taken as the difference between data and the nominal MC event generator after repeating the analysis with the derived calibration applied to data.
The total \etainter uncertainty is shown in \figref{EIC_uncert} as a function of \etadet for two jet \pt values.

\subsubsection{\Zjet and \gjet balance}
\label{sec:vjets}

An \insitu calibration of jets up to 950~\GeV and with $\abs\eta < 0.8$ is derived through the \pt balance of a jet against a $Z$ boson or a photon.
\Zgjet calibrations rely on the independent measurement and calibration of the energy of a photon or of the lepton decay products of a $Z$ boson,
through the decay channels of $ Z \rightarrow e^{+}e^{-}$ and $ Z \rightarrow \mu^{+}\mu^{-}$.
Bosons are ideal candidates for reference objects as they are precisely measured:  muons from tracks in the ID and MS
and photons and electrons through their relatively narrow showers in the electromagnetic calorimeter and the independent measurement of electron tracks in the ID.
The \Zjet calibration is limited to the statistically significant \pt range of $Z$ boson production of $20 < \pt < 500~\GeV$.
The \gjet calibration is limited by the small number of events at high \pt and by both dijet contamination
and an artificial reduction of the number of events due to the prescaled triggers at low \pt, limiting the calibration to $36 < \pt < 950~\GeV$.

Two techniques are used to derive the response with respect to the $Z$ boson and photon~\cite{PERF-2012-01}.
The Direct Balance (DB) technique measures the ratio of a fully reconstructed jet's \pt, calibrated up to the \etainter stage,
and a reference object's \pt.
The use of a fully reconstructed and calibrated jet allows the calibration to be applied to jets in a straightforward manner.
For a $2\rightarrow2$ \Zgjet event, the \pt of the jet can be expected to balance that of the reference object.
However, the DB technique can be affected by additional parton radiation contributing to the recoil of the boson, appearing as subleading jets.
This is mitigated through a selection against events with a second jet of significant \pt and a minimum requirement on $\Delta\phi$, the azimuthal angle between the \Zg boson and the jet, to ensure they are sufficiently back-to-back in $\phi$.
The component of the boson \pt perpendicular to the jet axis is also ignored, with the reference \pt defined as

\begin{equation*}
  \ptref = p_\text{T}^{\Zg} \times \cos\left(\Delta\phi\right).
\label{eq:VjetRef}
\end{equation*}
The DB technique is also affected by out-of-cone radiation, consisting of the
energy lost outside of the reconstructed jet's cone of $R=0.4$ due to fragmentation processes.
The out-of-cone radiation may lead to a \pt imbalance between a jet and the reference boson,
and is estimated by measuring the profile of tracks around the jet axis~\cite{PERF-2012-01}.

The Missing-$E_\text{T}$ Projection Fraction (MPF) technique instead derives a \pt balance
between the full hadronic recoil in an event and the reference boson.
The average MPF response is defined as

\begin{equation}
  \Response_\text{MPF} = \left\langle 1 +   \frac{\hat{n}_\text{ref}\cdot \vec{E}_\text{T}^\text{miss} }{\pt^\text{ref} } \right\rangle,
\label{eq:MPF}
\end{equation}
where $\Response_\text{MPF}$ is the calorimeter response to the hadronic recoil,
$\hat{n}_\text{ref}$ is the direction of the reference object, and $\pt^\text{ref}$ is the transverse momentum of the reference object.
The $\vec{E}_\text{T}^\text{miss}$ in \eref{MPF} is calculated directly from all the topo-clusters of calorimeter cells, calibrated at the EM scale,
and is corrected with the \pt of the minimum ionizing muons in \Zmm events.
No correction is needed for electrons or photons as their calorimeter response is nearly unity.

The MPF technique utilizes the full hadronic recoil of an event rather than a single reconstructed jet.
The MPF response is therefore less sensitive to the jet definition, radius parameter, and out-of-cone radiation than the DB response,
with reconstructed jets only explicitly used in the event selections.
The MPF technique is less sensitive to the generally $\phi$-symmetric pile-up and underlying-event activity.
As the MPF technique is not derived from a reconstructed jet the correction does not
directly reflect the energy within the reconstructed jet's cone.
The out-of-cone uncertainty derived for the DB technique is therefore applied as an estimate of the effect of showering and jet topology.
As the MPF technique does not use jets directly, the impact of the GSC is accounted for by applying a correction to the cluster-based $\vec{E}_\text{T}^\text{miss}$,
equal to the difference in momentum of the leading jet with and without the GSC.
The results from this method are compared with those using no GSC and those with the GSC applied to all jets in the event,
with negligible differences seen in the MC-to-data response ratio.

The response of the jet (DB) or hadronic recoil (MPF) is measured in both data and MC simulation,
and a residual correction is derived using the MC-to-data ratio.
The two methods are complementary and they are both pursued to check the compatibility of the measured response.
The results below present the \Zjet results using the MPF technique and the \gjet results using the DB technique.

For both techniques the average response is initially derived in bins of \ptref.
In each bin of \ptref, a maximum-likelihood fit is performed using a modified Poisson distribution extended to non-integer values.
The fit range is limited to twice the RMS of the response distribution
around the mean to minimize the effect of MC mismodeling in the tails of the distribution.
The average response is taken as the mean of the best-fit Poisson distribution.
For 2015 data, a new procedure is used to reparameterize the average balance from the reference object \pt to the corresponding jet \pt, better representing
the mismeasured jet to which the calibration is applied.
This procedure is used after the calibration is derived by finding the average jet \pt, without \Zgjet calibrations applied,
within each bin of reference \pt.

Events in the \Zjet selection are required to have a leading jet with $\pt > 10~\GeV$, and in the \gjet selection are required to have a leading jet with $\pt > 20~\GeV$.
In the \gjet DB (\Zjet MPF) technique, the leading jet must sufficiently balance the reference boson in the azimuthal plane,
requiring $\Delta\phi(\text{jet},Z(\gamma)) > 2.8~(2.9)$ rad.
To reduce contamination from events with significant hadronic radiation, a selection of
$\ptsecond < \max(15~\GeV, 0.1 \times \ptref)$ is placed on the second jet, ordered by \pt, in the \gjet DB technique.
For the \Zjet MPF technique, this selection is mostly looser as $\Response_\text{MPF}$  is less sensitive to QCD radiation, requiring
the second jet to have $\ptsecond < \max(12~\GeV, 0.3 \times \ptref)$.

Electrons~\cite{PERF-2013-05}
(muons~\cite{PERF-2015-10}) used in the \Zjet events are required to pass basic quality and isolation cuts, and to
fall within the range \abs\eta < 2.47 (2.4).
Events are selected based on the lowest-\pt unprescaled single-electron or single-muon trigger.
Electrons that fall in the transition region between the barrel and the endcap of the electromagnetic calorimeter (1.37 < \abs\eta < 1.52) are rejected.
Both leptons are required to have $\pt > 20~\GeV$, and the invariant mass of the opposite-charge
pairs must be consistent with the $Z$ boson mass, with $66 < m_{\ell\ell} < 116~\GeV$.
Photons~\cite{PERF-2013-05} used in the \gjet events must satisfy tight selection criteria and be within the range \abs\eta < 1.37 with \pt > 25~\GeV.
Events are selected based on a combination of fully efficient single-photon triggers.
Energy isolation criteria are applied to the photon showers to discriminate photons from $\pi^{0}$
decays and to maximize the suppression of jets misidentified as photons~\cite{STDM-2010-08}.
Jets within $\Delta R = 0.35$ of a lepton are removed from consideration in the \Zjet selection,
while jets within $\Delta R = 0.2$ of photons are similarly removed from consideration in both the \Zjet and \gjet selections.

\begin{figure}[!ht]
\begin{center}
  \subfigure[]{
    \label{fig:Vjet:Zjet}
    \includegraphics[width=0.48\textwidth]{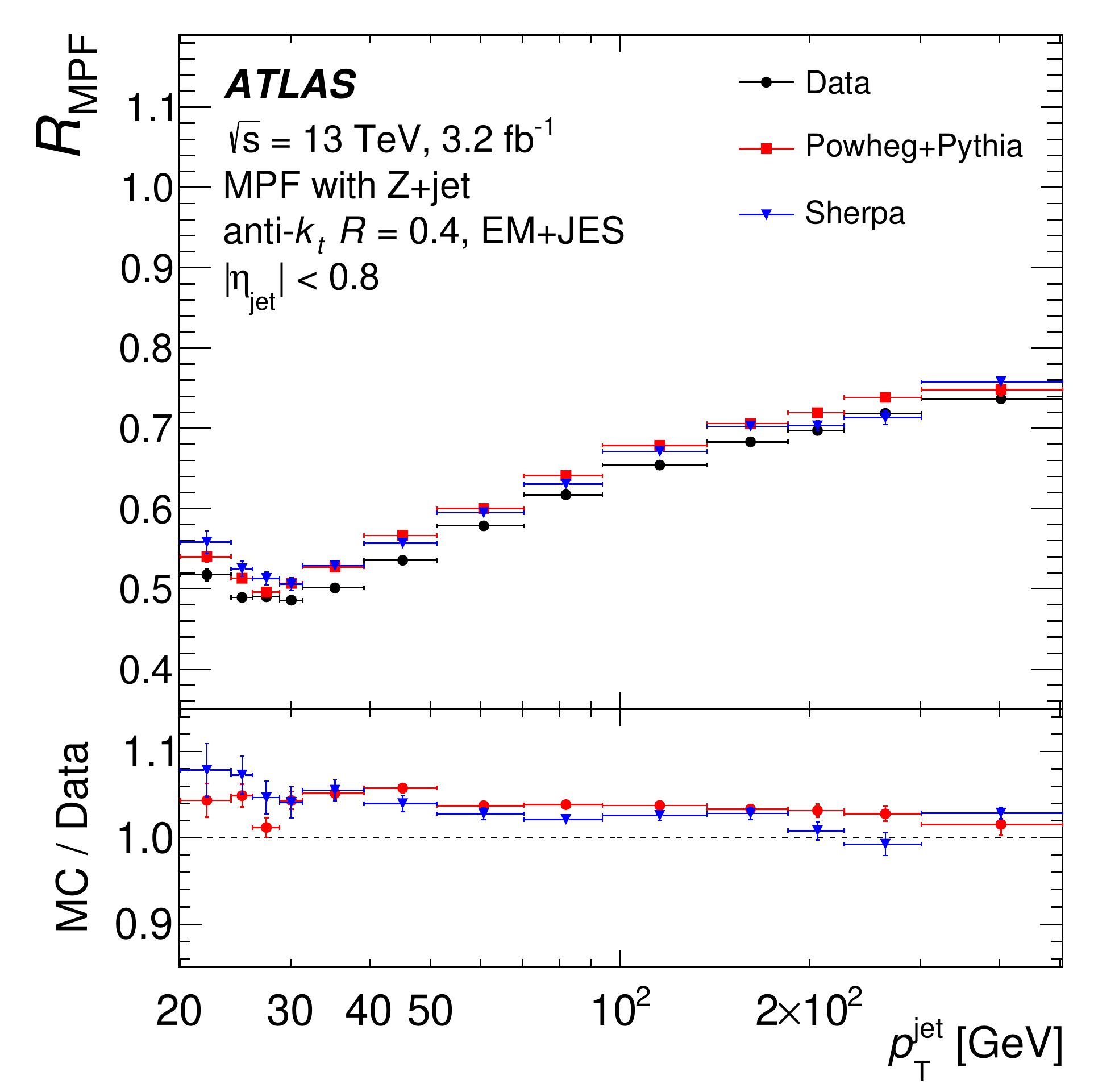}
  }
  \subfigure[]{
    \label{fig:Vjet:Gjet}
    \includegraphics[width=0.48\textwidth]{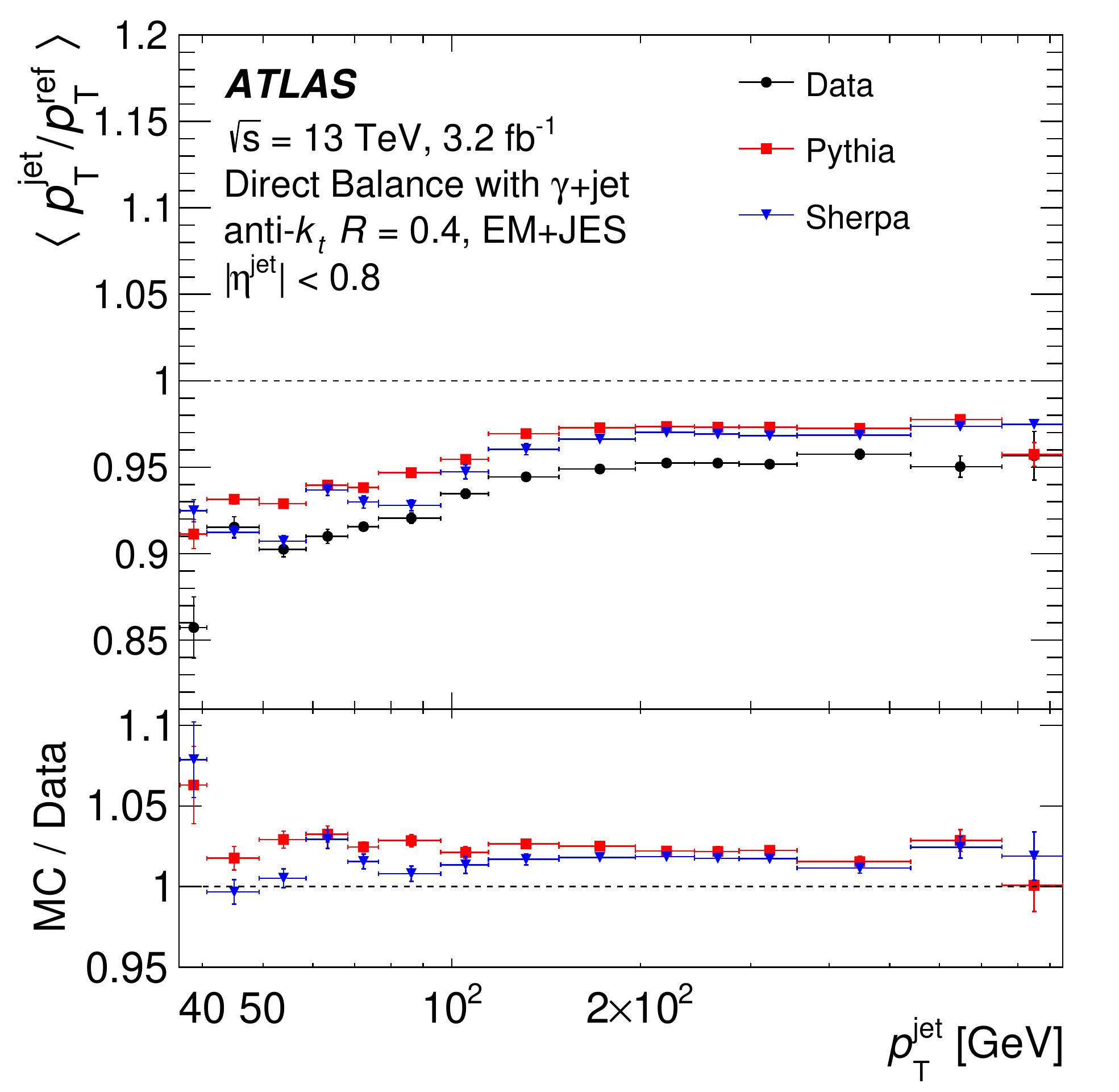}
  }
  \caption{ The average \subref{fig:Vjet:Zjet} MPF response in \Zjet events and 
    \subref{fig:Vjet:Gjet} Direct Balance jet \pt response in \gjet events as a function of jet \pt
    for EM+JES jets calibrated up to the \etainter.
    The response is given for data and two distinct MC samples,
    and the MC-to-data ratio plots in the bottom panels reflect the derived \insitu corrections.
    A dotted line is drawn at unity in the top-right panel and bottom panels to guide the eye.
  \label{fig:Vjet_Response}}
\end{center}
\end{figure}

The average response in \Zgjet
events as a function of jet \pt is shown in \figref{Vjet_Response} for data and two MC samples.
For the DB technique in \gjet events, the response is slightly below unity, reflecting
the fraction of \pt falling outside of the reconstructed jet cone.
For the MPF technique in \Zjet events, $\Response_\text{MPF}$ is significantly below unity, reflecting that the $Z$ boson is fully calibrated
while the topo-clusters used in calculating the hadronic recoil are at the EM scale.
However, in both cases the data and MC simulation are in agreement, with the MC-to-data ratio within $\sim$5\% of unity for both MC samples.
The rise in $\Response_\text{MPF}$ at low \pt in \figref{Vjet:Zjet} is caused by the jet reconstruction threshold.

\begin{figure}[!ht]
\begin{center}
  \subfigure[]{
    \label{fig:Vjet:Sys_Zjet}
    \includegraphics[width=0.47\textwidth]{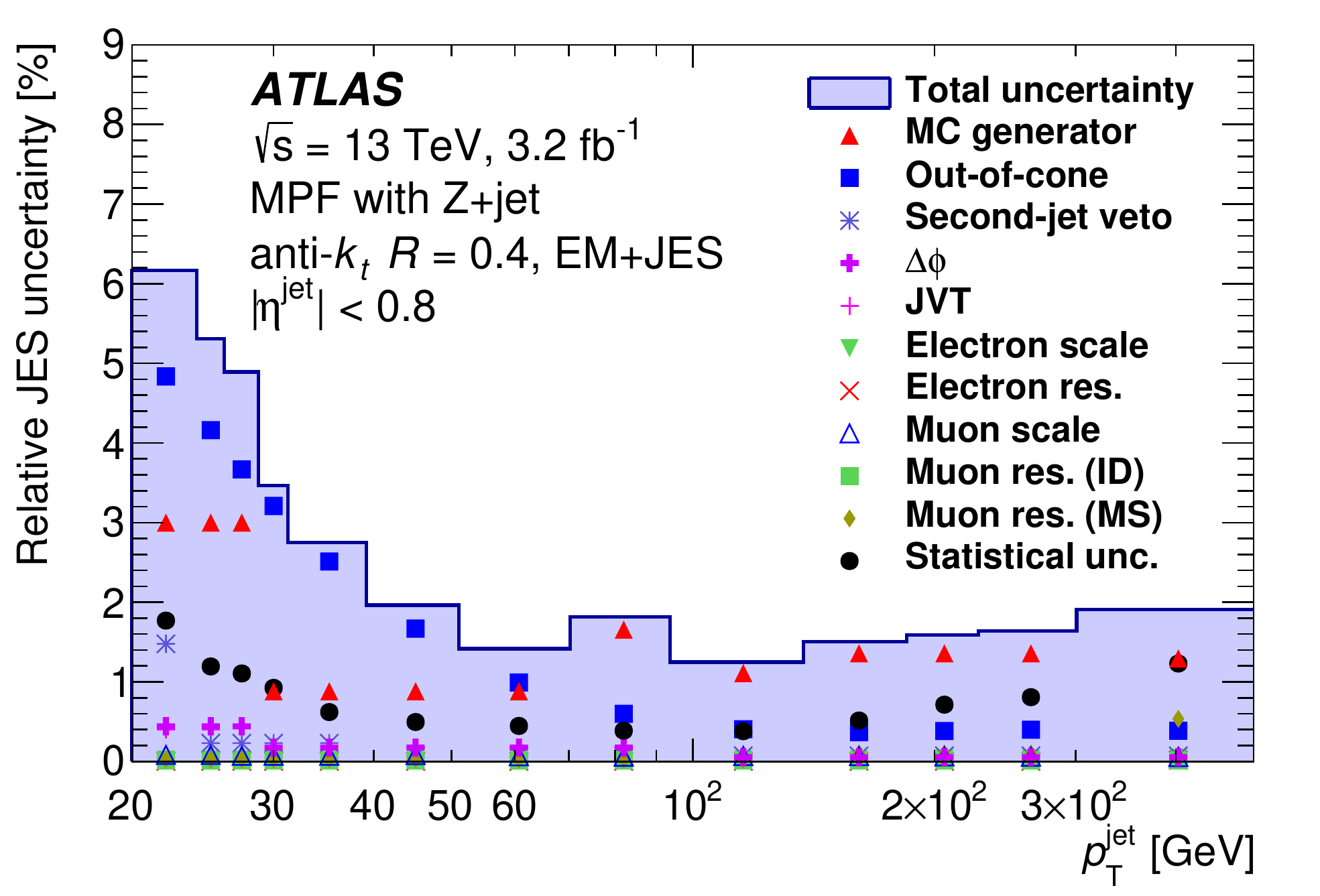}
  }
  \subfigure[]{
    \label{fig:Vjet:Sys_Gjet}
    \includegraphics[width=0.47\textwidth]{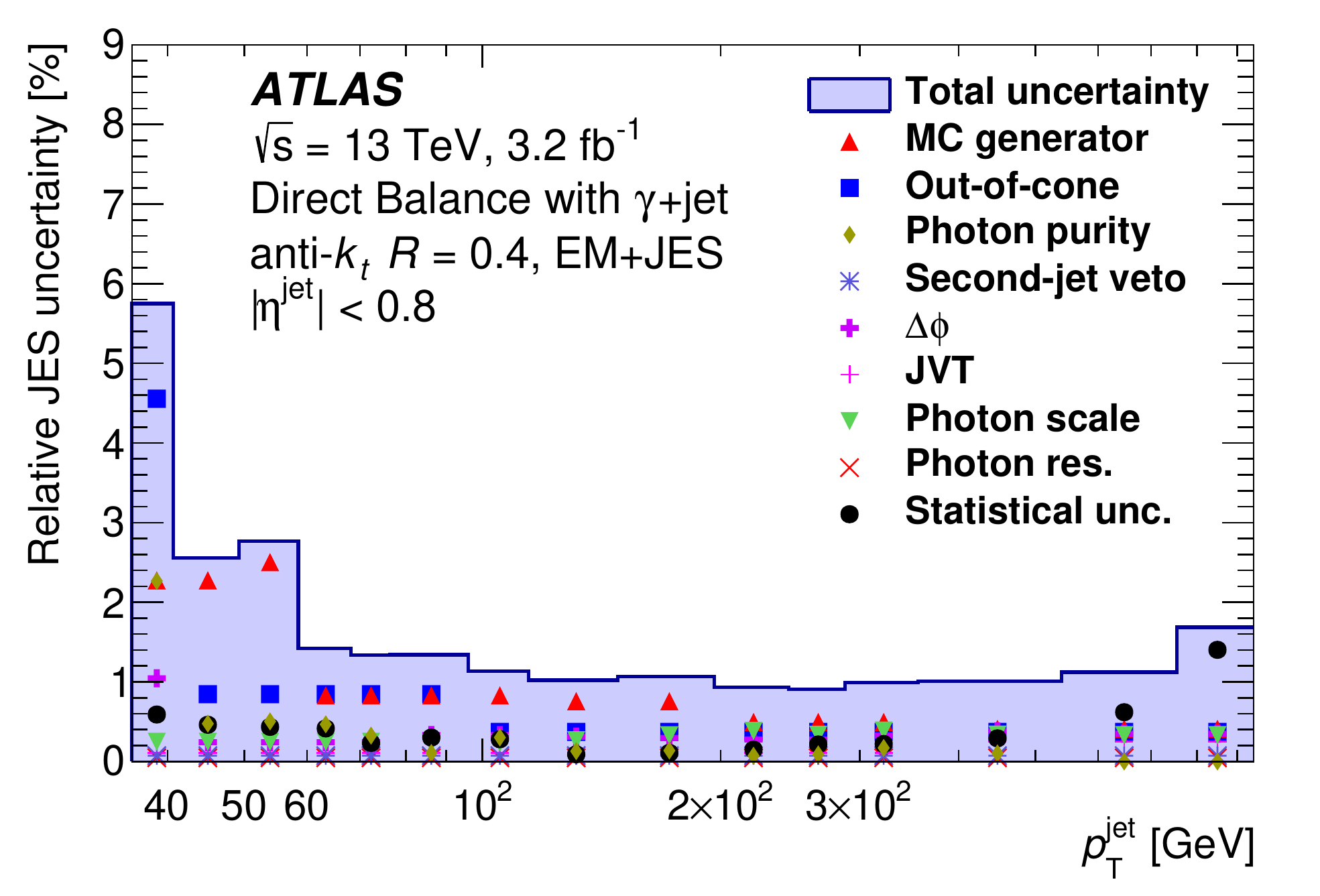}
  }
  \caption{ Systematic uncertainties of EM+JES jets, calibrated up to the \etainter, as a function of jet \pt for
    \subref{fig:Vjet:Sys_Zjet} \Zjet events using the MPF technique and
    \subref{fig:Vjet:Sys_Gjet} \gjet events using the Direct Balance technique.
    Uncertainties account for out-of-cone radiation and variations of the JVT, $\Delta\phi$, second-jet veto, and photon purity event selections.
    Uncertainties are also propagated from the electron and photon energy scale and resolution
    and the muon momentum scale and resolution in the ID and MS.
    Also shown are the statistical uncertainties of the MC-to-data response ratio and the uncertainty derived from an alternative MC event generator.
    Small fluctuations in the uncertainties are statistically significant and are smoothed in the combination, described in \secref{insituCombo}.
  }
  \label{fig:Vjet_Systematic}
\end{center}
\end{figure}

Systematic uncertainties in the MC-to-data response ratios are shown in \figref{Vjet_Systematic}.
In both the DB and MPF techniques the event selections are varied to estimate the impact of the choice of event topology on the MC mismodeling of the \pt response.
Variations are made to the selection criteria for the second-jet \pt and $\Delta\phi$ between the leading jet and reference object to assess the effect of additional parton radiation.
The effect of pile-up suppression is similarly studied by varying the JVT cut about its nominal value.
Potential MC event generator mismodeling is explored by repeating the analyses with alternative MC event generators, with the difference in
the MC-to-data response ratios taken as a systematic uncertainty.
Uncertainties in the energy (momentum) scale and resolution of electrons and photons (muons)
are estimated from studies of \Zee (\Zmm) measurements in data~\cite{PERF-2013-05,PERF-2015-10}.
Variations of $\pm1\sigma$ are propagated through the analyses to the MC-to-data response ratios.
A purity uncertainty in the \gjet balance accounts for the contamination from multijet events arising from jets appearing as fake photons.
The effect of this contamination on the MC-to-data response ratio is studied by relaxing the photon identification criteria.
The uncertainty due to out-of-cone radiation is derived from differences between data and MC simulation in the transverse momentum of
charged-particle tracks around the jet axis.
The bootstrapping procedure is used to ensure only statistically significant variations of the response are included in the uncertainties.

\subsubsection{Multijet balance}
\label{sec:MJB}

The multijet balance (MJB)~\cite{PERF-2012-01} is the last stage of the \insitu calibration and is used to extend the calibrations to a \pt of 2~\TeV.
Topologies with three or more jets are used to balance a high-\pt jet against a recoil system
composed of several lower-\pt jets.
The recoil jets are of sufficiently low \pt as to be in the range of \Zgjet calibrations
and are therefore fully calibrated.
The \Zgjet input calibrations are combined using the procedure outlined in \secref{insituCombo}.

The leading jet is taken as the highest-\pt jet of an event and the four-momenta of all other subleading jets are combined into a recoil-system four-momentum.
The leading jet is calibrated only up to the \etainter stage, and is therefore at the same scale as the jets explored by the \Zgjet methods.
A \pt limit of 950~\GeV is imposed on each subleading jet to ensure they are fully calibrated by the \Zgjet methods.
A consequence of this limit is the rejection of events with very high-\pt leading jets, which often have subleading jets with \pt above this limit.
These events are recovered through the use of multiple iterations of the MJB method,
with the previously derived MJB calibration being applied to higher-\pt subleading jets.
The new \pt limit on the subleading jets is determined by the statistical reach of the previous iteration of the MJB method.
Using 2015 data, the MJB method is able to cover a range of $300 < \pt < 2000~\GeV$ using two iterations.

The average response between the leading jet and recoil system, $\Response_\text{MJB}$, is defined as

\begin{equation*}
  \Response_\text{MJB} = \left\langle\frac{\ptleading}{\ptrecoil}\right\rangle,
\label{eq:MJBBalance}
\end{equation*}
where \ptleading is the transverse momentum of the highest-\pt jet and \ptrecoil is from the vectorial sum of all subleading jets.
The response is initially binned as a function of \ptrecoil, corresponding to the well-calibrated reference object.
As with the \Zgjet calibrations, a new procedure is used for 2015 data to reparameterize the response from \ptrecoil to \ptleading.
This procedure is applied after the calibration is derived by finding the average \ptleading, without \Zgjet or MJB calibrations applied,
within each bin of \ptrecoil.

Events entering the MJB calibration are recorded using a combination of fully efficient single-jet triggers used in distinct regions of \ptrecoil.
Events are required to have at least three jets with $\pt > 25~\GeV$ and $\abs\etadet < 2.8$, with the leading jet required to be central ($\abs\etadet < 1.2$).
Events dominated by a dijet \pt balance are rejected if the second jet's \pt is a considerable fraction of the recoil system's \pt,
with a \pt asymmetry requirement of $\pta = \ptsecond/\ptrecoil < 0.8$.
The azimuthal angle between the leading jet and the recoil system, $\alpha^\text{MJB}$, must satisfy the requirement $|\alpha^\text{MJB} - \pi| < 0.3$ rad,
ensuring the \pt of the recoil system is balanced against that of the leading jet.
Contamination of the leading jet from other jets is minimized by requiring the absolute value of the azimuthal angle, $\beta^\text{MJB}$, between the leading jet
and the nearest jet with $\pt > 0.25\ptleading$ to be greater than 1 rad.

\begin{figure}[!ht]
\begin{center}
  \subfigure[]{
    \label{fig:MJB:Balance_EM}
    \includegraphics[width=0.47\textwidth]{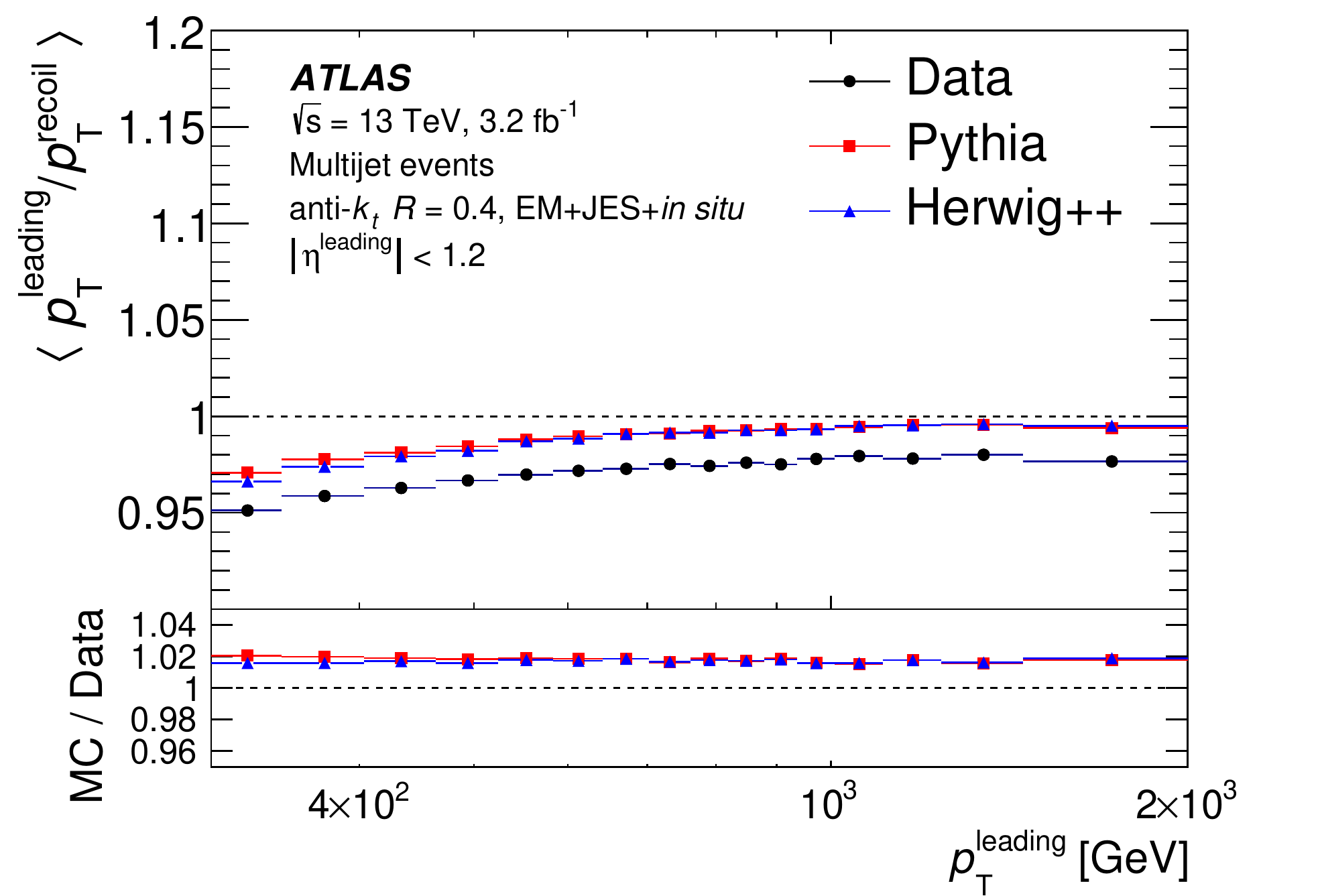}
  }
  \subfigure[]{
    \label{fig:MJB:Systematics_EM}
    \includegraphics[width=0.50\textwidth]{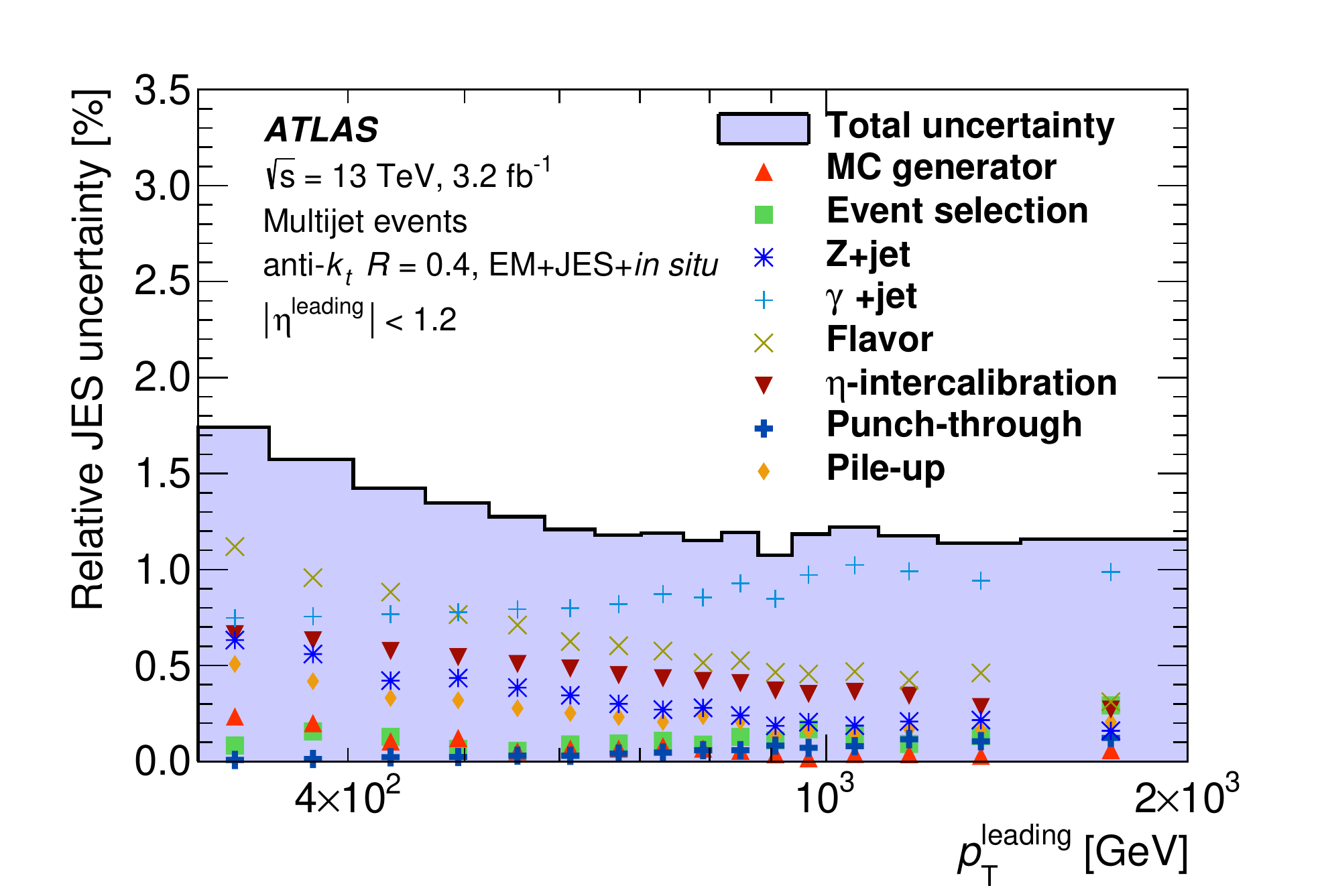}
  }
  \caption{ \subref{fig:MJB:Balance_EM} Average \pt response of the leading EM+JES jet, calibrated up to the \etainter,
    in multijet events as a function of \ptleading.
    The \Zgjet calibrations are applied only to subleading jets in the recoil system.
    The response is given for data and two distinct MC event generators,
    and the MC-to-data response ratio in the bottom panel reflects the derived \insitu correction.
    \subref{fig:MJB:Systematics_EM} Systematic uncertainties in the ratio account for variations of the event selection, MC mismodeling,
    and propagated uncertainties from the \Zjet, \gjet, \etainter, flavor, punch-through, and pile-up studies.
    Small fluctuations in the uncertainties are statistically significant and are smoothed during the combination, described in \secref{insituCombo}.
  \label{fig:MJB}}
\end{center}
\end{figure}

The response $\Response_\text{MJB}$ is shown for data and MC simulation in \figref{MJB:Balance_EM}.
As expected, an offset is seen between data and MC simulation, reflecting that the recoil system in data is fully calibrated to the \insitu stage
while the leading jet is only partially calibrated.
The response is below unity, particularly at low \pt, reflecting the bias in $\Response_\text{MJB}$ due to the leading-jet isolation requirement,
which is well modeled in MC simulation.
The MC-to-data ratio of $\Response_\text{MJB}$, given by \eref{doubleratio}, is seen in the bottom panel of \figref{MJB:Balance_EM}.
A fairly constant correction of 2\% is derived, up from 1\% in 2011.
This increase is partially due to changes in the simulation of hadronic showers in Geant4
as well as the response drift in the Tile calorimeter PMTs, which will be directly corrected in future data reprocessing.

Systematic uncertainties in the MC-to-data response ratio as a function of \ptleading are shown in \figref{MJB:Systematics_EM}.
They reflect $1\sigma$ uncertainties derived from the MJB event selection, MC modeling, and jet calibration.
Event selection uncertainties are derived by varying the event selections and
examining the impact on the MC-to-data ratio.
The $1\sigma$ variations were conservatively found in 2011 to be $\pm0.1$ rad for $\alpha^\text{MJB}$, $\pm0.5$ rad for $\beta^\text{MJB}$, $\pm5~\GeV$ for $\ptt$, and $\pm0.1$ for $\pta$.
The uncertainty due to MC modeling is taken as the difference in the MJB correction between the nominal \PYTHIA generator and \HERWIGpp.
Uncertainties in the calibration of subleading jets are taken from the input \insitu calibrations,
with each component individually varied by $\pm1\sigma$ and propagated through each MJB iteration.
The JES uncertainties related to the pile-up, punch-through, flavor composition, and flavor response are also propagated through each iteration in the 2015 calibration.
The bootstrapping procedure is used to ensure statistical significance for each systematic uncertainty, with each pseudo-experiment
independently propagated through the iterations.
The combined uncertainty is generally below 1.5\%, consistent with that from the 2011 calibration.

\begin{figure}[!ht]
\begin{center}
  \includegraphics[width=0.64\textwidth]{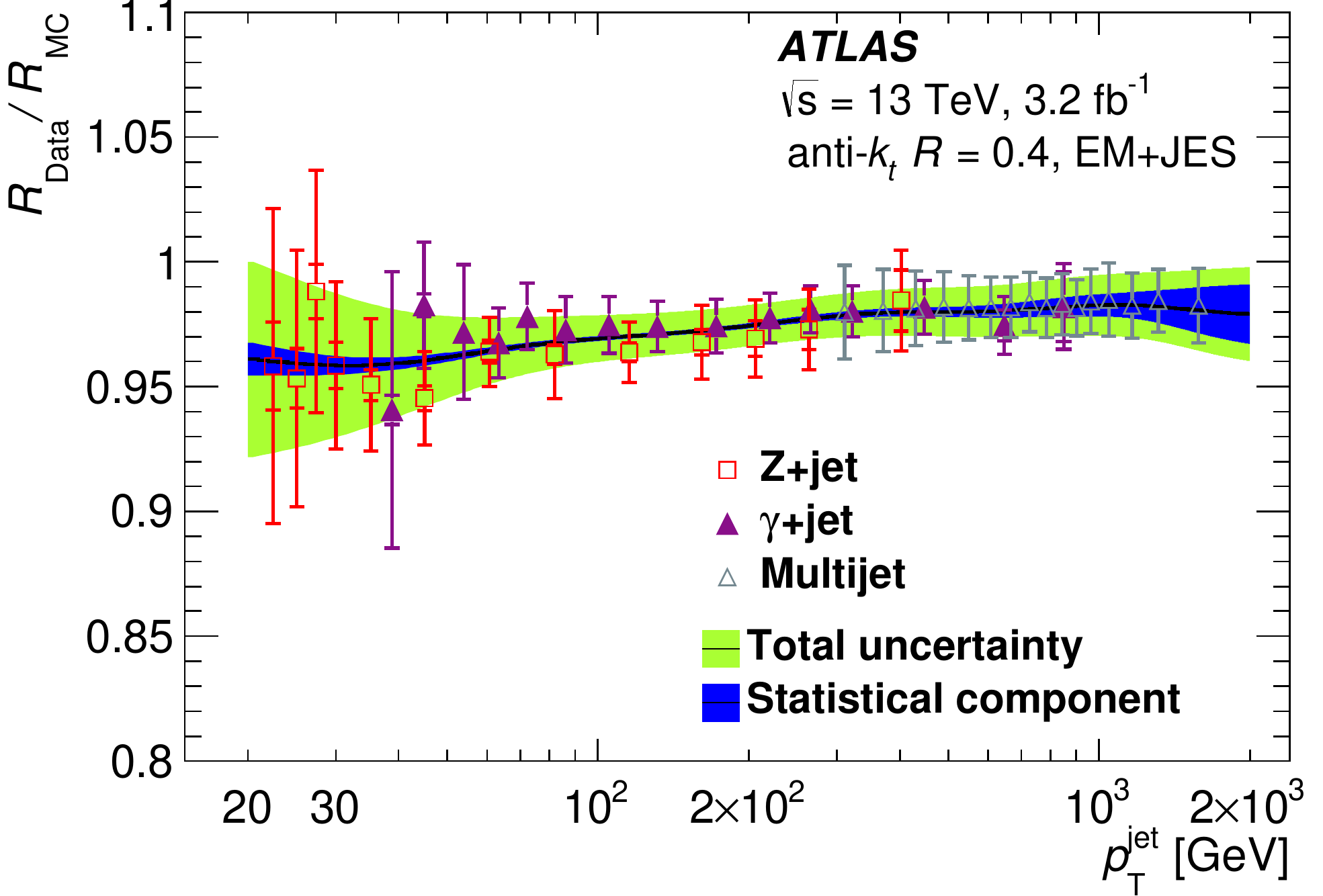}
  \caption{
    Ratio of the EM+JES jet response in data to that in the nominal MC event generator as a function of jet \pt for \Zjet, \gjet, and multijet
    \insitu calibrations.
    The final derived correction (black line) and its statistical (dark blue) and total (light green) uncertainty bands are also shown.
  }
  \label{fig:Comb_resp}
\end{center}
\end{figure}

\subsubsection{\Insitu combination}
\label{sec:insituCombo}

The data-to-MC ratio and the associated systematic uncertainties derived from the orthogonal \Zjet, \gjet, and MJB
calibrations are combined across overlapping regions of jet \pt~\cite{PERF-2012-01}.
For each method, the results are recast into a common, fine binning in \pt by interpolating second-order polynomial splines.
Each \insitu method is assigned a \pt-dependent weight through a $\chi^{2}$ minimization, using as inputs the response ratios and their uncertainties in each \pt bin.
A method's weight is therefore increased in \pt regions of smaller relative uncertainty and smaller bin size, with the combination favoring the method of greatest precision in each region.
The combined data-to-MC ratio is smoothed with a sliding Gaussian kernel to reduce statistical fluctuations.

The combined data-to-MC ratio is shown in \figref{Comb_resp} alongside the \Zjet, \gjet, and MJB ratios in their original binnings.
The inverse of the combined data-to-MC ratio is taken as the \insitu correction applied to data.
The combined correction is 4\% at low \pt and decreases to 2\% at 2~\TeV.
This is a larger correction than seen in 2011, but it is expected due to changes in the simulation of hadronic showers in Geant4
and the slight PMT down-drift in the Tile calorimeter.
The individual \insitu results show good agreement with one another in the various regions of overlapping \pt.
The differences between \insitu measurements are quantified with $\sqrt{\chi^{2}/N_\text{dof}}$, which is generally below 1.

The systematic uncertainties are averaged and smoothed with the same combination procedure through a linear transformation~\cite{PERF-2011-03,PERF-2012-01}.
One at a time, each uncertainty source of each \insitu method is shifted coherently by $1\sigma$, within the method's original binning.
The binning interpolation and combination are then repeated with the nominal weighting of the methods.
In this procedure, the various systematic uncertainties are treated independently of one another and as fully correlated across \pt.
Their independent treatment during the combination allows for alternative correlation assumptions at a later stage,
and the difference between treating correlations before and after the combination are found to be negligible.
The difference between the shifted combined correction factor and the nominal is taken as the $1\sigma$ variation for each uncertainty source.
The \Zgjet uncertainties have a one-to-one correlation with the corresponding uncertainties propagated through the MJB technique.
Therefore, for each of these uncertainties, the correction factors of the \insitu methods are shifted coherently by $1\sigma$, before the binning interpolation and combination steps.

If the nominal corrections from different \insitu methods disagree in a \pt bin,
such that the tension factor $\sqrt{\chi^{2}/N_\text{dof}}$ is above 1,
the uncertainty from each source is scaled by the tension factor in that bin.
In the 2015 calibration, a tension factor of $\sim$1.1 was found only in the narrow \pt region between 45 and 50~\GeV.
As with the nominal result, each systematic uncertainty component is smoothed using a sliding Gaussian kernel.

\section{Systematic uncertainties}
\label{sec:SysUncert}

The final calibration includes a set of 80 JES systematic uncertainty terms propagated from the individual calibrations and studies,
listed in \tabref{JESListTable}.
The majority (67) of uncertainties come from the \Zgjet and MJB \insitu calibrations and account for assumptions made in the
event topology, MC simulation, sample statistics, and propagated uncertainties of the electron, muon, and photon energy scales.
The remaining 13 uncertainties are derived from other sources.
Four pile-up uncertainties are included to account for potential MC mismodeling of \Npv, $\mu$, $\rho$, and the residual \pt dependence.
Three \etainter uncertainties account for potential physics mismodeling, statistical uncertainties, and the method non-closure
in the $2.0 < \abs\etadet < 2.6$ region.
Three additional uncertainties account for differences in the jet response and simulated jet composition of
light-quark, $b$-quark, and gluon-initiated jets.
As in the 2011 calibration, the flavor response uncertainties are derived by comparing the average jet response for each jet flavor using \PYTHIA and \HERWIGpp.
The flavor composition uncertainty is analysis dependent, and is either derived from MC samples in the relevant phase-space, 
or is assumed to be a 50\% quark and 50\% gluon composition with a conservative 100\% uncertainty.
An uncertainty in the GSC punch-through correction is also considered, derived as the maximum difference between the jet responses in
data and MC simulation as a function of the number of muon segments.
One AFII modeling uncertainty accounts for non-closure in the absolute JES calibration of fast-simulation jets, and is applied only to AFII MC samples.
A high-\pt jet uncertainty is derived from single-particle response studies~\cite{PERF-2015-05} and is applied to jets with $\pt > 2~\TeV$, beyond the reach of the \insitu methods.

\begin{table}
\centering
\footnotesize
  \begin{tabular}{|l|l|}
  \hline
  Name & Description \\
  \hline
  \textbf{\Zjet} & \\
  Electron scale          & Uncertainty in the electron energy scale   \\
  Electron resolution     & Uncertainty in the electron energy resolution  \\
  Muon scale            & Uncertainty in the muon momentum scale  \\
  Muon resolution (ID)      & Uncertainty in muon momentum resolution in the ID  \\
  Muon resolution (MS)      & Uncertainty in muon momentum resolution in the MS  \\
  MC generator      & Difference between MC event generators  \\
  JVT               & Jet vertex tagger uncertainty  \\
  $\Delta\phi$      & Variation of $\Delta\phi$ between the jet and $Z$ boson \\
  2nd jet veto      & Radiation suppression through second-jet veto  \\
  Out-of-cone       & Contribution of particles outside the jet cone  \\
  Statistical       & Statistical uncertainty over 13 regions of jet \pt  \\
  \hline
  \textbf{\gjet} & \\
  Photon scale        & Uncertainty in the photon energy scale  \\
  Photon resolution   & Uncertainty in the photon energy resolution  \\
  MC generator    & Difference between MC event generators  \\
  JVT             & Jet vertex tagger uncertainty  \\
  $\Delta\phi$    & Variation of $\Delta\phi$ between the jet and $\gamma$ \\
  2nd jet veto    & Radiation suppression through second-jet veto  \\
  Out-of-cone     & Contribution of particles outside the jet cone  \\
  Photon purity   & Purity of sample in \gjet balance  \\
  Statistical     & Statistical uncertainty over 15 regions of jet \pt  \\
  \hline
  \textbf{Multijet balance} & \\
  $\alpha^\text{MJB}$ selection  & Angle between leading jet and recoil system  \\
  $\beta^\text{MJB}$ selection   & Angle between leading jet and closest subleading jet  \\
  MC generator            & Difference between MC event generators  \\
  \pta selection          & Second jet's \pt contribution to the recoil system  \\
  Jet \pt threshold       & Jet \pt threshold  \\
  Statistical components  & Statistical uncertainty over 16 regions of \ptleading  \\
  \hline
  \textbf{\etainter} & \\
  Physics mismodeling   & Envelope of the MC, pile-up, and event topology variations \\
  Non-closure           & Non-closure of the method in the $2.0 < \abs\etadet < 2.6$ region \\
  Statistical component & Statistical uncertainty  \\
  \hline
  \textbf{Pile-up} & \\
  $\mu$ offset    & Uncertainty of the $\mu$ modeling in MC simulation  \\
  \Npv offset     & Uncertainty of the \Npv modeling in MC simulation  \\
  $\rho$ topology & Uncertainty of the per-event \pt density modeling in MC simulation  \\
  \pt dependence  & Uncertainty in the residual \pt dependence  \\
  \hline
  \textbf{Jet flavor} & \\
  Flavor composition  & Uncertainty in the jet composition between quarks and gluons \\
  Flavor response     & Uncertainty in the jet response of gluon-initiated jets  \\
  $b$-jet               & Uncertainty in the jet response of $b$-quark-initiated jets   \\
  \hline
  Punch-through       & Uncertainty in GSC punch-through correction  \\
  \hline
  AFII non-closure    & Difference in the absolute JES calibration using AFII  \\
  \hline
  Single-particle response & High-\pt jet uncertainty from single-particle and test-beam measurements \\
  \hline
 \end{tabular}
\caption{
  Summary of the systematic uncertainties in the JES, including those propagated from electron, photon, and muon
  energy scale calibrations~\cite{PERF-2013-05,PERF-2015-10}.
 \label{tab:JESListTable}}
\end{table}

\begin{figure}[!ht]
\begin{center}
  \subfigure[]{
    \label{fig:Comb_syst:pt}
    \includegraphics[width=0.47\textwidth]{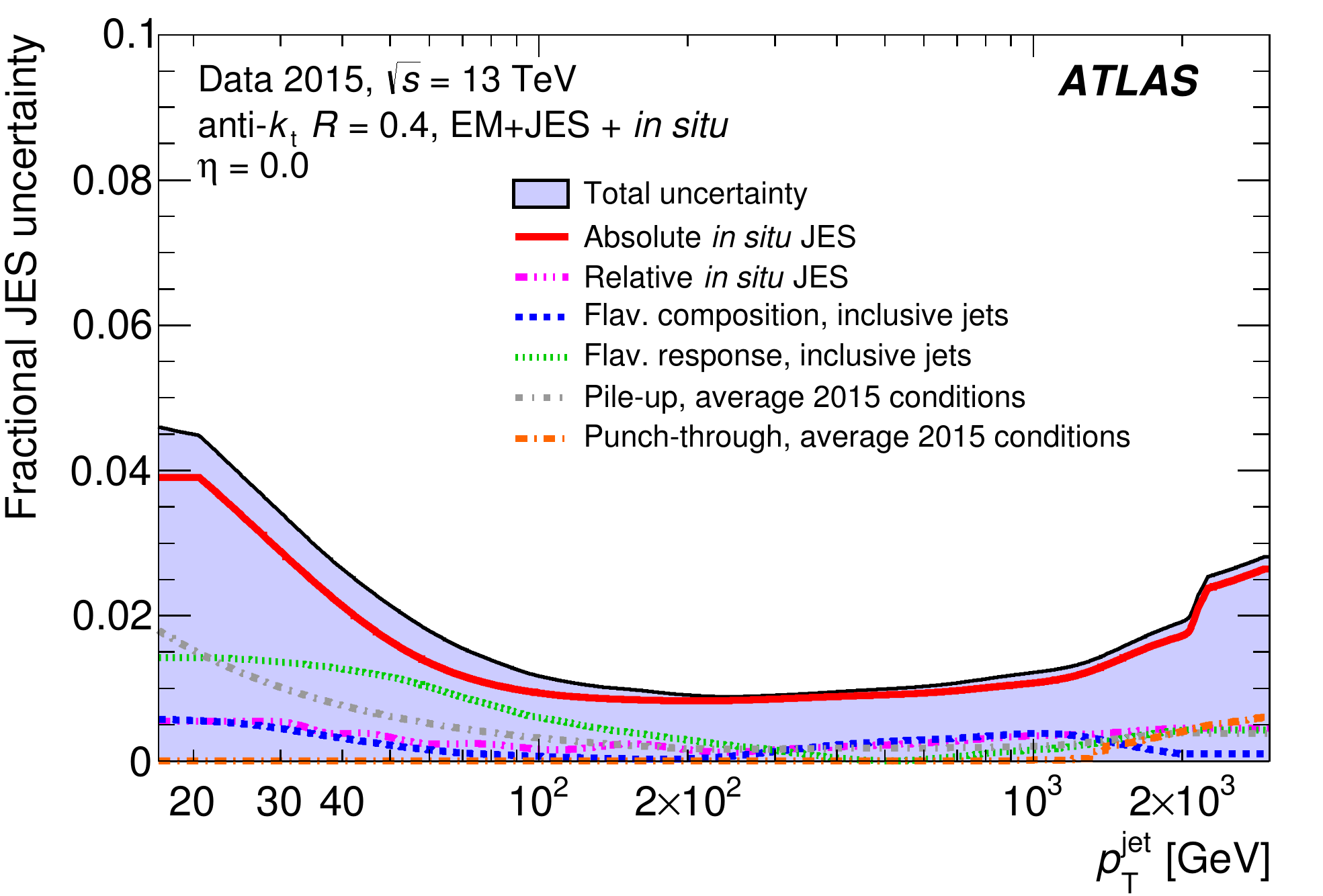}
  }
  \subfigure[]{
    \label{fig:Comb_syst:eta}
    \includegraphics[width=0.47\textwidth]{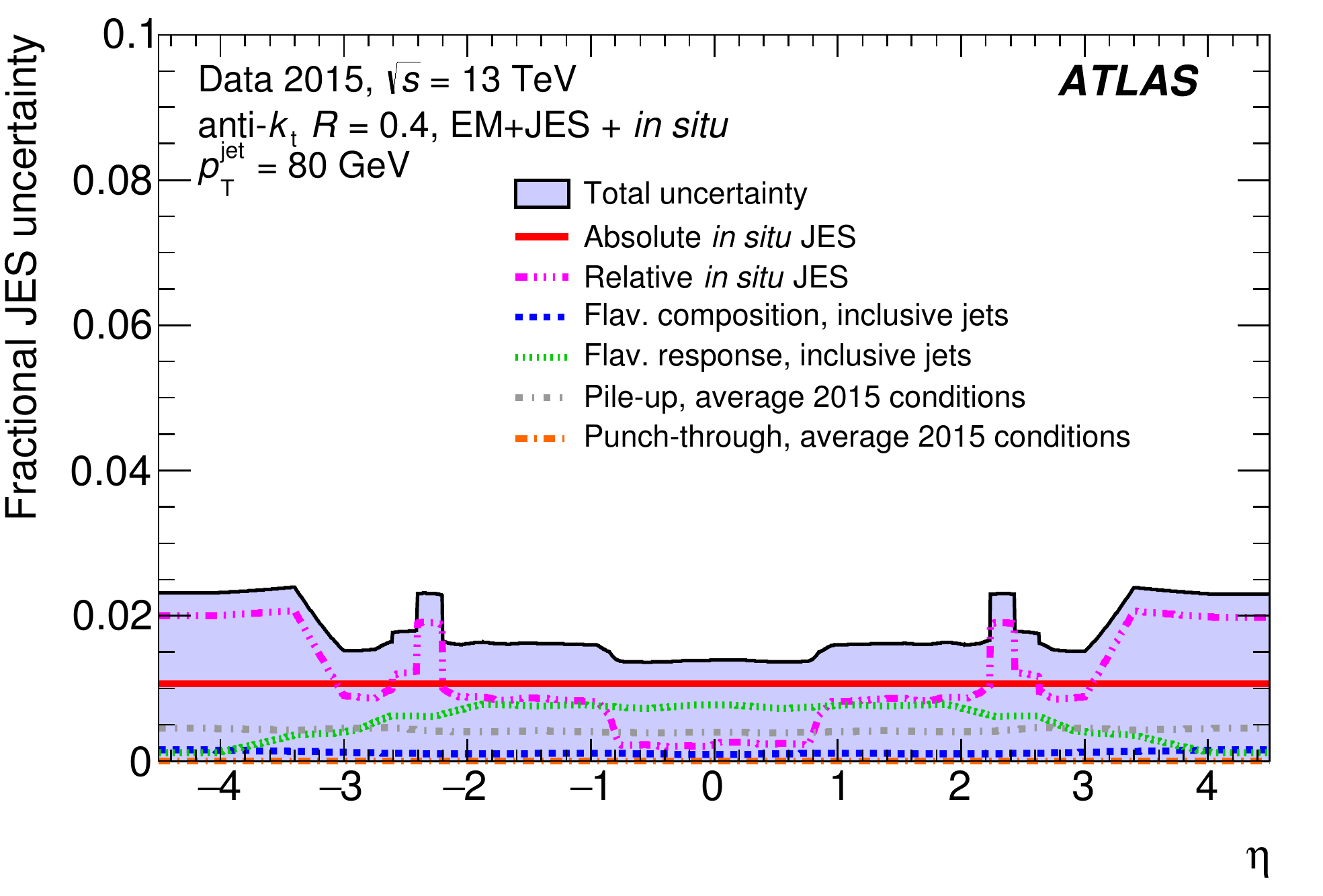}
  }
  \caption{Combined uncertainty in the JES of fully calibrated jets as a function of
    \subref{fig:Comb_syst:pt} jet \pt at $\eta = 0$ and \subref{fig:Comb_syst:eta} $\eta$ at $\pt = 80~\GeV$.
    Systematic uncertainty components include pile-up, punch-through, and uncertainties propagated from the \Zgjet and MJB (absolute \insitu JES)
    and \etainter (relative \insitu JES).
    The flavor composition and response uncertainties assume a quark and gluon composition taken from \PYTHIA dijet MC simulation (inclusive jets).
  \label{fig:Comb_syst}}
\end{center}
\end{figure}

The full combination of all uncertainties is shown in \figref{Comb_syst} as a function of \pt at $\eta=0$ and
as a function of $\eta$ at $\pt=80$~\GeV, assuming a flavor composition taken from the inclusive dijet selection in \PYTHIA.
Each uncertainty is generally treated independently of the others but fully correlated across \pt and $\eta$.
Exceptions are the electron and photon energy scale measurements, which are treated as fully correlated.
The uncertainty is largest at low \pt, starting at 4.5\% and decreasing to 1\% at 200~\GeV.
It rises after 200~\GeV due to the statistical uncertainties related to the \insitu calibrations, and
increases sharply after 2~\TeV where MJB measurements end and larger uncertainties are taken from the single-particle response.
The uncertainty is fairly constant as a function of $\eta$ and reaches a maximum of 2.5\% for the most forward jets.
A sharp feature can be seen in the region $2.0 < \abs\eta < 2.6$ due to the non-closure uncertainty of the \etainter.

The complete set of systematic uncertainties provides a detailed understanding of the many factors that influence the JES.
Uncertainties are generally derived in specific regions of jet \pt and $\eta$,
and the correlation of uncertainties between two jets with different kinematics can vary in strength.
For the set of variables $\chiNP$, the Pearson correlation coefficient ($C$) between two jets is used to quantify the correlations, and is defined as

\begin{equation}
  C(\chiNP_{1},\chiNP_{2}) = \frac{\mathrm{Cov}(\chiNP_{1},\chiNP_{2})}{\sqrt{ \mathrm{Cov}(\chiNP_{1},\chiNP_{1}) \times \mathrm{Cov}(\chiNP_{2},\chiNP_{2}) }},
\label{eq:Corr}
\end{equation}
where Cov is the covariance of the systematic uncertainties between the two sets of variables.

\begin{figure}[!ht]
\begin{center}
  \subfigure[]{
    \label{fig:NP:fullpt}
    \includegraphics[width=0.47\textwidth]{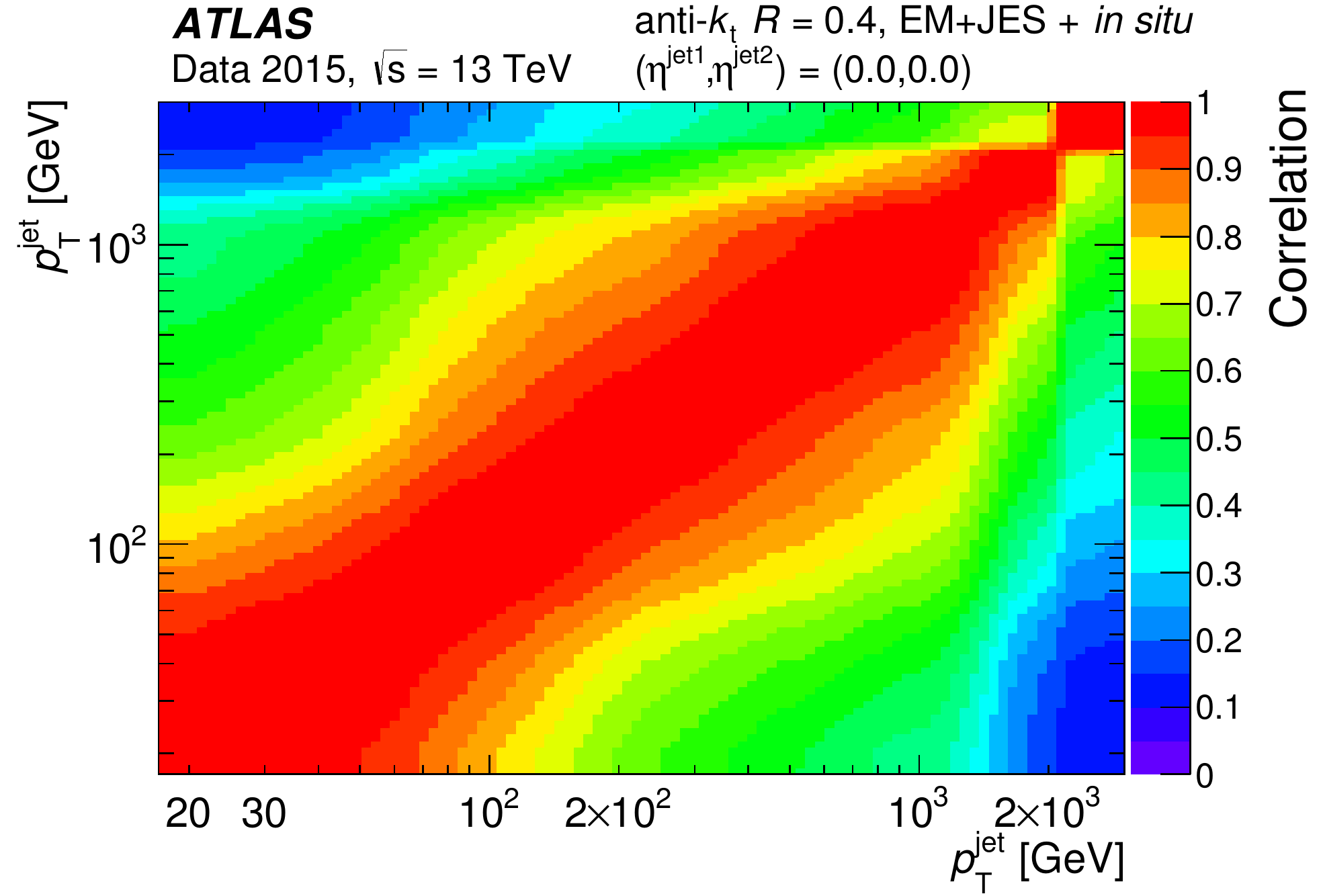}
  }
  \subfigure[]{
    \label{fig:NP:fulleta}
    \includegraphics[width=0.47\textwidth]{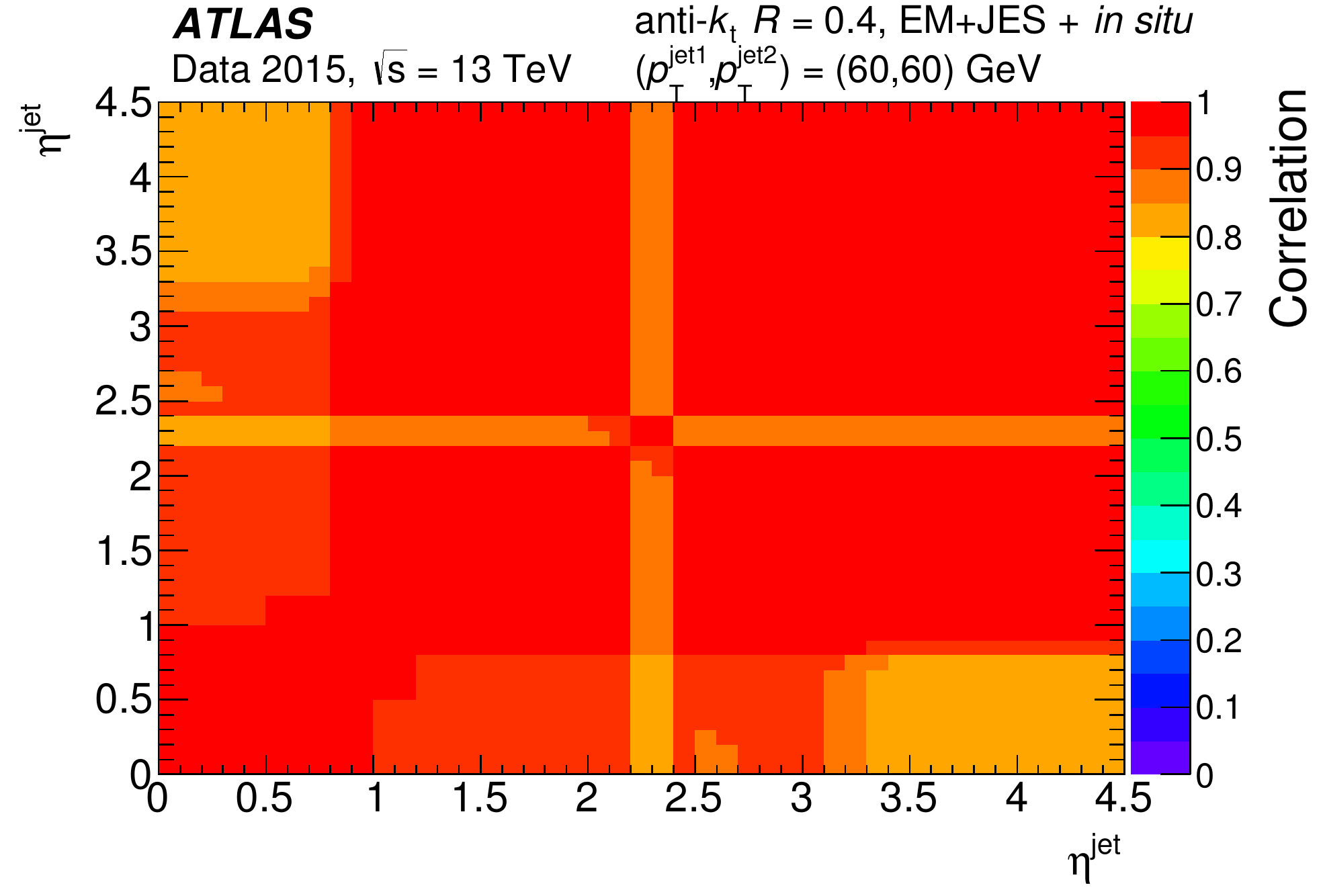}
  }
  \caption{
    The full correlation matrix between two jets using all 80 uncertainty sources as a
    function of \subref{fig:NP:fullpt} $\pt^{\text{jet}}$ for $\eta^{\text{jet1}}=\eta^{\text{jet2}}=0$
    and \subref{fig:NP:fulleta} $\eta^{\text{jet}}$ for $\pt^{\text{jet1}}=\pt^{\text{jet2}}=60$~\GeV.
    Regions of strong correlation are visible at low, medium, high, and very high \pt
    corresponding to the \Zjet, \gjet, and MJB calibrations and the single-particle response, as well as
    in the central and forward jet $\eta$ regions from the \etainter.
  }
  \label{fig:NP}
\end{center}
\end{figure}

The jet--jet correlation matrix, including all 80 uncertainties, is shown as a function of jet \pt ($\eta^\text{jet1}=\eta^\text{jet2}=0$) in \figref{NP:fullpt} and
as a function of jet $\eta$ ($\pt^\text{jet1}=\pt^\text{jet2}=60~\GeV$) in \figref{NP:fulleta}.
Regions of strong correlation ($C\sim1$) are shown in mid-tone red, and of weak correlation ($C\sim0$) in dark blue.
In the \pt correlation map, features are visible at low, medium, high, and very high \pt, corresponding to the kinematic phase space
of the \insitu \pt-balance calibrations and the single-particle response.
In the $\eta$ correlation map the correlation is strongest in the central and forward $\eta$ regions of the \etainter.
Strong jet--jet correlations are seen as a function of $\eta$ due to the dominance of the MC modeling term in the \etainter.
Correlations due to the non-closure uncertainty, being most significant for $2.2 < \abs\eta < 2.4$, are
seen to be localized in a narrow $\eta$ region, as expected.

While the 80 uncertainties provide the most accurate understanding of the JES uncertainty,
a number of physics analyses would be hampered by the implementation and evaluation of them all.
Furthermore, many would receive no discernible benefit from the rigorous conservation of all correlations.
For these cases a reduced set of nuisance parameters (NPs) is made available that seeks to
preserve as precisely as possible the correlations across jet \pt and $\eta$.

\begin{figure}[!ht]
\begin{center}
  \subfigure[]{
    \label{fig:NP:global}
    \includegraphics[width=0.47\textwidth]{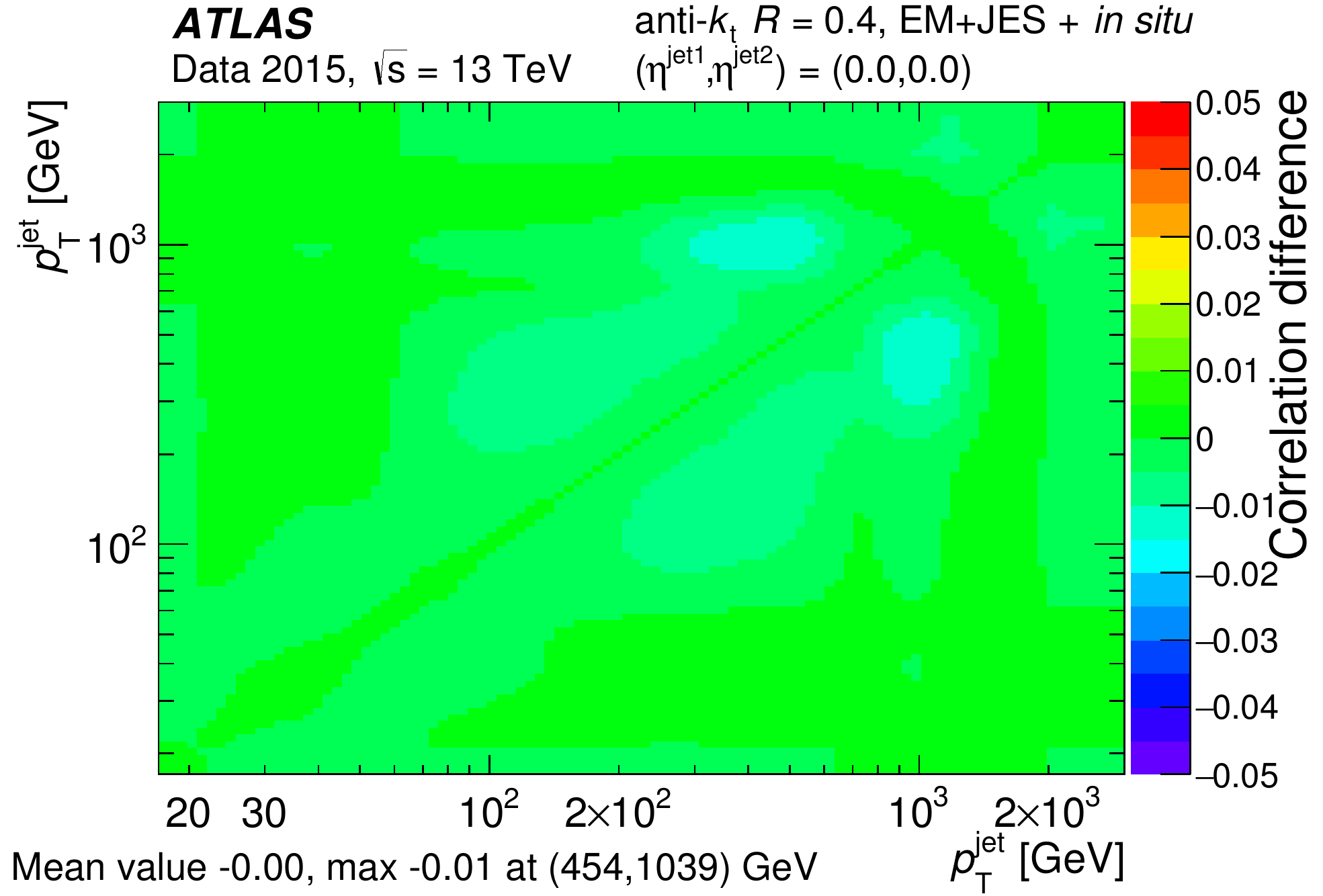}
  }
  \subfigure[]{
    \label{fig:NP:strong}
    \includegraphics[width=0.47\textwidth]{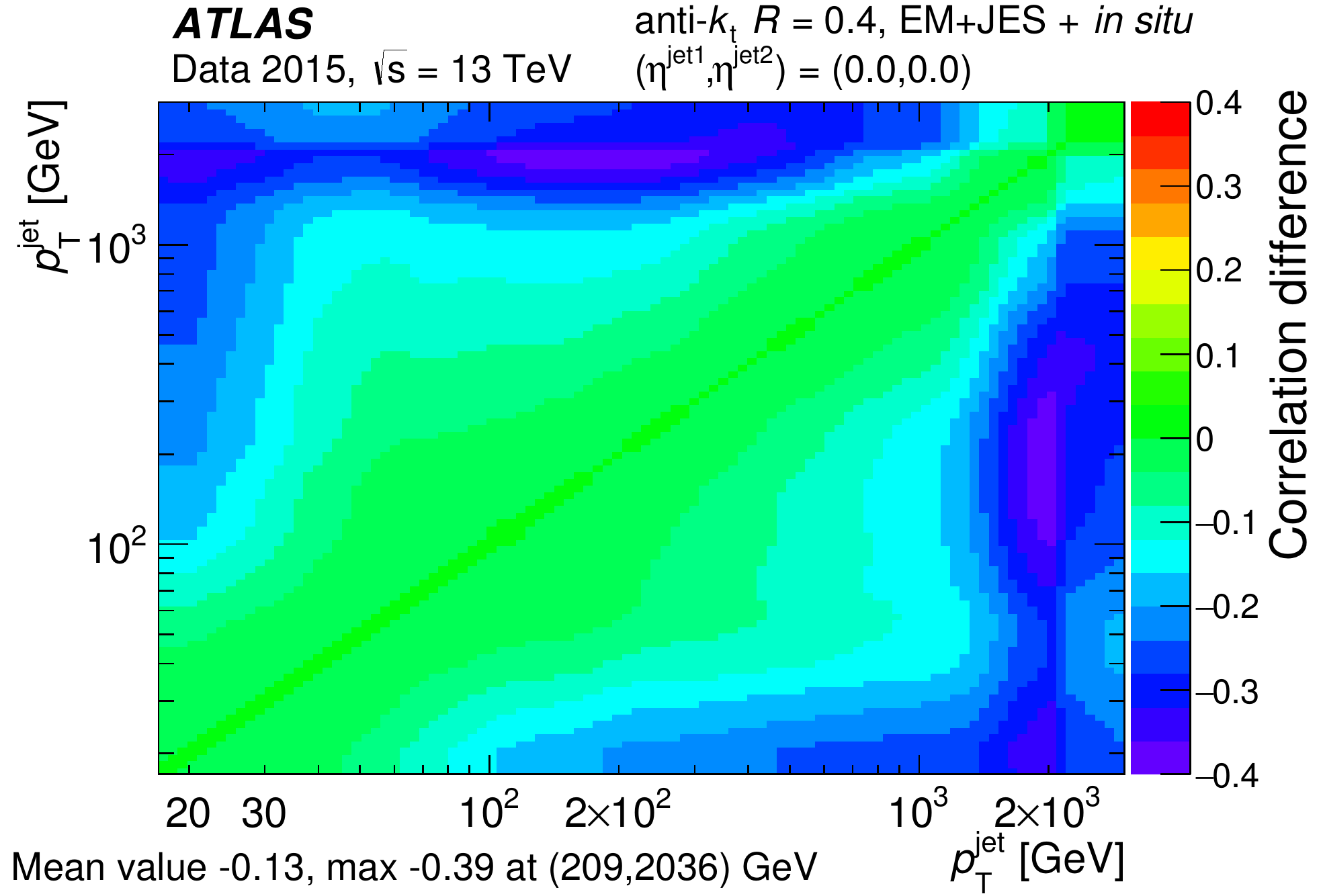}
  }
  \caption{
    Jet-jet correlation losses after applying \subref{fig:NP:global} the global reduction and \subref{fig:NP:strong} subsequent strong reduction
    as a function of $\pt^{\text{jet}}$ for $\eta^{\text{jet1}}=\eta^{\text{jet2}}=0$.
    The correlation loss is relatively minor from the global reduction and larger from the strong reduction in certain kinematic regions.
  }
  \label{fig:NPred}
\end{center}
\end{figure}

As a first step, the global reduction~\cite{PERF-2012-01} is performed through an eigen-decomposition of the 67 \pt-dependent \insitu uncertainties following from the \Zgjet and MJB calibrations.
The five principal components of greatest magnitude are kept separate and the remaining components are quadratically combined into a single NP,
treating them as independent of one another.
This reduces the number of independent \insitu uncertainty sources from 67 to 6 NPs, with only percent-level losses to the correlations between jets.
The difference in correlation, given by \eref{Corr}, between the full NP representation and the reduced representation as a function of jet \pt is given in \figref{NP:global},
showing the losses to be small and constrained in kinematic phase space.

A new procedure is introduced for 2015 data to further reduce the remaining 19 NPs (6 \insitu \pt-balance NPs and 13 others)
into a smaller, strongly reduced representation.
Various combinations of the remaining NPs into three components are attempted, and NPs within a single component are quadratically combined.
The combinations attempt to group NPs into \pt and $\eta$ regions where they are most relevant,
thereby minimizing the correlation loss and reducing the potential for artificial correlation structures across large regions of jet kinematic phase space.
Combinations that group NPs that are dominant in low-, medium-, and high-\pt kinematic regimes are therefore generally favored.
The \etainter non-closure uncertainty (\secref{etaInter}), being fairly large and localized,
and the AFII uncertainty, being specific to a certain type of MC simulation, are not included in this procedure.
This procedure is performed using \PYTHIA MC simulation, assuming a conservative 50\% quark and 50\% gluon composition with a 100\% uncertainty.

The correlation loss between a strongly reduced representation $\text{NP}_\text{red}$ and the full representation $\text{NP}_\text{full}$ is generally non-negligible,
as seen in the matrix of the jet--jet \pt correlation differences shown in \figref{NP:strong}.
For two jets with $\eta$ = 0, the maximum (mean) \pt correlation loss is $-0.39$ ($-0.13$), and is largest between jets
in very different kinematic phase space.
This simple mean is taken as the average correlation loss over the fine logarithmic \pt bins,
excluding bins in kinematically forbidden regions.
Sensitivity to this correlation loss is analysis dependent and is determined by the regions in jet \pt--$\eta$ phase space
where the analysis events fall.
To allow analyses to probe their sensitivity to this loss, a set of four different strongly reduced representations $\{\text{NP}_\text{red}\}$
is generated which varies the regions of greatest correlation loss between them.
Each $\text{NP}_\text{red}$ combines the components in a unique way, with different kinematic regions becoming better or worse
descriptions of the full correlation matrix.
The sensitivity of an analysis to the correlation loss can be quantified by examining the effect of each $\text{NP}_\text{red}$ on the final analysis observable.
The four $\text{NP}_\text{red}$ are each derived to focus on one of the following correlation scenarios:
\begin{enumerate}
\item the general representation with low-, medium-, and high-\pt kinematic regimes;
\item preservation of low-\pt vs medium-\pt correlation structure as well as $\eta$ dependencies;
\item preservation of medium-\pt vs high-\pt correlation structure;
\item preservation of very high-\pt correlation structure.
\end{enumerate}

When deriving a reduced representation, it can be useful to highlight exceptional uncertainties or vary the way in which they are combined.
An uncertainty may exhibit a large anti-correlation across \pt or $\eta$, and the anti-correlation
information is lost when summed in quadrature with other uncertainties to form a single NP.
If such an uncertainty is non-negligible, it is useful to isolate it as a single strongly reduced NP.
For uncertainties derived from the comparison of two MC event generators, the correlation structure is not well defined.
These NPs can be split into two identical components of complementary weight, such that their combination
sums to the original uncertainty for all points in the \pt--$\eta$ phase space.
The split NP can then be divided between two strongly reduced NPs, changing the correlation information in certain kinematic regions.
A reduced representation can also recover the correlation information from globally subdominant eigenvectors that were initially combined in the preceding eigen-decomposition.
These eigenvectors are smaller overall than others but may be dominant for specific kinematic regions.
By keeping these eigenvectors separate until the strong reduction procedure, the correlation structure in
kinematic regions of interest can be better probed, at the expense of an increased loss in the overall global correlation structure.

To ensure the set of four reduced representations $\{\text{NP}_\text{red}\}$ is suitable in bracketing the full correlation matrix, a metric is defined
to quantify the uncovered correlation loss of any derived set.
The metric measures the maximum correlation difference between any two reduced representations $\text{NP}_\text{red} \in \{\text{NP}_\text{red}\}$
and compares it with the smallest difference between the full representation $\text{NP}_\text{full}$ and any $\text{NP}_\text{red}$.
If the difference between any two $\text{NP}_\text{red}$ is larger than that of any $\text{NP}_\text{red}$ and $\text{NP}_\text{full}$,
then analyses that bracket their sensitivity to correlation loss with $\{\text{NP}_\text{red}\}$ are conservative with respect to any differences with the full representation.
The metric to quantify the uncovered correlation loss of any derived $\{\text{NP}_\text{red}\}$ is defined as

\begin{equation}
  \min\limits_{i \in \{\text{NP}_\text{red}\}}\left|C_\text{full} - C_\text{red}^{i}  \right| -  \max\limits^{i\neq j}_{i,j \in \{\text{NP}_\text{red}\}}\left|C_\text{red}^{i} - C_\text{red}^{j}\right|,
\label{eq:metric2}
\end{equation}
where $C_\text{full}$ is the correlation coefficient for the full NP set and $C_\text{red}$ is that of a reduced NP representation.
The metric is calculated throughout the jet--jet \pt--$\eta$ phase space, and is not allowed to be greater than zero.

\begin{figure}[!ht]
\begin{center}
    \includegraphics[width=0.9\textwidth]{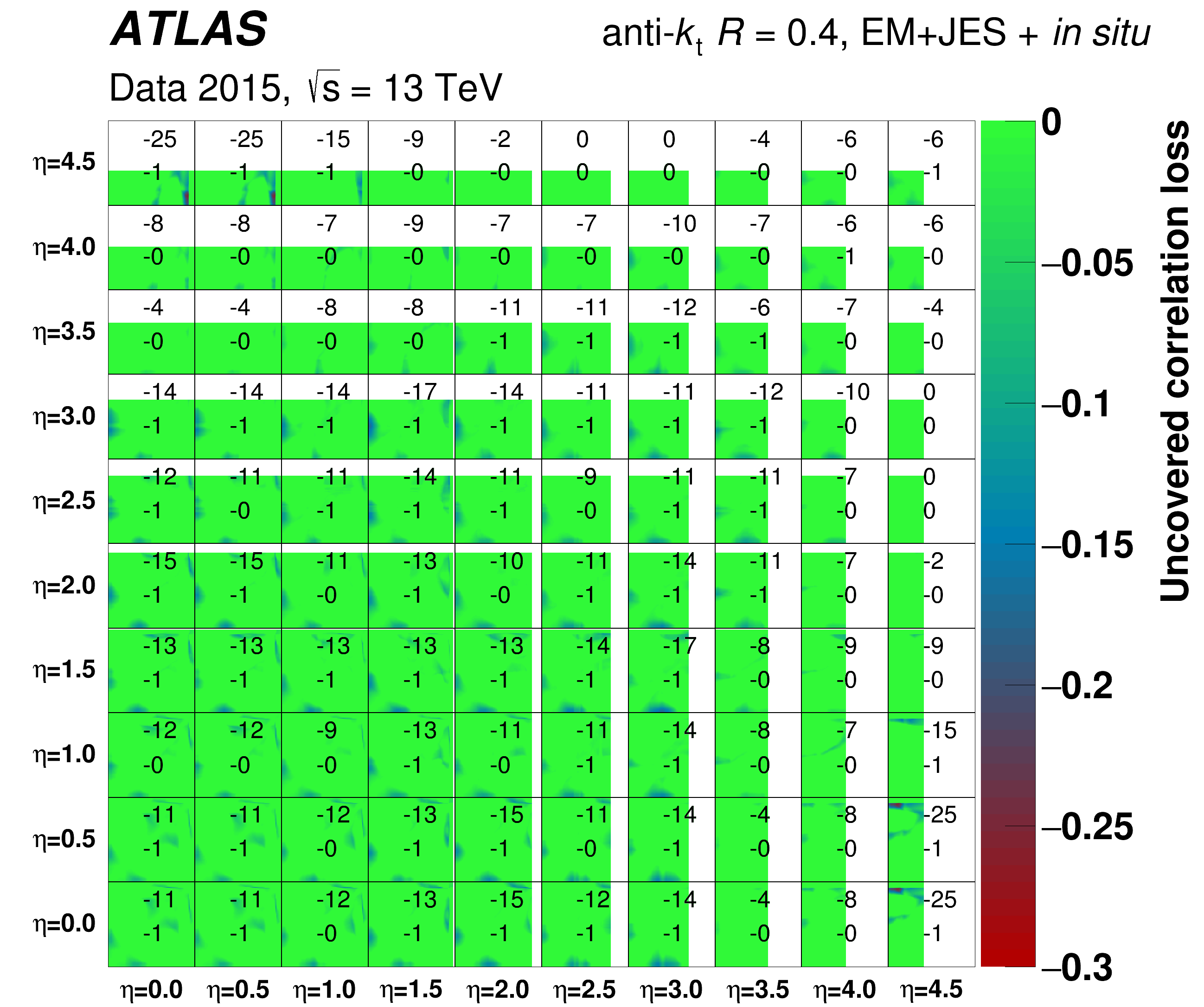}
  \caption{
  Uncovered jet--jet correlation loss between the full NP representation and the set of strongly reduced representations, showing
  regions which are not fully covered by the strongly reduced set of four representations.
  The uncovered correlation loss is calculated by the metric given in \eref{metric2}.
  The uncovered correlation loss is explored in the four-dimensional jet--jet \pt--$\eta$ phase space.
  Each subplot shows the uncovered correlation loss as a function of \pt, and subplots are shown for several regions of $\eta$ in steps of $\Delta\eta = 0.5$.
  White regions represent the kinematically forbidden phase space beyond the reach of $\sqrt{s}=13~\TeV$.
  The top (bottom) number in each subplot gives the maximum (mean) uncovered correlation loss, multiplied by a factor of 100 for visibility,
  with the mean excluding kinematically forbidden regions.
  }
  \label{fig:NP_4D}
\end{center}
\end{figure}

This uncovered correlation loss is shown in \figref{NP_4D} for several points in the four-dimensional jet--jet \pt--$\eta$ phase space.
It is shown as a function of \pt for several distinct regions of $\eta$ in steps of $\Delta\eta = 0.5$.
For each $\eta$ region, the maximum correlation loss not covered by differences between the reduced representations
is above the level of $-0.3$ with a mean at or above $-0.01$.
The regions of maximum difference are very limited in kinematic phase space and therefore have a minimal impact,
with the strongly reduced representation procedure probing almost all of the JES correlation structure.
The majority of ATLAS searches using 2015 data have been shown to be insensitive to this limited loss of correlation information and have used the strongly reduced NPs successfully,
such as the dijet~\cite{EXOT-2015-02} and multijet~\cite{EXOT-2015-09} resonance searches.

\FloatBarrier
\section{Conclusions}
\label{sec:conclusion}

The derivation of the 2015 ATLAS calibration of the jet energy scale is presented for EM-scale \antikt $R=0.4$ jets.
An area-based pile-up correction and a pile-up-sensitive residual correction are derived to
reduce contamination from the busy detector environment at a bunch spacing of 25~ns.
Absolute jet energy scale and $\eta$ calibrations are derived from Monte Carlo simulation to correct the jet four-momentum to the particle-level energy scale
and to improve the jet angular resolution.
The global sequential calibration is derived from \pt-sensitive observables to improve the jet resolution
and to account for the differing energy response between quark- and gluon-initiated jets.

\Insitu calibrations are derived using \intLumi of $\sqrt{s}=13~\TeV$ proton--proton collision data collected by ATLAS in 2015 at the LHC.
Dijet events are selected to measure the \pt- and $\eta$-dependent response of forward jets with respect to central jets.
A \pt-dependent correction is derived by balancing the \pt of jets against reference photons and $Z$ bosons decaying into electrons and muons.
A final correction is derived for higher-\pt jets through multijet events
in which the highest-\pt jet is significantly more energetic than the others.
The \insitu corrections are combined in their overlapping \pt ranges to provide a single consistent
calibration at a level of 4\% at 20~\GeV\ and 2\% at 2~\TeV.

The uncertainty in the jet energy scale is consistent with previous results in 2011 using 7~\TeV data,
and is at a level of 4.5\% at 20~\GeV, 1\% at 200~\GeV, and 2\% at 2~\TeV for an inclusive dijet sample.
The uncertainties are fairly constant with respect to $\eta$, and a dedicated uncertainty is introduced for 2.0 < \abs\eta < 2.6
to account for details in the calorimeter energy reconstruction.
A new method for combining systematic uncertainties into a strongly reduced set while preserving correlations is described.
The full set of 80 uncertainties is reduced to five, and the correlation information loss is probed through a set of four unique combination scenarios.

\section*{Acknowledgements}

We thank CERN for the very successful operation of the LHC, as well as the
support staff from our institutions without whom ATLAS could not be
operated efficiently.

We acknowledge the support of ANPCyT, Argentina; YerPhI, Armenia; ARC, Australia; BMWFW and FWF, Austria; ANAS, Azerbaijan; SSTC, Belarus; CNPq and FAPESP, Brazil; NSERC, NRC and CFI, Canada; CERN; CONICYT, Chile; CAS, MOST and NSFC, China; COLCIENCIAS, Colombia; MSMT CR, MPO CR and VSC CR, Czech Republic; DNRF and DNSRC, Denmark; IN2P3-CNRS, CEA-DSM/IRFU, France; SRNSF, Georgia; BMBF, HGF, and MPG, Germany; GSRT, Greece; RGC, Hong Kong SAR, China; ISF, I-CORE and Benoziyo Center, Israel; INFN, Italy; MEXT and JSPS, Japan; CNRST, Morocco; NWO, Netherlands; RCN, Norway; MNiSW and NCN, Poland; FCT, Portugal; MNE/IFA, Romania; MES of Russia and NRC KI, Russian Federation; JINR; MESTD, Serbia; MSSR, Slovakia; ARRS and MIZ\v{S}, Slovenia; DST/NRF, South Africa; MINECO, Spain; SRC and Wallenberg Foundation, Sweden; SERI, SNSF and Cantons of Bern and Geneva, Switzerland; MOST, Taiwan; TAEK, Turkey; STFC, United Kingdom; DOE and NSF, United States of America. In addition, individual groups and members have received support from BCKDF, the Canada Council, CANARIE, CRC, Compute Canada, FQRNT, and the Ontario Innovation Trust, Canada; EPLANET, ERC, ERDF, FP7, Horizon 2020 and Marie Sk{\l}odowska-Curie Actions, European Union; Investissements d'Avenir Labex and Idex, ANR, R{\'e}gion Auvergne and Fondation Partager le Savoir, France; DFG and AvH Foundation, Germany; Herakleitos, Thales and Aristeia programmes co-financed by EU-ESF and the Greek NSRF; BSF, GIF and Minerva, Israel; BRF, Norway; CERCA Programme Generalitat de Catalunya, Generalitat Valenciana, Spain; the Royal Society and Leverhulme Trust, United Kingdom.

The crucial computing support from all WLCG partners is acknowledged gratefully, in particular from CERN, the ATLAS Tier-1 facilities at TRIUMF (Canada), NDGF (Denmark, Norway, Sweden), CC-IN2P3 (France), KIT/GridKA (Germany), INFN-CNAF (Italy), NL-T1 (Netherlands), PIC (Spain), ASGC (Taiwan), RAL (UK) and BNL (USA), the Tier-2 facilities worldwide and large non-WLCG resource providers. Major contributors of computing resources are listed in Ref.~\cite{ATL-GEN-PUB-2016-002}.

\printbibliography

\newpage
\begin{flushleft}
{\Large The ATLAS Collaboration}

\bigskip

M.~Aaboud$^\textrm{\scriptsize 137d}$,
G.~Aad$^\textrm{\scriptsize 88}$,
B.~Abbott$^\textrm{\scriptsize 115}$,
J.~Abdallah$^\textrm{\scriptsize 8}$,
O.~Abdinov$^\textrm{\scriptsize 12}$$^{,*}$,
B.~Abeloos$^\textrm{\scriptsize 119}$,
S.H.~Abidi$^\textrm{\scriptsize 161}$,
O.S.~AbouZeid$^\textrm{\scriptsize 139}$,
N.L.~Abraham$^\textrm{\scriptsize 151}$,
H.~Abramowicz$^\textrm{\scriptsize 155}$
% ...
\bigskip
\\
% ...
$^{8}$ Department of Physics, The University of Texas at Arlington, Arlington TX, United States of America\\
% ...
$^{12}$ Institute of Physics, Azerbaijan Academy of Sciences, Baku, Azerbaijan\\
% ...
$^{88}$ CPPM, Aix-Marseille Universit{\'e} and CNRS/IN2P3, Marseille, France\\
% ...
$^{115}$ Homer L. Dodge Department of Physics and Astronomy, University of Oklahoma, Norman OK, United States of America\\
% ...
$^{119}$ LAL, Univ. Paris-Sud, CNRS/IN2P3, Universit{\'e} Paris-Saclay, Orsay, France\\
% ...
$^{137}$ $^{(a)}$ Facult{\'e} des Sciences Ain Chock, R{\'e}seau Universitaire de Physique des Hautes Energies - Universit{\'e} Hassan II, Casablanca; $^{(b)}$ Centre National de l'Energie des Sciences Techniques Nucleaires, Rabat; $^{(c)}$ Facult{\'e} des Sciences Semlalia, Universit{\'e} Cadi Ayyad, LPHEA-Marrakech; $^{(d)}$ Facult{\'e} des Sciences, Universit{\'e} Mohamed Premier and LPTPM, Oujda; $^{(e)}$ Facult{\'e} des sciences, Universit{\'e} Mohammed V, Rabat, Morocco\\
% ...
$^{139}$ Santa Cruz Institute for Particle Physics, University of California Santa Cruz, Santa Cruz CA, United States of America\\
% ...
$^{151}$ Department of Physics and Astronomy, University of Sussex, Brighton, United Kingdom\\
% ...
$^{155}$ Raymond and Beverly Sackler School of Physics and Astronomy, Tel Aviv University, Tel Aviv, Israel\\
% ...
$^{161}$ Department of Physics, University of Toronto, Toronto ON, Canada\\
% ...
$^{a}$ Also at Department of Physics, King's College London, London, United Kingdom\\
$^{b}$ Also at Institute of Physics, Azerbaijan Academy of Sciences, Baku, Azerbaijan\\
$^{c}$ Also at Novosibirsk State University, Novosibirsk, Russia\\
$^{d}$ Also at TRIUMF, Vancouver BC, Canada\\
$^{e}$ Also at Department of Physics {\&} Astronomy, University of Louisville, Louisville, KY, United States of America\\
% ...
$^{*}$ Deceased
\end{flushleft}


\end{document}